\RequirePackage{fix-cm}
\documentclass[smallextended]{svjour3}       
\smartqed  
\usepackage{graphicx}
\usepackage{xcolor}
\usepackage{amsmath}
\usepackage{amssymb}
\usepackage{xspace}
\usepackage{physics}
\usepackage{bm}
\usepackage{hyperref}
\usepackage{mathtools}
\usepackage[toc]{appendix}
\usepackage{mathrsfs}
\newcommand{\bmxi}{{\bm{\xi}}}
\newcommand{\bmxizero}{{\bm{\xi} = 0}}
\newcommand{\bfD}{{\bf{D}}}
\newcommand{\bfr}{\mathbf{r}}

\newcommand{\bfkap}{\bm{\kappa}}

\newcommand{\bfw}{\mathbf{w}}
\newcommand{\ttw}{{\tt{w}}}
\newcommand{\xim}{{\xi_{-}}}
\newcommand{\xip}{{\xi_{+}}}
\newcommand{\xipm}{{\xi_{\pm}}}
\newcommand{\Nm}{{N-1}}
\newcommand{\Np}{{N+1}}
\newcommand{\Npm}{{N\pm 1}}

\newcommand{\dwI}[1]{\ensuremath{\dfrac{\partial #1}{\partial \ttw_I}}}

\newcommand{\dxip}[1]{\ensuremath{\dfrac{\partial #1}{\partial \xip}}}
\newcommand{\dxim}[1]{\ensuremath{\dfrac{\partial #1}{\partial \xim}}}
\newcommand{\dxipm}[1]{\ensuremath{\dfrac{\partial #1}{\partial \xipm}}}

\newcommand{\Hxc}{{\rm Hxc}}
\newcommand{\xc}{{\rm xc}}

\newcommand{\ddroit}{{\rm d}}
\newcommand{\be}{\begin{eqnarray}}
\newcommand{\ee}{\end{eqnarray}}
\newcommand{\ie}{{\it{i.e.}}}
\DeclarePairedDelimiter\ceil{\lceil}{\rceil}

\DeclareMathOperator*{\argmin}{arg\,min}
\begin{document}

\title{Ensemble Density Functional Theory of Neutral and Charged Excitations
}
\subtitle{Exact formulations, standard approximations, and open questions}


\author{Filip Cernatic         \and
        Bruno Senjean \and Vincent Robert \and Emmanuel
Fromager$^*$ 
}


\institute{Filip Cernatic \at
              Laboratoire de Chimie Quantique, Institut de Chimie,
CNRS/Universit\'{e} de Strasbourg, 4 rue Blaise Pascal, 67000 Strasbourg, France \\
              \email{filip.cernatic@etu.unistra.fr}           
           \and
           Bruno Senjean \at
              ICGM, Univ Montpellier, CNRS, ENSCM, Montpellier, France
 \\
              \email{bruno.senjean@umontpellier.fr}           
           \and
           Vincent Robert \at
              Laboratoire de Chimie Quantique, Institut de Chimie,
CNRS/Universit\'{e} de Strasbourg, 4 rue Blaise Pascal, 67000 Strasbourg, France \\
              \email{vrobert@unistra.fr}           
           \and
           $^*$Corresponding author: Emmanuel Fromager  \at
              Laboratoire de Chimie Quantique, Institut de Chimie,
CNRS/Universit\'{e} de Strasbourg, 4 rue Blaise Pascal, 67000 Strasbourg, France \\
              \email{fromagere@unistra.fr}           
           \and
}

\date{Received: date / Accepted: date}

\maketitle

\begin{abstract}

Recent progress in the field of (time-independent)
ensemble density-functional theory (DFT) for excited states are reviewed. Both
Gross--Oliveira--Kohn (GOK) and $N$-centered ensemble formalisms, which
are mathematically very similar and allow for an in-principle-exact
description of neutral and charged electronic excitations, respectively, are
discussed. Key exact results like, for example, the
equivalence between the infamous derivative discontinuity problem and
the description of weight dependencies in the ensemble
exchange-correlation density functional, are highlighted. The variational evaluation of
orbital-dependent ensemble Hartree-exchange (Hx) energies is discussed in
detail. We show in passing that state-averaging individual exact Hx
energies can lead to severe (solvable though) $v$-representability
issues. Finally, we explore the possibility to use the concept of density-driven correlation,
which has been recently introduced and does not exist in regular ground-state
DFT, for 
improving state-of-the-art correlation density-functional approximations
for ensembles. The present review  reflects the efforts of
a growing community to turn ensemble DFT into a rigorous and reliable
low-cost computational method for excited states. We hope
that, in the near future, this contribution will stimulate new formal and practical
developments in the field.

\keywords{Density-functional theory \and Excited states \and Many-body ensembles}
\end{abstract}

\setcounter{tocdepth}{4}
\setcounter{secnumdepth}{4}
\tableofcontents


\section{Introduction}
\label{sec:intro}

Kohn--Sham density-functional theory (KS-DFT)~\cite{KS} has become over the last two decades
the method of choice for modeling the electronic structure of molecules
and materials. This success originates from the relatively low
computational cost of the method and its relatively good
accuracy in the description of ground-state properties
such as equilibrium structures and activation barriers. KS-DFT is
an in-principle-exact ground-state theory. As
such, it cannot be used straightforwardly for calculating excited-state
properties. The formal beauty of KS-DFT lies in its universal description of electronic
ground states. Indeed, in KS-DFT, all the quantum many-electron effects are
encoded into a system-independent exchange-correlation (xc) density
functional $E_{\rm xc}[n]$ which, if it were known, would allow us to
compute the  
exact ground-state energy of any electronic system, simply by solving
self-consistent one-electron
equations, instead of the many-electron Schr\"{o}dinger equation. This universality is somehow lost, at least partially, when
turning to the excited states. An obvious reason is that there are
different types of electron
excitations. First we should distinguish charged excitations, where the number of electrons in
the system is
modified, from neutral excitations, which occur at a fixed electron
number~\cite{bredas2014mind}. In the context of DFT,
these two processes are usually approached very differently. In the
case of charged excitations, one traditionally refers to the extension of DFT to fractional
electron
numbers~\cite{perdew1982density,cohen2011challenges,mori2008localization,cohen2008insights,cohen2008fractional,stein2010fundamental,zheng2011improving}. Its implementation
at the simplest (semi-) local xc density functional
level of approximation usually yields too small fundamental
gaps for solids~\cite{perdew2017understanding}. This can be related to the 
discontinuities that the density functional derivative of the xc energy should
in principle exhibit, when crossing an integer electron number, but
that are completely absent from standard approximations. This is the reason why
hybrid functionals (where a fraction of orbital-dependent exchange
energy is
combined with density
functionals)~\cite{imamura2011linearity,atalla2016enforcing,stein2010fundamental,stein2012curvature}
or even more involved frequency-dependent post KS-DFT approaches like $GW$~\cite{onida2002electronic,sottile2007efficient,bruneval2012ionization,bruneval2012benchmarking,jiang2015first,pacchioni2015first,ou2016comparison,reining2018gw} are usually
employed for improving the description of fundamental gaps, which
implies a substantial increase in computational cost. If we now turn to neutral
excitation processes, the most popular approach is (linear response) time-dependent-DFT 
(TD-DFT)~\cite{runge1984density,Casida_tddft_review_2012}. 
In TD-DFT, the excitation energies are determined by searching for the poles of
the KS linear response function. The success of TD-DFT lies in the fact
that in many (but not all) cases the (rather crude) adiabatic
approximation performs relatively well with a moderate 
computational cost. In the latter approximation, the
time-dependent density-functional xc energy (it is in fact an {\it
action}~\cite{PRA08_Vignale_TDVP,PRA11_Vignale_TDVP_reply_Schirmer}, to be more precise) is
evaluated, in the time range $t_0\leq t\leq t_1$, from the regular
(time-independent)
ground-state xc functional and the density $n(t)\equiv n(\bfr,t)$ at
time $t$ as follows, $\mathcal{A}_{\rm xc}[n]\approx
\int^{t_1}_{t_0}dt\,E_{\rm
xc}[n(t)]$~\cite{PRA08_Vignale_TDVP,PRA11_Vignale_TDVP_reply_Schirmer,TDDFTfromager2013,Casida_tddft_review_2012}.
Nevertheless, the description of charge-transfer
excitations becomes problematic when semi-local functionals are
employed~\cite{Casida_tddft_review_2012,fuks2014challenging,dreuw2004failure,maitra2021double}.
Moreover, multiple excitations are completely absent from
standard TD-DFT spectra, precisely because of the adiabatic
approximation~\cite{maitra2004double,Casida_tddft_review_2012,maitra2021double}. Let us finally mention that
TD-DFT can be seen as a single-reference post-DFT method, because it
relies on the single-configuration KS ground-state wave function. This
can be problematic, for example, when the system under study has near-degenerate
low-lying states, like in the vicinity of a conical intersection, for
example~\cite{Casida_tddft_review_2012}.\\

The various limitations (in terms of computational cost, accuracy, or
physics) of the above-mentioned frequency-dependent post-DFT approaches
explain why, in recent years, time-independent
formulations of DFT for
excited states have attracted an increasing attention. For charged
excitations, the extended Koopman's theorem turns out to be an appealing
alternative~\cite{filatov2020computation_IP,filatov2020computation_EA}.
DFT for fractional electron numbers has also been further developed in the last
decade (see, for example,
Refs.~\cite{chan1999fresh,kraisler2013piecewise,gould2014kohn,kraisler2014fundamental,kraisler2015elimination,kraisler2015effect,li2015local,li2017extending}).
It has also been argued that restricting to integer electron numbers, in the calculation of charged excitations, is a completely valid alternative~\cite{gorling2015exchange,baerends2017kohn}.
For neutral excitations, various state-specific approaches have been
explored at the formal~\cite{gorling1999density,levy1999variational} but
also practical levels. In the latter case, 
orbital optimizations must be performed under proper constraints in
order to avoid 
variational collapses. This leads to various variational computational
schemes such as the $\Delta$ {\it self-consistent field} ($\Delta$SCF)
approach~\cite{ziegler1977calculation,gavnholt2008delta,kowalczyk2011assessment}, the maximum overlap method 
(MOM)~\cite{levi2020variational,hait2020excited,carter2020state,gilbert2008self,ivanov2021method}, orthogonally-constrained DFT~\cite{evangelista2013orthogonality,ramos2018low,roychoudhury2020neutral,karpinski2020capturing}, and
constricted DFT~\cite{ziegler2009relation,cullen2011formulation,ziegler2012implementation,krykunov2013self,park2016self}.
Interestingly, when a system has a 
Coulombic one-electron external local potential, which is the case for
any real molecule, an excited state can be identified directly from its
density~\cite{ayers2012time,ayers2015communication,glushkov2016highly,ayers2018time}. This fundamental property can be used for constructing an
in-principle-exact DFT for individual excited states. The practical
implementation of such a theory is not straightforward though, in particular because density
functionals must be defined also for non-Coulombic densities, so that
functional derivatives can be evaluated. Another strategy, which is the
main topic of this review, is ensemble DFT (eDFT). The ensemble
formalism is often referred to in DFT for mathematical purposes like, for
example, extending the
domain of definition of density functionals~\cite{LFTransform-Lieb} or describing strict
degeneracies~\cite{Kohn2001Kohn,Yang2000Degenerate}. It has probably be
underestimated as a potential alternative to standard time-dependent methods for the
practical
calculation of (charged or neutral) excitations~\cite{filatov1999spin,kazaryan2008excitation,filatov2015spin,filatov2015ensemble,filatov2017analytical,liu2021analytical,filatov2021description}. A clear advantage
of eDFT over time-dependent approaches is that its computational cost is
essentially that of a standard KS-DFT calculation. The only
difference is that, in an ensemble, orbitals can be fractionally occupied.
Moreover, like in TD-DFT, regular ground-state xc functionals can be
recycled in eDFT~\cite{filatov2015spin}. Note that,
unlike in
thermal
DFT~\cite{PRL11_Pittalis_exact_conds_thermalDFT,PRB16_Pribram-Jones_AC_thermalDFT,pastorczak2013calculation}, the fractional orbital occupation numbers are actually known before
the eDFT calculation starts. They are determined by the ensemble weights that the user
(arbitrarily) assigns to the $M$ ($M=1,2,\ldots$) lowest excited
states she/he wants to study. In (say canonical) thermal DFT~\cite{pastorczak2013calculation}, the ensemble weights are
determined not only from the temperature (that is arbitrarily fixed by the user) but
also from the (to-be-calculated) KS orbital energies. Another important feature of eDFT is that it can in principle describe any
kind of excitation, including the double
excitations~\cite{marut2020weight,gould2021double} that standard
approximate TD-DFT
misses. The eDFT formalism for neutral excitations is often referred to
as Gross--Oliveira--Kohn DFT (GOK-DFT) because it relies on the GOK
variational principle~\cite{gross1988rayleigh,gross1988density}. A
similar formalism, referred to as $N$-centered eDFT, has been
derived recently by Senjean and Fromager~\cite{senjean2018unified} for
the description of charged
excitations. The present review aims at highlighting recent progress in
eDFT, with a particular focus on the exact theory and the development of
approximations from first principles.\\ 

The chapter is organized as follows. An introduction to GOK-DFT and $N$-centered eDFT
is given in Sec.~\ref{sec:theory}. Even though the two theories describe completely different physical
processes, their mathematical
formulations are very similar, as highlighted in the section. We also explain how individual energy levels
(which give access to excited-state properties) can
be extracted, in principle exactly, from these theories. Then, in
Sec.~\ref{sec:DD}, we discuss the equivalence between the xc derivative
discontinuity, which is a fundamental concept in DFT, and the ensemble
weight derivative of the xc density functional, which is central in
eDFT. Strategies for developing weight-dependent xc density-functional
approximations (DFAs) for ensembles, which is the most challenging task in
eDFT, are then reviewed. Key concepts will be illustrated with the prototypical asymmetric Hubbard dimer
model~\cite{carrascal2015hubbard,carrascal2016corrigendum}. The rigorous construction of hybrid functionals
for ensembles is discussed in Sec.~\ref{sec:EXX}. We reveal 
that using state-averaged exact exchange
energies, which is common in computational
eDFT studies~\cite{filatov2021description}, can lead to  
severe (solvable though) $v$-representability issues. Finally, we
discuss in Sec.~\ref{sec:corr_energy} the concept of ensemble
density-driven correlation, which was recently introduced by Gould and
Pittalis~\cite{gould2019density}, and how it could be used in the design of
correlation DFAs for ensembles. Conclusions and perspectives are given
in Sec.~\ref{sec:conclu}. Detailed derivations of some key equations are
provided in the appendices.

\section{Unified ensemble DFT formalism for neutral and charged excitations}
\label{sec:theory}

In this chapter, we are interested in the evaluation of neutral ($E_I^N - E_0^N$) and charged ($E^{N\pm 1}_0 - E^N_0$) excitation energies,
where the $I$th lowest energy $E_I^M$ of the $M$-electron system under
study is in principle obtained by solving the following Schr\"odinger equation, 
\begin{eqnarray}
\hat{H}\ket{\Psi_I^M} = E_I^M \ket{\Psi^M_I},
\end{eqnarray}
where
\begin{eqnarray}
\label{eq:Hamil_elec}
\hat{H} = \hat{T} + \hat{W}_{\rm ee} + \hat{V}_{\rm ext}
\end{eqnarray}
is the electronic Hamiltonian within the Born--Oppenheimer approximation
and
\begin{eqnarray}\label{eq:ab_initio_hamil}
\begin{split}
\hat{T} &\equiv - \sum_{i = 1}^M \dfrac{1}{2} \nabla_{{\bfr}_i}^2 \\
\hat{W}_{\rm ee} &\equiv  \sum_{i=1}^M \sum_{j>i}^M \dfrac{1}{|\bfr_i - \bfr_j |} \times \\
\hat{V}_{\rm ext} &\equiv  \sum_{i=1}^M v_{\rm ext}(\bfr_i) \times 
\end{split}
\end{eqnarray}
are the $M$-electron kinetic
energy, Coulomb repulsion, and local (\ie, multiplicative) external potential operators, respectively.
Both neutral and charged excitations can be described within a
unified eDFT formalism. The calculations will simply differ in the type of excited states
(charged or neutral) that
is included into the ensemble. On the one hand, the GOK-DFT
formalism~\cite{gross1988density}
will be employed for neutral excitations while, for charged excitations,
we will use the more recent $N$-centered eDFT formalism~\cite{senjean2018unified}. 
In this section, we derive key
equations for each theory, and we show how individual excited-state
properties (energy levels and densities) can be extracted, in principle exactly,
from the KS ensemble. Real algebra will be used throughout this work. 
For the sake of clarity, derivations
will be detailed only for ensembles consisting of
non-degenerate states. The theory obviously applies to more general
cases~\cite{gross1988density,yang2017direct}. 

\subsection{DFT of neutral excitations}
\label{subsec:GOK-DFT}

GOK-DFT has been formulated in the end of the 1980's by Gross, Oliveira, and Kohn~\cite{gross1988rayleigh,gross1988density,oliveira1988density}
and is a generalization of the equiensemble DFT of Theophilou~\cite{theophilou1979energy}.
In contrast to standard DFT, which is a ground-state theory, GOK-DFT
can describe both 
ground and (neutral) excited states. In this context, 
the ensemble density is used as a basic variable (in place of the ground-state
density).

\subsubsection{GOK ensembles}\label{subsubsec:GOK}

Before deriving the main equations of GOK-DFT, let us introduce the exact ensemble theory.
We start with the ensemble GOK energy
expression~\cite{gross1988rayleigh}
\begin{eqnarray}
\label{eq:Ens_energy}
E^\bfw = \sum_{I} \ttw_I E_I,
\end{eqnarray}
which is simply a state-averaged energy
where $\bfw=(\ttw_1,\ttw_2,\ldots)$ denotes the collection of ensemble
weights that are assigned to
the {\it excited} states, and $E_I\equiv E^N_I$ are the energies of the $N$-electron ground
($I=0$) and excited ($I>0$) states $\ket{\Psi^N_I}$. We assumed in
Eq.~(\ref{eq:Ens_energy}) that the
full set of weights (which includes the weight $\ttw_0$ assigned to the ground state) is
normalized, \ie, $\ttw_0 =  1 - \sum_{I>0}
\ttw_I$, so that 
\begin{eqnarray}
\label{eq:Ens_energy_normalized}
\begin{split}
E^\bfw &= \left( 1 - \sum_{I>0} \ttw_I \right)E_0 + \sum_{I>0} \ttw_I E_I
\\
&=E_0+\sum_{I>0} \ttw_I\left(E_I-E_0\right).
\end{split}
\end{eqnarray}
For ordered weights $\ttw_I \geq \ttw_{I+1} \geq 0$, with $I\geq 0$,
the following (so-called GOK) variational principle holds~\cite{gross1988rayleigh},
\begin{eqnarray}
E^\bfw \leq \sum_I \ttw_I \bra{\tilde{\Psi}_I} \hat{H} \ket{\tilde{\Psi}_I},
\end{eqnarray}
where $\left\{\tilde{\Psi}_I\right\}$ is a trial set of orthonormal
$N$-electron wave functions.
Note that the lower bound $E^\bfw$, which is the exact ensemble energy, is not an
observable. It is just an (artificial) auxiliary quantity from which
properties of interest, such as the excitation energies, can be extracted. Since it varies {\it linearly}
with the ensemble weights, the extraction of individual energy levels is
actually trivial. Indeed, combining the following two relations
[see Eq.~(\ref{eq:Ens_energy_normalized})],
\begin{eqnarray}
\label{eq:derivEw_dw}
\dwI{E^\bfw} = E_I - E_0,
\end{eqnarray}
and
\be
\begin{split}
E_K&\underset{K\geq 0}{=} E_0+\sum_{I>0}\delta_{IK}(E_I-E_0)
\\&= E^\bfw + \sum_{I>0} (\delta_{IK} - \ttw_I)(E_I-E_0),
\end{split}
\ee
leads to
\begin{eqnarray}
\label{eq:indiv_EI}
E_K = E^\bfw + \sum_{I>0} (\delta_{IK} - \ttw_I) \dwI{E^\bfw}.
\end{eqnarray}
Despite its simplicity the above expression has not been used until very recently
for extracting excited-state energies from a GOK-DFT
calculation~\cite{deur2019ground,loos2020weight}. Further details will be given in the next section.

\subsubsection{DFT of GOK ensembles}\label{subsubsec:GOK-DFT}

In GOK-DFT, the ensemble energy is obtained variationally as
follows~\cite{gross1988density}, 
\begin{eqnarray}\label{eq:Ew_GOK-DFT}
E^\bfw = \min_{n\rightarrow N} \left \lbrace F^\bfw[n] + \int \ddroit \bfr \, v_{\rm ext}(\bfr) n(\bfr) \right\rbrace,
\end{eqnarray}
where the minimization is restricted to $N$-electron densities, \ie, $\int d\bfr\,n(\bfr)=N$, and the universal GOK density functional
\be\label{eq:universal_GOK_func}
F^\bfw[n]:=\sum_I\ttw_I\mel{\Psi_I^\bfw[n]}{\hat{T}+\hat{W}_{\rm
ee}}{\Psi_I^\bfw[n]},
\ee
which is evaluated from the density-functional eigenfunctions
$\left\{\Psi_I^\bfw[n]\right\}$ that fulfill the density constraint
$\sum_I\ttw_I n_{\Psi_I^\bfw[n]}(\bfr)=n(\bfr)$, is the analog for GOK
ensembles of the universal Hohenberg--Kohn functional. Its construction relies on a
potential-ensemble-density map that is established for a {\it given and
fixed} set $\bfw$ of
ensemble weight values. Therefore, the universality of the functional
implies, like in ground-state DFT, that it does not depend on the local external
potential. However, it does {\it not} mean that it is  
ensemble-independent and therefore applicable to any excited state.
As discussed in further detail in Secs.~\ref{sec:EXX} and
\ref{sec:corr_energy}, encoding ensemble dependencies into density functionals is probably
the most challenging task in eDFT.\\

In the
standard KS formulation of GOK-DFT~\cite{gross1988density}, the GOK
functional is split into non-interacting kinetic and Hartree-xc (Hxc)
ensemble energy contributions, by analogy with regular KS-DFT: 
\begin{eqnarray}\label{eq:F_KS_GOK-DFT}
F^\bfw [n] = T^\bfw_{\rm s}[n] + E_\Hxc^\bfw[n].
\end{eqnarray}
The non-interacting
ensemble kinetic energy functional can be expressed more explicitly as
follows within the constrained-search
formalism~\cite{levy1979universal},
\begin{eqnarray}\label{eq:T_KS_GOK-DFT}
T^\bfw_{\rm s}[n] &=& \min_{\hat{\gamma}^\bfw \rightarrow n} \left\lbrace {\rm Tr}\left[\hat{\gamma}^\bfw  \hat{T} \right] \right\rbrace
\\
\label{eq:explicit_Tsw_func}
&\equiv&\sum_{I} \ttw_I \bra{\Phi_I^\bfw[n]} \hat{T}
\ket{\Phi_I^\bfw[n]},
\end{eqnarray}
where $\Tr$ denotes the trace,
$\hat{\gamma}^\bfw = \sum_{I} \ttw_I  \ket{{\Phi}_I}\bra{{\Phi}_I}$ is a
trial ensemble density matrix operator that fulfills the density
constraint
$n_{\hat{\gamma}^\bfw}(\bfr) \equiv {\rm Tr} \left[ \hat{\gamma}^\bfw
\hat{n}(\bfr)\right] = \sum_{I} \ttw_I n_{\Phi_I}(\bfr) = n(\bfr)$, and
$\hat{n}(\bfr)\equiv \sum^N_{i=1}\delta(\bfr-\bfr_i)$ is the electron density operator at position $\bfr$.
Combining Eqs.~(\ref{eq:Ew_GOK-DFT}), (\ref{eq:F_KS_GOK-DFT}) and
(\ref{eq:T_KS_GOK-DFT}) leads to the final GOK-DFT variational energy
expression
\begin{eqnarray}
\label{eq:Ew_GOK-DFT_VP}
E^\bfw &=& \min_{\lbrace \varphi_p \rbrace} \left\lbrace {\rm Tr}\left[\hat{\gamma}^\bfw \left( \hat{T} + \hat{V}_{\rm ext}\right) \right] + E_\Hxc^\bfw \left[ n_{\hat{\gamma}^\bfw}\right] \right\rbrace \nonumber \\
&=&{\rm Tr}\left[\hat{\gamma}_{\rm KS}^\bfw \left( \hat{T} + \hat{V}_{\rm ext}\right) \right] + E_\Hxc^\bfw \left[ n_{\hat{\gamma}_{\rm KS}^\bfw}\right],
\end{eqnarray}
where the minimization can be restricted to single-configuration wave
functions (determinants or configuration state functions), hence the
minimization over orbitals $\left\{\varphi_p\right\}$ on the first line of
Eq.~(\ref{eq:Ew_GOK-DFT_VP}). The minimizing KS orbitals $\lbrace
\varphi_p^\bfw \rbrace$, from which the KS wave functions $\lbrace
\Phi_I^\bfw[n^\bfw] \equiv \Phi_I^\bfw\rbrace$
in the minimizing density matrix operator $\hat{\gamma}_{\rm KS}^\bfw$
are constructed,
fulfill the following self-consistent GOK-DFT equations,
\begin{eqnarray}
\label{eq:ensemble_KS_eq}
\left( - \dfrac{\nabla_{\bfr}^2}{2} + v_{\rm ext}(\bfr) + v_\Hxc^\bfw[n^\bfw](\bfr)  \right)\varphi_p^\bfw (\bfr) = \varepsilon_p^\bfw \varphi^\bfw_p(\bfr)
,
\end{eqnarray}
where
\begin{eqnarray}
v_\Hxc^\bfw[n] (\bfr) = \dfrac{\delta E^\bfw_\Hxc [n]}{\delta n(\bfr)}
\end{eqnarray}
is the ensemble Hxc density-functional potential.
In the exact theory, the ensemble KS orbitals reproduce the exact (interacting)
ensemble density, \ie, 
\begin{eqnarray}
\label{eq:nphi_npsi}
 \sum_{I} \ttw_I n_{\Phi_I^\bfw}(\bfr) = \sum_{I} \ttw_I n_{\Psi_I}(\bfr) = n^\bfw(\bfr),
\end{eqnarray}
where the individual KS densities read as 
\begin{eqnarray}
\label{eq:n_Phi}
n_{\Phi_I^\bfw}(\bfr) = \sum_p n_p^I \abs{\varphi_p^\bfw(\bfr)}^2,
\end{eqnarray}
and $n_p^I$ is the (weight-independent) occupation number of the orbital $\varphi_p^\bfw$ in
the single-configuration wave function $\Phi^\bfw_I$.\\

Let us now focus on the ensemble Hxc density functional. By analogy with
regular KS-DFT, it can decomposed into Hx and correlation energy
contributions: $E_\Hxc^\bfw[n] = E_{\rm Hx}^\bfw [n] + E_{\rm c}^\bfw[n]$.
In the original formulation of GOK-DFT~\cite{gross1988density}, the Hx
functional is further decomposed as follows, 
\be\label{eq:original_decomp_H_and_xc}
E_{\rm Hx}^\bfw[n]=E_{\rm H}[n]+E_{\rm x}^\bfw[n],
\ee
where 
\be
E_{\rm H}[n]= \dfrac{1}{2} \int \ddroit\bfr\int \ddroit \bfr' \dfrac{n(\bfr)n(\bfr')}{| \bfr - \bfr' |}
\ee
is the standard {\it weight-independent} Hartree functional,
and 
\be
E^\bfw_{\rm x}[n]=\sum_{I} \ttw_I \bra{\Phi_I^\bfw[n]} \hat{W}_ {\rm ee}
\ket{\Phi_I^\bfw[n]}-E_{\rm H}[n]
\ee
is the exact (complementary and weight-dependent) ensemble exchange
functional. Note that $\Phi_I^\bfw[n]$, which describes one of the
configurations included into the ensemble, may not be a pure Slater
determinant~\cite{gould2017hartree}. 
Other (weight-dependent) definitions for the ensemble Hartree energy, where the explicit
dependence on the ensemble density $n$ is lost, have been
explored~\cite{gould2020ensemble}.
In the most intuitive one, the
ensemble
Hartree energy is evaluated as the  
weighted sum of the individual KS Hartree energies: 
\be
E^\bfw_{\rm H}[n]:=\sum_I\ttw_IE_{\rm H}\left[n_{\Phi_I^\bfw[n]}\right].
\ee
For the sake of generality, we will keep in the following both Hartree and exchange energies into a single functional
$E_{\rm Hx}^\bfw [n]$ which is defined as
\begin{eqnarray}
\label{eq:def_exact_ens_Hx_func}
E_{\rm Hx}^\bfw [n] = \sum_{I} \ttw_I \bra{\Phi_I^\bfw[n]} \hat{W}_ {\rm ee} \ket{\Phi_I^\bfw[n]}.
\end{eqnarray}
The remaining weight-dependent correlation energy can then be expressed
as follows, according to Eqs.~(\ref{eq:universal_GOK_func}) and (\ref{eq:explicit_Tsw_func}),
\be
\label{eq:ens_corr_fun_components}
E^\bfw_{\rm c}[n]&=&F^\bfw[n]-T^\bfw_{\rm s}[n]-E^\bfw_{\rm Hx}[n]
\nonumber \\
&=&\sum_I\ttw_I\left(\mel{\Psi_I^\bfw[n]}{\hat{T}+\hat{W}_{\rm
ee}}{\Psi_I^\bfw[n]}-\mel{\Phi_I^\bfw[n]}{\hat{T}+\hat{W}_{\rm
ee}}{\Phi_I^\bfw[n]}\right),
\ee
where the non-interacting KS $\lbrace \Phi_I^\bfw[n]\rbrace$ and
interacting $\lbrace \Psi_I^\bfw[n]\rbrace$ wave functions, which both
reproduce the (weight-independent here) trial ensemble density $n$, {\it
whatever} the ensemble weight values $\bfw$, are in principle weight-dependent~\cite{franck2014generalised,deur2017exact}.
Interestingly, the interacting density-functional wave functions lose their
weight dependence when the trial density $n$ matches 
the exact physical ensemble density $n^\bfw$, \ie,
$\Psi_I^\bfw[n^\bfw]=\Psi_I\equiv \Psi^N_I$. However, as shown in Sec.~\ref{subsec:difference_KS_true_dens}, the KS wave
functions remain weight-dependent, even in this special case.\\

As readily seen from Eqs.~(\ref{eq:F_KS_GOK-DFT}),
(\ref{eq:Ew_GOK-DFT_VP}) and (\ref{eq:ensemble_KS_eq}),
the only (but crucial) difference between regular ground-state
KS-DFT and GOK-DFT is the weight dependence in the ensemble
density-functional Hxc energy and potential.
The computational cost should essentially be the same in both
approaches. The challenge lies in the proper description of the
weight-dependent ensemble Hxc density functional. 
Different approximations have been considered, such as the use of
(weight-independent) regular ground-state
functionals~\cite{pastorczak2013calculation,senjean2015linear}, or the
use of an ensemble
exact-exchange energy~\cite{gould2017hartree,gould2018charge,deur2018exploring}
with or without approximate weight-dependent correlation
functionals~\cite{deur2018exploring,loos2020weight,marut2020weight}.
Note that the expected linearity-in-weight of the ensemble energy is not
always reproduced in (approximate) practical GOK-DFT
calculations~\cite{senjean2015linear}. As a
result, different weights can give different excitation energies, which
is a serious issue. This lead to different computation strategies, such
as trying to find an optimal value for the weights~\cite{nagy1996local},
using Boltzmann weights instead~\cite{pastorczak2013calculation}, restricting
to equiensembles~\cite{loos2020weight,marut2020weight},
or considering the ground-state $\bfw=0$ limit of the theory, like in
the {\it direct ensemble correction} (DEC)
scheme~\cite{yang2017direct,sagredo2018can}. A linear interpolation
method has also been
proposed~\cite{senjean2015linear,senjean2016combining}.\\

Designing weight-dependent ensemble DFAs that systematically reduce the curvature in
weight of the ensemble energy, while providing at the same time accurate excitation
energies, is an important and challenging
task. Recent progress in this matter will be extensively discussed in
Secs.~\ref{sec:EXX} and \ref{sec:corr_energy}.

\subsubsection{Extraction of individual state properties}
\label{subsubsec:indivGOK}

In Sec.~\ref{subsubsec:GOK-DFT}, we have shown that both exact ensemble
energy and density can be calculated, in principle exactly, within
GOK-DFT. At this point we should stress that the KS and true physical
densities are not expected to match individually, even though they both
reproduce the same ensemble density [see Eq.~(\ref{eq:nphi_npsi})]. This
subtle point
will be discussed in more detail in
Sec.~\ref{subsec:difference_KS_true_dens}. Nevertheless, in complete
analogy with Eq.~(\ref{eq:indiv_EI}), the exact individual densities can
be extracted from the ensemble density as follows~\cite{fromager2020individual},  
\begin{eqnarray}
\label{eq:indiv_dens}
n_{\Psi_J}(\bfr) = n^\bfw(\bfr) + \sum_{I>0} \left(\delta_{IJ} - \ttw_I\right) \dwI{n^\bfw(\bfr)},
\end{eqnarray}
which, by inserting the expression in Eq.~(\ref{eq:nphi_npsi}), leads to
the key result~\cite{fromager2020individual} 
\begin{eqnarray}
\label{eq:indiv_dens_response}
n_{\Psi_J}(\bfr) = n_{\Phi_J^\bfw}(\bfr) + \sum_{I>0} \sum_{K\geq 0} \left(\delta_{IJ} - \ttw_I \right) \ttw_K \dwI{n_{\Phi_K^\bfw}(\bfr)}
,
\end{eqnarray}
where, according to Eq.~(\ref{eq:n_Phi}), the weight derivative of the
individual KS densities 
\begin{eqnarray}\label{eq:dnphi_dw}
\dfrac{\partial n_{\Phi_K^\bfw}(\bfr) }{ \partial \ttw_I} = 2\sum_p n_p^K \varphi_p^\bfw (\bfr) \dfrac{ \partial \varphi^\bfw_p(\bfr) }{ \partial \ttw_I}
\end{eqnarray}
can be evaluated from the (static) linear response of the KS orbitals.
This can be done, in practice, by solving an ensemble coupled-perturbed
equation~\cite{filatov2017analytical,fromager2020individual}, for
example.\\

Turning to the excitation energies, we obtain from the variational
GOK-DFT ensemble energy expression and the Hellmann--Feynman theorem the
following expression, where the derivatives of the minimizing (and
therefore stationary) KS wave functions do
not contribute, 
\begin{eqnarray}
\dwI{E^\bfw} &=& {\rm Tr} \left[ \Delta \hat{\gamma}_{{\rm KS},I}^\bfw \left( \hat{T} + \hat{V}_{\rm ext} \right) \right] + \left. \dwI{E_{\rm Hxc}^\bfw[n]} \right|_{n = n_{\hat{\gamma}_{\rm KS}^\bfw}} \nonumber \\
&&+ \int \ddroit \bfr \, \dfrac{\delta E_\Hxc^\bfw[n_{\hat{\gamma}_{\rm KS}^\bfw}]}{\delta n(\bfr)}
{\rm Tr}\left[ \Delta \hat{\gamma}^\bfw_{{\rm KS},I} \hat{n}(\bfr) \right],
\end{eqnarray}
with $\Delta \hat{\gamma}^\bfw_{{\rm KS},I} = \ket{\Phi_I^\bfw}\bra{\Phi_I^\bfw} - \ket{\Phi_0^\bfw}\bra{\Phi_0^\bfw}$.
This expression can be further simplified as follows~\cite{deur2019ground}:
\begin{eqnarray}\label{eq:dEwdw_GOK-DFT}
\dwI{E^\bfw} = E_I - E_0 = \mathcal{E}_I^\bfw - \mathcal{E}_0^\bfw + \left. \dwI{E_{\rm Hxc}^\bfw[n]} \right|_{n = n_{\hat{\gamma}_{\rm KS}^\bfw}}
,
\end{eqnarray}
where $\mathcal{E}_ I^\bfw$ denotes the $I$th (weight-dependent) KS
energy which is obtained by summing up the energies $\lbrace
\varepsilon_p^\bfw \rbrace$ of the KS orbitals that are occupied in $\Phi_I^\bfw$.
Hence, the excitation energies can all be determined, in principle
exactly, from a {\it single} GOK-DFT calculation.\\

As shown by Deur and Fromager~\cite{deur2019ground}, individual energy
levels can also be extracted (from the KS ensemble) and written in a
compact form. For that purpose, we will use the exact expression of
Eq.~(\ref{eq:indiv_EI}) where we see, in the light of Eq.~(\ref{eq:dEwdw_GOK-DFT}), that it is convenient to express
the total ensemble energy [first term on the right-hand side of
Eq.~(\ref{eq:indiv_EI})] in terms of total KS energies. Levy and
Zahariev (LZ) made such a suggestion in the context of regular ground-state DFT~\cite{levy2014ground}. For that
purpose, they introduced a shift in the Hxc potential that can be
trivially
generalized to GOK ensembles as follows~\cite{deur2019ground}, 
\begin{eqnarray}
\label{eq:LZshift_GOK-DFT}
\dfrac{\delta E_\Hxc^\bfw [n] }{\delta n(\bfr)} \rightarrow \overline{v}_\Hxc^\bfw[n](\bfr) = \dfrac{\delta E_\Hxc^\bfw [n] }{\delta n(\bfr)}  + \dfrac{
E_\Hxc^\bfw[n] - \int \ddroit \bfr \, \dfrac{\delta E_\Hxc^\bfw [n] }{\delta n(\bfr)} n(\bfr)}{
\int \ddroit \bfr \, n(\bfr)}.
\end{eqnarray}
Note that, if the exact LZ-shifted Hxc potential were known, we would be
able to
evaluate exact ensemble density-functional Hxc energies as follows,
\begin{eqnarray}
E_\Hxc^\bfw[n] = \int \ddroit \bfr \, \overline{v}_\Hxc^\bfw [n](\bfr)
n(\bfr).
\end{eqnarray}
Once the LZ shift has been applied to the ensemble Hxc potential, the
(total $N$-electron) KS
energies will be modified as follows, 
\begin{eqnarray}\label{eq:EKS_LZshifted}
\mathcal{E}_I^\bfw \rightarrow \overline{\mathcal{E}}_I^\bfw = \mathcal{E}_I^\bfw + 
E_\Hxc^\bfw[n_{\hat{\gamma}_{\rm KS}^\bfw}] - \int \ddroit \bfr \, \dfrac{\delta E_\Hxc^\bfw [n_{\hat{\gamma}_{\rm KS}^\bfw}] }{\delta n(\bfr)} n_{\hat{\gamma}_{\rm KS}^\bfw}(\bfr),
\end{eqnarray}
and the true ensemble energy will simply read as a weighted sum of
(LZ-shifted) KS energies:
\begin{eqnarray}\label{eq:Ew_LZshifted}
E^\bfw = \left( 1 - \sum_{I>0} \ttw_I \right) \overline{\mathcal{E}}_0^\bfw + \sum_{I>0} \ttw_I \overline{\mathcal{E}}_I^\bfw.
\end{eqnarray}
Note that the KS excitation energies are not affected by the shift:  
\begin{eqnarray}\label{eq:KSdiff_GOK-DFT_LZ}
\overline{\mathcal{E}}_I^\bfw - \overline{\mathcal{E}}_0^\bfw =
\mathcal{E}_I^\bfw -  \mathcal{E}_0^\bfw.
\end{eqnarray}
Thus, by combining Eqs.~(\ref{eq:indiv_EI}), (\ref{eq:dEwdw_GOK-DFT}),
(\ref{eq:Ew_LZshifted}) and (\ref{eq:KSdiff_GOK-DFT_LZ}), we recover the
exact expression of Ref.~\cite{deur2019ground} for ground- and
excited-state energy levels:
\begin{eqnarray}
\label{eq:Ek_GOK-DFT}
E_K = \overline{\mathcal{E}}_K^\bfw + \sum_{I>0}\left(\delta_{IK} -
\ttw_I\right)\left.\dwI{E_\Hxc^\bfw[n]}\right\vert_{n=n_{\hat{\gamma}_{\rm
KS}^\bfw}}
.
\end{eqnarray}
As readily seen from Eq.~(\ref{eq:Ek_GOK-DFT}), applying the LZ shift is not
sufficient for reaching an exact energy level. The ensemble weight
derivatives of the Hxc density functional are also needed for that purpose.\\     

Finally, it is instructive to consider the general expression of
Eq.~(\ref{eq:Ek_GOK-DFT}) in the ground-state ${\bfw} =0$ limit of the
theory, which gives 
\begin{eqnarray}\label{eq:Ek_GOK-DFT_w0}
E_I = 
\overline{\mathcal{E}}_I^{\bfw=0} + (1 - \delta_{I0}
)\left.\dwI{E_\Hxc^\bfw[n_{\Psi_0}]}\right\vert_{\bfw = 0},
\end{eqnarray}
where $n_{\Psi_0}$ is the exact ground-state density. 
As readily seen from Eq.~(\ref{eq:Ek_GOK-DFT_w0}), as we start from a
pure $I=0$ ground-state theory (we recover the energy expression of Levy
and Zahariev in this case~\cite{levy2014ground}), the inclusion of a
given $I>0$ excited state into the ensemble induces an additional shift in
the Hxc potential, which corresponds to the weight derivative
$\left.\partial {E_\Hxc^\bfw[n_{\Psi_0}]}/\partial\ttw_I\right\vert_{\bfw = 0}$
and can be interpreted as a derivative discontinuity, as shown in
Ref.~\cite{levy1995excitation} and extensively discussed
in Sec.~\ref{sec:DD}, in the context of charged excitations.

\subsection{DFT of charged excitations: The $N$-centered ensemble formalism}
\label{subsec:Ncentered}

A recent adaptation of GOK-DFT to charged excitations, which is referred to as
$N$-centered eDFT~\cite{senjean2018unified}, is introduced in the present
section. 

\subsubsection{$N$-centered ensembles}
\label{subsubsec:Ncentered}

The $N$-centered ensemble~\cite{senjean2018unified} can be seen
as the ``grand canonical'' ground-state version of GOK ensembles. It is
constructed from the $M$-electron ground states where the three possible
values of the {\it integer} $M \in \lbrace \Nm, N, \Np \rbrace$ are,
like the corresponding ensemble density (see below), centered in $N$,
hence the name ``$N$-centered''.
The exact $N$-centered ensemble energy is defined as follows~\cite{senjean2018unified}, 
\begin{eqnarray}
\label{eq:ens_E_two_weights}
E_0^\bmxi =
\xim E_0^{\Nm} + \xip E_0^\Np 
  + \left(1 - 
\xim\dfrac{\Nm}{N} - \xip\dfrac{\Np}{N}
\right)E_0^{N},
\end{eqnarray}
where the two $N$-centered ensemble weights $\xi_-$ and $\xi_+$, which
describe the removal/addition of an electron from/to the $N$-electron
system, respectively, are collected in 
\be\label{eq:Nc_weights_notation}
{\bm\xi}\equiv 
\left( \xi_- , \xi_+ \right).
\ee
Similarly, the $N$-centered ensemble density reads as 
\begin{eqnarray}\label{eq:Nc_ens_dens_Bruno}
n^\bmxi_0(\bfr) =  \xim n_{\Psi_0^{\Nm}}(\bfr) + \xip n_{\Psi_0^{\Np}}(\bfr) 
  + \left(1 - 
\xim\dfrac{\Nm}{N} - \xip\dfrac{\Np}{N}
\right)n_{\Psi_0^{N}}(\bfr). \nonumber \\
\end{eqnarray}
Designed by analogy with GOK-DFT (which describes {\it
neutral} excitations), the $N$-centered ensemble density integrates to the central {\it integer} number
$N$ of electrons:
\begin{eqnarray}
\int \ddroit \bfr \, n^\bmxi_0(\bfr) = N.
\end{eqnarray}
In other words, even though we describe charged excitation processes,
the number of electrons remains {\it fixed and equal to the integer
$N$} whatever the value of the ensemble weights $\bmxi$. This major difference with the conventional DFT for fractional
electron numbers~\cite{perdew1982density} has important implications
that will be discussed extensively in Sec.~\ref{sec:DD}.\\

In this context, the ensemble energy can be determined
variationally, as a  
direct consequence of the conventional Rayleigh--Ritz
variational principle for a fixed number of electrons, \ie,
\begin{eqnarray}\label{eq:ens_E_two_weights_VP}
E_0^\bmxi &\leq &
\xim \mel{\tilde{\Psi}^{\Nm}}{\hat{H}}{\tilde{\Psi}^\Nm} + \xip \mel{\tilde{\Psi}^{\Np}}{\hat{H}}{\tilde{\Psi}^\Np}
 \nonumber \\
 && + \left(1 - 
\xim\dfrac{\Nm}{N} - \xip\dfrac{\Np}{N}
\right)\mel{\tilde{\Psi}^{N}}{\hat{H}}{\tilde{\Psi}^N},
\end{eqnarray}
where $\lbrace \tilde{\Psi}^M \rbrace$ are trial $M$-electron normalized wave
functions, provided that the (so-called convexity) conditions
$\xi_-\geq 0$, $\xi_+\geq 0$, and 
$\xi_- (\Nm) +\xi_+(\Np)\leq N$ are fulfilled. Like in GOK-DFT, the
ensemble energy $E_0^\bmxi$ varies linearly with the ensemble weights.
As a result, charged excitation energies can be extracted through
differentiation with respect to the $N$-centered ensemble weights. For
example, since 
\begin{eqnarray}
\dxipm{E^\bmxi_0} = E_0^\Npm  - \left(\dfrac{\Npm}{N}\right)E_0^N,
\end{eqnarray}
the exact fundamental gap can be determined as follows,
\begin{eqnarray}
\label{eq:derivExi_xi_fundgap}
\dxim{E^\bmxi_0} + \dxip{E^\bmxi_0}  = E_0^{N -1} + E_0^{N+1} - 2E_0^{N} = E_{\rm gap}^{\rm fund}.
\end{eqnarray}
We can also extract the individual cationic,
anionic, and neutral energies, respectively, as follows,
\begin{eqnarray}\label{eq:ENm_from_ensE}
E_0^{\Nm}=\dfrac{\Nm}{N}
\left(E_0^\bmxi
-\xip
\dfrac{\partial E_0^\bmxi}{\partial
\xip}
+
\left(
\dfrac{N}{\Nm}
-\xim
\right)\dfrac{\partial E_0^\bmxi}{\partial
\xim}\right),
\end{eqnarray}
\begin{eqnarray}\label{eq:ENp_from_ensE}
E_0^{\Np}=\dfrac{\Np}{N}
\left(
E_0^\bmxi
-\xim
\dfrac{\partial E_0^\bmxi}{\partial
\xim}
+
\left(
\dfrac{N}{\Np}
-\xip
\right)\dfrac{\partial E_0^\bmxi}{\partial
\xip}
\right),
\end{eqnarray}
and
\begin{eqnarray}\label{eq:EN_from_ensE}
E_0^{N}=E_0^\bmxi
-\xim\dfrac{\partial E_0^\bmxi}{\partial
\xim}
-\xip\dfrac{\partial E_0^\bmxi}{\partial
\xip}.
\end{eqnarray}
Eqs.~(\ref{eq:ENm_from_ensE})--(\ref{eq:EN_from_ensE}) will be used in
the following for deriving exact ionization potential and
electron affinity theorems.\\

\subsubsection{DFT of $N$-centered ensembles}
\label{subsubsec:Ncentered_eDFT}

In complete analogy with GOK-DFT, the $N$-centered ensemble energy can
be determined variationally as follows, 
\begin{eqnarray}
\label{eq:ENxi_VP}
E_0^\bmxi &=& \min_{n\rightarrow N} \left \lbrace F^\bmxi [n] + \int \ddroit \bfr \, v_{\rm ext}(\bfr)n(\bfr) \right \rbrace,
\end{eqnarray}
where, in the KS formulation of the theory~\cite{senjean2018unified}, the universal $N$-centered ensemble
functional reads as 
\begin{eqnarray}\label{eq:F_KS_NeDFT}
F^\bmxi[n] = T_{\rm s}^\bmxi[n] + E_\Hxc^\bmxi[n].
\end{eqnarray}
The non-interacting kinetic energy functional
\begin{eqnarray}\label{eq:T_KS_NeDFT}
T_{\rm s}^\bmxi[n] = \min_{\hat{\gamma}^\bmxi \rightarrow n} \left\lbrace 
{\rm
Tr}\left[
\hat{\gamma}^\bmxi\hat{T}\right] \right\rbrace
\end{eqnarray}
is now determined through a minimization over $N$-centered density matrix operators
\begin{eqnarray}
\hat{\gamma}^\bmxi &\equiv& 
\xi_{-}  \ket{{\Phi}^{N - 1}}\bra{{\Phi}^{N - 1}} 
+
\xi_{+}  \ket{{\Phi}^{N + 1}}\bra{{\Phi}^{N + 1}}\nonumber \\ 
&&+ \left(1 - \xim\dfrac{\Nm}{N} - \xip\dfrac{\Np}{N}
\right) \ket{{\Phi}^N}\bra{{\Phi}^N}
\end{eqnarray}
that fulfill the density constraint $n_{\hat{\gamma}^\bmxi}(\bfr) = {\rm Tr} \left[ \hat{\gamma}^\bmxi \hat{n}(\bfr)\right] = n(\bfr)$.
Combining Eqs.~(\ref{eq:ENxi_VP}), (\ref{eq:F_KS_NeDFT}) and
(\ref{eq:T_KS_NeDFT}) leads to the final ensemble energy expression,
\begin{eqnarray}
\label{eq:ENxi_VP_KS}
E_0^\bmxi &=& \min_{\lbrace \varphi_p \rbrace} \left\lbrace 
{\rm
Tr}\left[
\hat{\gamma}^\bmxi
\left(\hat{T}+\hat{V}_{\rm ext}\right)\right]
+
E^\bmxi_{\rm
Hxc}\left[n_{\hat{\gamma}^\bmxi}\right] \right\rbrace \nonumber \\
&=& {\rm
Tr}\left[
\hat{\gamma}_{\rm KS}^\bmxi
\left(\hat{T}+\hat{V}_{\rm ext}\right)\right]
+
E^\bmxi_{\rm
Hxc}\left[n_{\hat{\gamma}_{\rm KS}^\bmxi}\right],
\end{eqnarray}
which is mathematically identical to its analog in GOK-DFT
[see Eq.~(\ref{eq:Ew_GOK-DFT_VP})], even though the physics it describes
is completely different. The orbitals $\lbrace \varphi_p^\bmxi\rbrace$, from which the minimizing 
single-configuration KS wave functions $\left\{\Phi_0^{M,\bmxi}\right\}$ in $\hat{\gamma}_{\rm KS}^\bmxi$
are constructed, fulfill self-consistent KS equations that are similar
to those of regular ($N$-electron ground-state) KS-DFT: 
\begin{eqnarray}
\label{eq:ensemble_KS_eq_xi}
\left( - \dfrac{\nabla_{\bfr}^2}{2} + v_{\rm ext}(\bfr) +
v_\Hxc^\bmxi[n^\bmxi](\bfr)  \right)\varphi_p^\bmxi (\bfr) =
\varepsilon_p^\bmxi \varphi^\bmxi_p(\bfr)
.
\end{eqnarray}
The only difference is that the $N$-centered ensemble Hxc potential $v_\Hxc^\bmxi[n] (\bfr)={\delta E^\bmxi_\Hxc [n]}/{\delta
n(\bfr)}$ is now employed. In the exact theory, the ensemble KS orbitals are
expected to reproduce the interacting $N$-centered ensemble density,
\ie, 
\begin{eqnarray}
\label{eq:nNxi_KSorbs}
 n_0^\bmxi(\bfr) &=& n_{\hat{\gamma}_{\rm KS}^\bmxi}(\bfr) 
\\
&=&
 \xi_{-}  n_{\Phi_0^{\Nm,\bmxi}}(\bfr)
+
\xi_{+} n_{\Phi_0^{\Np,\bmxi}}(\bfr)
\nonumber
\\
&&
+ \left(1 - \xim\dfrac{\Nm}{N} - \xip\dfrac{\Np}{N}
\right) n_{\Phi_0^{N,\bmxi}}(\bfr), 
\end{eqnarray}
or, equivalently~\cite{senjean2018unified}, 
\begin{eqnarray}
n_0^\bmxi(\bfr)= \left(1 + \dfrac{\xi_- - \xi_+}{N}\right) \sum_{p=1}^N
|{\varphi}_p^\bmxi(\bfr)|^2- \xi_- | {\varphi}_{N}^\bmxi(\bfr) |^2 +
\xi_+ | {\varphi}_{N+1}^\bmxi (\bfr) |^2.
\end{eqnarray}

Turning to the $N$-centered ensemble Hxc density functional, it can be decomposed as $E_\Hxc^\bmxi[n] = E_{\rm
Hx}^\bmxi [n] + E_{\rm c}^\bmxi[n]$, where, by analogy with GOK-DFT, the
exact Hx energy is expressed in terms of the $N$-centered ensemble
density-functional KS wave functions as follows,
\begin{eqnarray}
\label{eq:def_exact_ens_Hx_func_NeDFT}
\begin{split}
E_{\rm Hx}^\bmxi [n] &= \xi_{-}  \mel{{\Phi}_0^{N - 1,\bmxi}[n]}{\hat{W}_{\rm ee}}{{\Phi}_0^{N - 1,\bmxi}[n]} 
\\
&\quad+
\xi_{+}  \mel{{\Phi}_0^{N + 1,\bmxi}[n]}{\hat{W}_{\rm ee}}{{\Phi}_0^{N + 1,\bmxi}[n]} \\ 
&\quad+ \left(1 - \xim\dfrac{\Nm}{N} - \xip\dfrac{\Np}{N}
\right) \mel{{\Phi}_0^{N,\bmxi}[n]}{\hat{W}_{\rm ee}}{{\Phi}_0^{N,\bmxi}[n]} 
,
\end{split}
\end{eqnarray}
and the complementary correlation functional reads as
\be
\label{eq:ens_corr_fun_components_NeDFT}
E^\bmxi_{\rm c}[n]&=&F^\bmxi[n]-T^\bmxi_{\rm s}[n]-E^\bmxi_{\rm Hx}[n]
\nonumber \\
&=&
 \xi_{-} \left(\expval{\hat{T} + \hat{W}_{\rm ee}}_{{\Psi}_0^{N -
1,\bmxi}[n]} - \expval{\hat{T} + \hat{W}_{\rm ee}}_{{\Phi}_0^{N - 1,\bmxi}[n]}\right) \nonumber \\
&&+
\xi_{+}  \left( \expval{\hat{T} + \hat{W}_{\rm ee}}_{{\Psi}_0^{N +
1,\bmxi}[n]} - \expval{\hat{T} + \hat{W}_{\rm ee}}_{{\Phi}_0^{N + 1,\bmxi}[n]} \right) \nonumber \\ 
&&+ \left(1 - \xim\dfrac{\Nm}{N} - \xip\dfrac{\Np}{N}
\right) \nonumber \\
&&\times \left[
\expval{\hat{T} + \hat{W}_{\rm ee}}_{{\Psi}_0^{N,\bmxi}[n]} 
-
\expval{\hat{T} + \hat{W}_{\rm ee}}_{{\Phi}_0^{N,\bmxi}[n]} 
\right],
\ee
where $\lbrace \Psi_0^{M,\bmxi}[n]\rbrace$ denotes the interacting
density-functional $N$-centered ensemble.\\ 

When comparison is made with Sec.~\ref{subsubsec:GOK-DFT}, it becomes
clear that $N$-centered and GOK
eDFTs are essentially the same theory (they only differ in the definition
of the ensemble). From that point of view, we now have a unified eDFT for charged
and neutral electronic excitations. As a result, $N$-centered eDFT can
benefit from progress made in GOK-DFT, and {\it vice versa}.

\subsubsection{Exact ionization potential and electron affinity theorems}
\label{subsubsec:Ncentered_indiv}

We have shown in Sec.~\ref{subsubsec:Ncentered} that neutral, anionic,
and cationic ground-state energies can be extracted exactly from the $N$-centered ensemble
energy [see Eqs.~(\ref{eq:ENm_from_ensE}), (\ref{eq:ENp_from_ensE}), and
(\ref{eq:EN_from_ensE})]. We can now use the variational
density-functional expression of Eq.~(\ref{eq:ENxi_VP_KS}) to obtain
expressions for the fundamental gap, the ionization potential (IP), and
the electron affinity (EA). Note that these quantities are traditionally derived in the context of
DFT for fractional electron numbers~\cite{perdew1982density} (see Sec.~\ref{sec:DD} for a
detailed comparison).   
According to the Hellmann--Feynman theorem, we can express the weight
derivatives of the ensemble energy as follows, 
\begin{eqnarray}\label{eq:dENxidxi_KS}
\dxipm{E_0^\bmxi} &=& {\rm Tr}\left[ \Delta_\pm \hat{\gamma}^\bmxi_{\rm KS}\left( \hat{T} + \hat{V}_{\rm ext} \right) \right]
+ \left.\dxipm{E_\Hxc^\bmxi[n]}\right|_{n = n_{\hat{\gamma}_{\rm KS}^\bmxi}} \nonumber \\
&&+ \int \ddroit \bfr \dfrac{\delta E_\Hxc^\bmxi[n_{\hat{\gamma}_{\rm KS}^\bmxi}]}{\delta n(\bfr)}{\rm Tr}\left[\Delta_\pm \hat{\gamma}_{\rm KS}^\bmxi \hat{n}(\bfr) \right],
\end{eqnarray}
where $\Delta_\pm \hat{\gamma}^\bmxi_{\rm KS} = \ket{\Phi_0^{\Npm,\bmxi}}\bra{\Phi_0^{\Npm,\bmxi}} - \frac{\Npm}{N}\ket{\Phi_0^{N,\bmxi}}\bra{\Phi_0^{N,\bmxi}}$.
Since the single-configuration $M$-electron KS wave functions $\Phi_0^{M,\bmxi}$ are constructed from orbitals
that fulfill the KS Eq.~(\ref{eq:ensemble_KS_eq_xi}), the above energy
derivative can be rewritten in terms of the KS orbital energies
as~\cite{senjean2018unified}  
\begin{eqnarray}\label{eq:derivExi_xipm_KS}
\dxipm{E_0^\bmxi} 
=
\pm \dfrac{1}{N} \sum_{p=1}^N \left( \varepsilon^\bmxi_{N + \frac{1}{2} \pm \frac{1}{2}} - \varepsilon_p^\bmxi \right)
+
\left.
\dxipm{E_\Hxc^\bmxi[n]}\right|_{n = n_{\hat{\gamma}_{\rm KS}^\bmxi}}.
\end{eqnarray}
By plugging Eq.~(\ref{eq:derivExi_xipm_KS}) into
Eq.~(\ref{eq:derivExi_xi_fundgap}), we immediately obtain the following exact
expression for the fundamental gap:
\begin{eqnarray}\label{eq:fundgap_Ncentered_bmxi}
E_{\rm gap}^{\rm fund} = \varepsilon_{N+1}^\bmxi -
\varepsilon_{N}^\bmxi + \left. \left(\dxip{E_\Hxc^\bmxi[n]}+ \dxim{E_\Hxc^\bmxi[n]}\right)\right|_{n = n_{\hat{\gamma}_{\rm KS}^\bmxi}}
.
\end{eqnarray}

If we now apply the LZ shift-in-potential
procedure~\cite{levy2014ground}, by analogy with GOK-DFT (see
Sec.~\ref{subsubsec:indivGOK}), \ie,
\begin{eqnarray}
\label{eq:LZshift_NeDFT}
\dfrac{\delta E_\Hxc^\bmxi [n] }{\delta n(\bfr)} \rightarrow \overline{v}_\Hxc^\bmxi[n](\bfr) = \dfrac{\delta E_\Hxc^\bmxi [n] }{\delta n(\bfr)}  + \dfrac{
E_\Hxc^\bmxi[n] - \int \ddroit \bfr \, \dfrac{\delta E_\Hxc^\bmxi [n] }{\delta n(\bfr)} n(\bfr)}{
\int \ddroit \bfr \, n(\bfr)}
,
\nonumber \\
\end{eqnarray}
we can express both the ensemble energy and its derivatives in terms of
the LZ-shifted KS orbital energies $\overline{\varepsilon}_p^\bmxi$,
thus leading to the following compact
expressions for the ensemble and individual energies~\cite{senjean2018unified}, respectively:
\begin{eqnarray}
\label{eq:ENxi_KSorbs_LZ}
E_0^\bmxi = \left(1 + \dfrac{\xi_- - \xi_+}{N}\right) \sum_{p=1}^N
\overline{\varepsilon}_p^\bmxi - \xi_- \overline{\varepsilon}_{N}^\bmxi
+ \xi_+ \overline{\varepsilon}_{N+1}^\bmxi,
\end{eqnarray}
\begin{eqnarray}\label{eq:ind_ener_from_N-eDFT_m}
E_0^{\Nm}&=&
\sum^{\Nm}_{p=1}\overline{\varepsilon}_p^\bmxi
+
\left(1
-\dfrac{(\Nm)\xim}{N}
\right)
\left.\dfrac{\partial E^\bmxi_{\rm
Hxc}[n]}{\partial\xim}\right|_{n=n_{\hat{\gamma}_{\rm KS}^\bmxi}}
\nonumber
\\
&&-
\dfrac{(\Nm)\xip}{N}
\left.\dfrac{\partial E^\bmxi_{\rm
Hxc}[n]}{\partial\xip}\right|_{n=n_{\hat{\gamma}_{\rm KS}^\bmxi}}
,\nonumber \\
\end{eqnarray}
\begin{eqnarray}\label{eq:ind_ener_from_N-eDFT_p}
E_0^{\Np}&=&
\sum^{\Np}_{p=1}\overline{\varepsilon}_p^\bmxi
+
\left(
1
-\dfrac{(\Np)\xip}{N}
\right)
\left.\dfrac{\partial E^\bmxi_{\rm
Hxc}[n]}{\partial\xip}\right|_{n=n_{\hat{\gamma}_{\rm KS}^\bmxi}}
\nonumber
\\
&&-\dfrac{(\Np)\xim}{N}
\left.\dfrac{\partial E^\bmxi_{\rm
Hxc}[n]}{\partial\xim}\right|_{n=n_{\hat{\gamma}_{\rm KS}^\bmxi}}
,\nonumber \\
\end{eqnarray}
and
\begin{eqnarray}\label{eq:ind_ener_from_N-eDFT}
E_0^{N}=
\sum^{N}_{p=1}\overline{\varepsilon}_p^\bmxi
-\xim
\left.\dfrac{\partial E^\bmxi_{\rm
Hxc}[n]}{\partial\xim}\right|_{n=n_{\hat{\gamma}_{\rm KS}^\bmxi}}
-\xip
\left.\dfrac{\partial E^\bmxi_{\rm
Hxc}[n]}{\partial\xip}\right|_{n=n_{\hat{\gamma}_{\rm KS}^\bmxi}}
.\nonumber \\
\end{eqnarray}
By subtraction, we immediately obtain in-principle-exact IP and EA
theorems: 
\begin{eqnarray}\label{eq:IP_general}
I_0^N &=& E_0^{N-1} - E_0^N \nonumber \\
&=& -\overline{\varepsilon}_{N}^\bmxi + \left( 1 + \dfrac{\xim}{N}\right)
\left.\dxim{E_\Hxc^\bmxi[n]}\right|_{n = n_{\hat{\gamma}_{\rm KS}^\bmxi}} + \dfrac{\xip}{N} \left.\dxip{E_\Hxc^\bmxi[n]}\right|_{n = n_{\hat{\gamma}_{\rm KS}^\bmxi}},
\end{eqnarray}
and
\begin{eqnarray}\label{eq:EA_general}
A_0^N &=& E_0^{N} - E_0^\Np \nonumber \\
&=& -\overline{\varepsilon}_{N+1}^\bmxi - \left( 1 - \dfrac{\xip}{N}\right)
\left.\dxip{E_\Hxc^\bmxi[n]}\right|_{n = n_{\hat{\gamma}_{\rm KS}^\bmxi}} + \dfrac{\xim}{N} \left.\dxim{E_\Hxc^\bmxi[n]}\right|_{n = n_{\hat{\gamma}_{\rm KS}^\bmxi}}.
\end{eqnarray}
Interestingly, in the regular ground-state $N$-electron limit (\ie, when $\bmxi =
0$), the expression of Levy and Zahariev~\cite{levy2014ground} is
recovered for the IP,
\begin{eqnarray}\label{eq:IP}
I_0^N &=& - \overline{\varepsilon}_{N}^\bmxizero + \left.\dfrac{\partial E_\Hxc^\bmxi [n_{\Psi_0}]}{\partial \xi_-}\right|_{\bmxi = 0}, 
\end{eqnarray}
where the asymptotic value of the LZ-shifted Hxc potential away from the system [see Ref.~\cite{levy2014ground} and Eq.~(\ref{eq:IP_theo_Nc-eDFT})]
can now be expressed explicitly, within the $N$-centered ensemble formalism, as
$\left.{\partial E_\Hxc^\bmxi [n_{\Psi_0}]}/{\partial
\xi_-}\right|_{\bmxi = 0}$.
Similarly, we obtain the following expression for the EA:
\begin{eqnarray}\label{eq:EA}
A_0^N &=& - \overline{\varepsilon}_{N+1}^\bmxizero - \left.\dfrac{\partial E_\Hxc^\bmxi [n_{\Psi_0}]}{\partial \xi_+}\right|_{\bmxi = 0}.
\end{eqnarray}
As readily seen from the above expressions, neutral and charged
systems cannot be described with the same (LZ-shifted) Hxc potential.
As shown in Sec.~\ref{sec:DD}, the additional ensemble weight derivative
correction [second term on the right-hand side of Eqs.~(\ref{eq:IP}) and
(\ref{eq:EA})] is actually connected to the concept of derivative
discontinuity which manifests in conventional DFT for fractional
electron numbers, when crossing an integer~\cite{perdew1983physical}.   



\section{Equivalence between weight derivatives and xc derivative discontinuities}
\label{sec:DD}

The concept of derivative discontinuity originally appeared in the
context of DFT for fractional electron numbers~\cite{perdew1982density},
which is the conventional theoretical framework for the description of
charged excitations. The (xc functional) derivative discontinuities play
a crucial role in the evaluation of fundamental
gaps~\cite{perdew1983physical}. More specifically, they correct the
bare KS gap which is only an approximation to the true
interacting gap. It is well known that standard (semi)-local DFAs do not
contain such discontinuities, which explains why post-DFT methods based
on Green functions, for example, are preferred for the computation of
accurate gaps~\cite{onida2002electronic,sottile2007efficient,bruneval2012ionization,bruneval2012benchmarking,jiang2015first,pacchioni2015first,ou2016comparison,reining2018gw}.
Their substantially higher computational cost is
a motivation for exploring simpler (frequency-independent)
strategies. The recently proposed $N$-centered ensemble
formalism~\cite{senjean2018unified,senjean2020n}, which has been
introduced in Sec.~\ref{subsec:Ncentered}, is (among
others~\cite{kraisler2013piecewise,kraisler2014fundamental,baerends2017kohn,hodgson2017interatomic,Guandalini2019Fundamental,Guandalini2021Density,Rauch2020Accurate,Rauch2020Accurate_erratum}) promising in this respect.\\

From a more fundamental point of view, it is important to clarify the
similarities and differences between $N$-centered eDFT and the standard
formulation of DFT for charged excitations, which is often referred to
as Perdew--Parr--Levy--Balduz (PPLB) DFT~\cite{perdew1982density}. More
specifically, we should explain what the derivative discontinuity, which
is central in PPLB, becomes when switching to the $N$-centered
formalism. This is the purpose of this section.\\

After a brief review in Sec.~\ref{subsec:PPLB} of the  
PPLB formalism and its implications,
we will show (in Sec.~\ref{subsec:comparison_Ncentered}), on the basis of Ref.~\cite{hodgson2021exact}, that
derivative discontinuities exist also in $N$-centered eDFT and that they
are
directly connected to the ensemble weight derivatives of the xc
functional, like in GOK-DFT~\cite{levy1995excitation}. Finally, we will
explain in Sec.~\ref{subsec:removing_dd_xc_pot} why these
discontinuities can essentially be
removed from the theory, unlike in PPLB, and discuss the practical
implications.   



\subsection{Review of the regular PPLB approach to charged excitations}\label{subsec:PPLB}


\subsubsection{Ensemble formalism for open systems}

The key idea in PPLB is to describe electron ionization or affinity
processes through a continuous variation of the electron number, hence
the need for an extension of DFT to fractional electron numbers.
For that purpose, the energy of an artificial (zero-temperature) grand-canonical-type ensemble, which
should not be confused with physical finite-temperature
grand-canonical ensembles of statistical physics~\cite{MP20_Baerends_problem_PPLB}, is
constructed as follows,
\be
\mathcal{G}(\mu)=\min_{M}\left\{E^M_0-\mu M\right\},
\ee
where we minimize over {\it integer} numbers $M$ of
electrons and $E^M_0$ denotes the exact $M$-electron ground-state energy of the system. In this formalism, the number of
electrons in the system can be arbitrarily fixed by tuning the chemical potential
$\mu$. 
For example, if the following inequalities are fulfilled, 
\be
E^{N-1}_0-\mu(N-1)>E^N_0-\mu N<E^{N+1}_0-\mu(N+1),
\ee 
or, equivalently,
\be\label{eq:constraint_chem_pot_int_e_nbr}
-I_0^N<\mu<-A_0^N,
\ee
then the system contains an integer number $N$ of electrons (it is
assumed that the $N$-electron fundamental gap $E^N_g=I_0^N-A_0^N$ is
positive, so that Eq.~(\ref{eq:constraint_chem_pot_int_e_nbr}) can be fulfilled).
In the special case where one of the inequality becomes a strict
equality, say 
\be
E^{N-1}_0-\mu(N-1)=E^N_0-\mu N,
\ee
which means that the chemical potential is exactly equal to minus the $N$-electron ionization potential,
\be\label{eq:fixed_chem_pot_interacting}
\mu=E^N_0-E^{N-1}_0=-I_0^N,
\ee
the $N$- and $(N-1)$-electron solutions are {\it degenerate}
(grand-canonical energy wise). Therefore, they can be {\it mixed} as follows,
\be
\mathcal{G}\left(\mu\right)&\overset{\mu=-I_0^N}{=}&(1-\alpha)\Big(E^{N-1}_0-\mu(N-1)\Big)+\alpha\Big(E^N_0-\mu N\Big)
\\
\label{eq:grand_can_ener_alpha}
&=&\Big((1-\alpha)E^{N-1}_0+\alpha E^N_0\Big)-\mu\Big(N-1+\alpha\Big)
\\
\label{eq:grand_can_ener_frac_N}
&\equiv& E_0^{\mathcal{N}}-\mu\mathcal{N},
\ee
where $0\leq \alpha\leq 1$,
thus allowing for a {\it continuous} variation of the electron number
$\mathcal{N}$ (which now becomes fractional)
from $N-1$ to $N$:
\be\label{eq:frac_N_PPLB}
\mathcal{N}\equiv N-1+\alpha.
\ee
This is the central idea in PPLB for describing the ionization of an
$N$-electron system. Ionizing the $(N+1)$-electron system gives access to
the $N$-electron affinity. Interestingly, we recover from
Eqs.~(\ref{eq:grand_can_ener_alpha}), (\ref{eq:grand_can_ener_frac_N}), and (\ref{eq:frac_N_PPLB}) the well-known piecewise linearity of the energy with respect to the electron number~\cite{perdew1982density}:
\be
E_0^{\mathcal{N}}\equiv(1-\alpha)E^{N-1}_0+\alpha
E^N_0=(N-\mathcal{N})E^{N-1}_0+(\mathcal{N}-N+1)E^N_0.
\ee
In order to establish a clearer connection between the PPLB and $N$-centered
formalisms, we follow the approach of Kraisler and
Kronik~\cite{kraisler2013piecewise} where the ensemble weight $\alpha$
is used as a
variable, in place of the electron number $\mathcal{N}$. Therefore, in
PPLB, the ensemble energy reads as 
\be\label{eq:PPLB_ens_ener}
E^\alpha&=&(1-\alpha)E^{N-1}_0+\alpha E^N_0
\\
&=&(1-\alpha)\mel{\Psi^{N-1}_0}{\hat{T}+\hat{W}_{\rm ee}}{\Psi^{N-1}_0}+\alpha\mel{\Psi^{N}_0}{\hat{T}+\hat{W}_{\rm ee}}{\Psi^{N}_0}
\nonumber
\\
&&+\int d\bfr\,v_{\rm ext}(\bfr)n^\alpha_0(\bfr),
\ee
where 
\be\label{eq:ens_dens_PPLB}
n^\alpha_0(\bfr)=(1-\alpha)n_{\Psi^{N-1}_0}(\bfr)+\alpha n_{\Psi^{N}_0}(\bfr)
\ee
is the exact ground-state ensemble density. Note that, if we introduce
the ensemble density matrix operator 
\be\label{eq:ensemble_alpha_PPLB}
\hat{\Gamma}_0^\alpha=(1-\alpha)\ket{\Psi_0^{N-1}}\bra{\Psi_0^{N-1}}+\alpha\ket{\Psi_0^{N}}\bra{\Psi_0^{N}},
\ee
the ensemble energy and density can be expressed in a compact way as
follows,
\be
E^\alpha=\Tr\left[\hat{\Gamma}_0^\alpha\hat{H}\right]
\ee
and
\be
n^\alpha_0(\bfr)=\Tr\left[\hat{\Gamma}_0^\alpha\hat{n}(\bfr)\right],
\ee
respectively.
 
\subsubsection{DFT for fractional electron numbers}

On the basis of the ``grand-canonical'' ensemble formalism introduced in
the previous section, we can extend the domain of definition of the
universal Hohenberg--Kohn functional $F[n]$ to densities $n$ that integrate to
fractional electron numbers, \ie,
\be\label{eq:int_dens_PPLB}
\int d\bfr\,n(\bfr)=N-1+\alpha,
\ee
as follows,
\be\label{eq:HK_PPLB_fun}
\begin{split}
F[n]&=(1-\alpha)\expval{\hat{T}+\hat{W}_{\rm ee}}_{\Psi^{N-1}_0[n]}
+\alpha\expval{\hat{T}+\hat{W}_{\rm ee}}_{\Psi^{N}_0[n]},
\end{split}
\ee
where the ground-state density-functional wave functions fulfill the
density constraint 
\be
(1-\alpha)n_{\Psi^{N-1}_0[n]}(\bfr)+\alpha n_{\Psi^{N}_0[n]}(\bfr)=n(\bfr).
\ee
From now on we will take $\alpha$ in the range 
\be
0<\alpha\leq 1,
\ee
so that the {\it integer} electron number case systematically
corresponds to $\alpha=1$. Therefore, in the present density-functional PPLB ensemble, the
$N$-electron state will always contribute (even infinitesimally), and 
\begin{align}\label{eq:ceil_value_dens}
N=\ceil{\smallint d\bfr\,n(\bfr)}.
\end{align}
At this point it is essential to realize that, unlike in $N$-centered
eDFT, the ensemble weight $\alpha$ is {\it not} an independent variable.
Indeed, according to Eqs.~(\ref{eq:int_dens_PPLB}) and (\ref{eq:ceil_value_dens}),
it is an explicit functional of the density: 
\be
\alpha\equiv \alpha[n]=\int d\bfr\,n(\bfr)-\ceil{\smallint d\bfr\,n(\bfr)}+1.
\ee 
Therefore, in PPLB, the ensemble is fully determined from the density.
The latter
remains, like in regular DFT for integer electron numbers, the sole
basic variable in the theory. Following Levy and
Lieb~\cite{levy1979universal,levy1982electron,LFTransform-Lieb}, the
extended universal functional of Eq.~(\ref{eq:HK_PPLB_fun}) can be expressed in a compact way as
follows,
\be
F[n]=\min_{\hat{\gamma}^\alpha\rightarrow
n}\Tr\left[\hat{\gamma}^\alpha\left(\hat{T}+\hat{W}_{\rm
ee}\right)\right],
\ee
where we minimize over grand-canonical ensemble density matrix operators 
\be
\hat{\gamma}^\alpha\equiv(1-\alpha)\ket{\Psi^{N-1}}\bra{\Psi^{N-1}}+\alpha\ket{\Psi^{N}}\bra{\Psi^{N}}
\ee
that fulfill the following density contraint:
\be
\Tr\left[\hat{\gamma}^\alpha\hat{n}(\bfr)\right]=n_{\hat{\gamma}^\alpha}(\bfr)=(1-\alpha)n_{\Psi^{N-1}}(\bfr)+\alpha n_{\Psi^{N}}(\bfr)=n(\bfr).
\ee  

\subsubsection{Kohn--Sham PPLB}

The commonly used KS formulation of PPLB is recovered when introducing
the non-interacting kinetic energy functional
\be\label{eq:Ts_PPLB}
T_{\rm s}[n]=\min_{\hat{\gamma}^\alpha\rightarrow n}\Tr\left[\hat{\gamma}^\alpha\hat{T}\right]
\ee
and the in-principle-exact decomposition
\be\label{eq:KS_decomp_PPLB}
F[n]=T_{\rm s}[n]+E_{\rm Hxc}[n],
\ee
where the Hxc functional now applies to fractional electron
numbers. Let us stress that, unlike in $N$-centered eDFT, the Hxc
functional has no ensemble weight
dependence because the weight is determined from the density $n$. 
Any dependence in $\alpha$ is incorporated into the functional through
the density. This is a major difference with $N$-centered eDFT where the
ensemble weight and the density are {\it independent} variables, like in
GOK-DFT.     
This
subtle point will be central later on when comparing the two theories.\\

According to the variational principle, the exact ensemble energy can be
determined, for a given and {\it fixed} value of $\alpha$, as follows, 
\be
E^\alpha=\min_{n\rightarrow N-1+\alpha}\left\{F[n]+\int d\bfr\,v_{\rm
ext}(\bfr)n(\bfr)\right\},
\ee
where we minimize over densities that integrate to the desired number
$N-1+\alpha$ of electrons. According to Eqs.~(\ref{eq:Ts_PPLB})
and (\ref{eq:KS_decomp_PPLB}), the ensemble energy can be rewritten as
\be\label{eq:VP_KS_frac_N}
\begin{split}
E^\alpha&=
\min_{n\rightarrow
N-1+\alpha}\left\{\min_{\hat{\gamma}^\alpha\rightarrow n}\left\{\Tr\left[\hat{\gamma}^\alpha\left(\hat{T}+\hat{V}_{\rm ext}\right)\right]+E_{\rm Hxc}[n_{\hat{\gamma}^\alpha}]\right\}\right\}
\\
&=\min_{\hat{\gamma}^\alpha}\left\{\Tr\left[\hat{\gamma}^\alpha\left(\hat{T}+\hat{V}_{\rm ext}\right)\right]+E_{\rm Hxc}[n_{\hat{\gamma}^\alpha}]\right\}
\\
&\equiv\Tr\left[\hat{\gamma}_{\rm KS}^\alpha\left(\hat{T}+\hat{V}_{\rm
ext}\right)\right]+E_{\rm Hxc}[n_{\hat{\gamma}_{\rm KS}^\alpha}],
\end{split}
\ee
where the minimizing KS density matrix operator
\be\label{eq:mini_KS_DM_PPLB}
\hat{\gamma}_{\rm KS}^\alpha=(1-\alpha)\ket{\Phi_0^{N-1,\alpha}}\bra{\Phi_0^{N-1,\alpha}}+\alpha\ket{\Phi_0^{N,\alpha}}\bra{\Phi_0^{N,\alpha}}
\ee
reproduces the exact ensemble density of Eq.~(\ref{eq:ens_dens_PPLB}):
\be
n_{\hat{\gamma}_{\rm KS}^\alpha}(\bfr)=\Tr\left[\hat{\gamma}^\alpha\hat{n}(\bfr)\right]=n^\alpha_0(\bfr).
\ee
The orbitals from which $\Phi_0^{N-1,\alpha}$ and $\Phi_0^{N,\alpha}$
are constructed fulfill self-consistent KS equations,
\be\label{eq:SC_KS_eq_PPLB}
\left(-\dfrac{1}{2}\nabla^2_{\bfr}+v_{\rm ext}(\bfr)+\dfrac{\delta E_{\rm Hxc}[n_{\hat{\gamma}_{\rm KS}^\alpha}]}{\delta n(\bfr)}\right)
\varphi_i^\alpha(\bfr)=\varepsilon_i^\alpha\varphi_i^\alpha(\bfr),
\ee  
where, as readily seen from the following ensemble density expression, 
\be\label{eq:KS_dens_frac_occ}
\begin{split}
n_{\hat{\gamma}_{\rm KS}^\alpha}(\bfr)&=(1-\alpha)\sum^{N-1}_{i=1}\abs{\varphi_i^\alpha(\bfr)}^2+\alpha \sum^{N}_{i=1}\abs{\varphi_i^\alpha(\bfr)}^2
\\
&=\sum^{N-1}_{i=1}\abs{\varphi_i^\alpha(\bfr)}^2+\alpha\abs{\varphi_N^\alpha(\bfr)}^2
,
\end{split}
\ee
the {\it highest occupied molecular orbital} (HOMO) [\ie,
$\varphi_{N}^\alpha$] is {\it fractionally} occupied. This is
the main difference with conventional DFT calculations for integer electron
numbers.\\

\subsubsection{Janak's theorem and its implications}

Once the ensemble energy $E^\alpha$ has been determined (variationally), we can evaluate the IP, which is the
quantity we are interested in, by differentiation with respect to the
ensemble weight $\alpha$ [see Eq.~(\ref{eq:PPLB_ens_ener})], \ie, 
\be
\dfrac{dE^\alpha}{d\alpha}=-I^N_0,
\ee
which, according to the Hellmann--Feynman theorem and
Eqs.~(\ref{eq:VP_KS_frac_N}), (\ref{eq:mini_KS_DM_PPLB}), and
(\ref{eq:SC_KS_eq_PPLB}), can be written more explicitly as follows, 
\be\label{eq:deriv_alpha_ener_PPLB}
\dfrac{dE^\alpha}{d\alpha}&=&\mel{\Phi_0^{N,\alpha}}{\hat{T}+\hat{V}_{\rm ext}}{\Phi_0^{N,\alpha}}-\mel{\Phi_0^{N-1,\alpha}}{\hat{T}+\hat{V}_{\rm ext}}{\Phi_0^{N-1,\alpha}}
\nonumber
\\
&&\quad+\int d\bfr\dfrac{\delta E_{\rm Hxc}[n_{\hat{\gamma}_{\rm KS}^\alpha}]}{\delta n(\bfr)}\left(n_{\Phi_0^{N,\alpha}}(\bfr)-n_{\Phi_0^{N-1,\alpha}}(\bfr)\right)
\\
&\equiv& \sum^N_{i=1}\varepsilon_i^\alpha-\sum^{N-1}_{i=1}\varepsilon_i^\alpha
\nonumber
\\
&=&\varepsilon_N^\alpha,
\ee
thus leading to the famous Janak's theorem~\cite{janak1978proof}:
\be\label{eq:Janak_theorem}
I^N_0=-\varepsilon_N^\alpha,\hspace{0.2cm}\forall\alpha\in]0,1].
\ee     
As readily seen from Eq.~(\ref{eq:Janak_theorem}), the energy $\varepsilon_N^\alpha$ of the KS HOMO does
not vary with the 
fraction $\alpha>0$ of electron that is introduced into the
$(N-1)$-electron system. Therefore, it matches the $N$-electron KS HOMO
energy that we simply denote $\varepsilon_N^N$: 
\be\label{eq:IP_theorem}
\varepsilon_N^\alpha=\varepsilon_N^{\alpha=1}\equiv \varepsilon_N^N=-I^N_0.
\ee 
At this point it is important to mention that, unlike in $N$-centered
eDFT [see Eq.~(\ref{eq:IP})], there is no ensemble weight derivative of the Hxc functional
involved in Janak's theorem. Such a quantity does not exist in PPLB, simply because the ensemble
weight $\alpha$ and the density $n$ {cannot} vary independently.
However, while the number of electrons is artificially held constant in
the $N$-centered formalism, it is not the case in PPLB.
Indeed, variations in $\alpha$ induce a change in density [see the third contribution on the right-hand
side of Eq.~(\ref{eq:deriv_alpha_ener_PPLB})] that does not integrate to
zero:
\be\label{eq:int_dn_dalpha_beta}
1=\int d\bfr \left(n_{\Phi_0^{N,\alpha}}(\bfr)-n_{\Phi_0^{N-1,\alpha}}(\bfr)\right)
\neq 0.
\ee
Therefore, it is crucial, when evaluating the functional derivative of
the Hxc energy ${\delta E_{\rm Hxc}[n_{\hat{\gamma}_{\rm
KS}^\alpha}]}/{\delta n(\bfr)}$ (\ie, the Hxc potential), to consider
variations of the density $\delta n(\bfr)$ that do not integrate to zero. This is
unnecessary in $N$-centered eDFT. 
In PPLB, however, the proper modeling of the xc potential is essential for
describing charged excitations. This is clearly
illustrated by the fact that the exact xc potential exhibits derivative
discontinuities 
when crossing an integer
electron number, as discussed further in Sec.~\ref{subsec:comparison_Ncentered}.\\
 

Let us finally discuss the unicity of the xc potential. 
We recall that, in the present review, the external
potential is simply the (Coulomb) nuclear potential
of the molecule under study. It is {\it fixed} and it vanishes away from the system:
\be
v_{\rm ext}(\bfr)
\underset{\abs{\bfr}\rightarrow +\infty}{\rightarrow}0,
\ee   
which we simply denote $v_{\rm
ext}(\infty)=0$ in the following.
As readily seen from Eq.~(\ref{eq:Janak_theorem}), when describing
a continuous
variation of the
electron number $\mathcal{N}$ in the range $N-1<\mathcal{N}<N$, the KS potential becomes truly
unique, not anymore up to a constant. This can be related to the unicity of the
chemical potential which allows for fractional electron numbers, as
discussed previously in the interacting case [see
Eq.~(\ref{eq:fixed_chem_pot_interacting})]. As a result, the xc
potential is truly unique. More precisely, as illustrated in
Appendix~\ref{app:v_xc_infty} for a one-dimensional (1D) system, Janak's theorem 
implies that~\cite{Levy1984_asymptotic_dens}         
\be
\left.\dfrac{\delta E_{\rm xc}[n_{\hat{\gamma}_{\rm KS}^\alpha}]}{\delta
n(\bfr)}\right|_{\abs{\bfr}\rightarrow +\infty}\equiv
v^\alpha_{\rm xc}(\infty)=0.
\ee

\subsubsection{Fundamental gap problem}

According to Janak's theorem, the fundamental gap can be evaluated in PPLB, in
principle exactly, from the HOMO energies as follows,
\be
E^N_g=I_0^N-I_0^{N+1}=\varepsilon_{N+1}^{N+1}-\varepsilon_N^N.
\ee
What is truly challenging in practice, in particular in solids~\cite{perdew2017understanding},
is the extraction of this gap from a single $N$-electron calculation.
Indeed, the HOMO energy $\varepsilon_{N+1}^{N+1}$ of the
$(N+1)$-electron system has no reason to match the {\it lowest
unoccupied molecular orbital} (LUMO) energy
$\varepsilon_{N+1}^N$ of the $N$-electron system, simply because the
infinitesimal addition of an electron to the latter system will affect
the density [see Eq.~(\ref{eq:KS_map_dens_PPLB_asymptotic})] and,
consequently, the xc potential. The impact of an electron addition on
the xc potential will be scrutinized in
Sec.~\ref{subsec:comparison_Ncentered}, in the context of $N$-centered eDFT. If we denote
\be
\Delta^N_{\rm xc}=\varepsilon_{N+1}^{N+1}-\varepsilon_{N+1}^N
\ee 
the deviation in energy between the above-mentioned HOMO and LUMO,
we recover the usual expression~\cite{perdew1983physical}
\be\label{eq:PPLB_exp_gap_with_DDy}
E^N_g=\varepsilon_{N+1}^N-\varepsilon_N^N+\Delta^N_{\rm xc},
\ee
where $\Delta^N_{\rm xc}$ can now be interpreted as the difference in
gap between the physical and KS systems. As readily seen from
the key Eq.~(\ref{eq:fundgap_Ncentered_bmxi}) of $N$-centered eDFT, that we take
in the regular $N$-electron ground-state DFT limit (\ie, $\xi_+=\xi_-=0$), $\Delta^N_{\rm xc}$ is indeed a nonzero
correction to the KS gap that can be expressed more explicitly as follows,
\be\label{eq:DDty_from_Nc-eDFT_weight0}
\Delta^N_{\rm xc}=\left.
\dxim{E_\xc^{(\xi_-,0)}[n_{\Psi^N_0}]}\right|_{\xi_-=0}
+\left.
\dxip{E_\xc^{(0,\xi_+)}[n_{\Psi^N_0}]}\right|_{\xi_+=0}.
\ee
Note that we used in Eq.~(\ref{eq:DDty_from_Nc-eDFT_weight0}) the
in-principle-exact decomposition 
\be\label{eq:H_andxc_decomp_Nc-eDFT}
E^{(\xi_-,\xi_+)}_\Hxc[n]=E_{\rm
H}[n]+E^{(\xi_-,\xi_+)}_{\rm xc}[n],
\ee
where the regular
(weight-independent) Hartree functional is employed. The practical
disadvantage of such a decomposition will be extensively discussed in
Sec.~\ref{sec:EXX}. We focus here on the exact theory.\\

In the language of
$N$-centered eDFT, $\Delta^N_{\rm xc}$ describes the variation in
ensemble weights (while holding the ensemble density fixed and equal to
the $N$-electron ground-state density $n_{\Psi^N_0}$) of the $N$-centered ensemble xc
energy due to the infinitesimal removal/addition of an electron from/to
the $N$-electron system.
Evidently, standard (local or semi-local) DFAs do not
incorporate such a weight dependence because they were not designed for 
$N$-centered eDFT calculations (we recall that the concept of
$N$-centered ensemble has been proposed quite
recently~\cite{senjean2018unified,senjean2020n,hodgson2021exact}). Therefore, when such DFAs are used, the physical gap
is systematically
approximated by the (also approximate) KS one. Note that the resulting
underestimation of the fundamental gap is highly problematic, for
example, when computing transport
properties~\cite{PRB12_Kummel_transport_hydrogen,PRB06_Evers_transport_DD_problem}.
The interpretation that is given in PPLB for $\Delta^N_{\rm xc}$ is
completely different. The latter actually
originates from the discontinuity that the xc potential (which
is the functional derivative of the xc energy) exhibits when
crossing an integer electron number, hence the name {\it derivative
discontinuity}. In the language of PPLB, $\Delta^N_{\rm xc}$ is not
described at all when (semi-) local xc functionals are employed, simply because the latter do not
incorporate functional derivative discontinuities. The connection
between these two very different interpretations
will be made in Sec.~\ref{subsec:comparison_Ncentered}.\\

\subsubsection{Exchange-only derivative discontinuity}
 
Let us finish the previous discussion with a detailed comment on the use
of (orbital-dependent) exact exchange energies, which
is often recommended for improving the description of fundamental
gaps~\cite{perdew2017understanding}. From the perspective of
$N$-centered eDFT, using an exact exchange energy (or a fraction of it) is a way to incorporate weight dependencies
into the ensemble exchange density functional. Indeed, according to Eq.~(\ref{eq:H_andxc_decomp_Nc-eDFT}),
the exact exchange-only derivative discontinuity can be rewritten as
\be\label{eq:DDy_x_simp}
\Delta^N_{\rm x}=\left.\dfrac{\partial E^{(\xi,\xi)}_{\rm
x}[n_{\Psi^N_0}]}{\partial \xi}\right|_{\xi=0}
=\left.\dfrac{\partial E^{(\xi,\xi)}_{\rm
Hx}[n_{\Psi^N_0}]}{\partial \xi}\right|_{\xi=0},
\ee
or, equivalently,
\be\label{eq:DDy_x_other_exp}
\begin{split}
\Delta^N_{\rm x}&=\left.\dfrac{\partial E^{(\xi,\xi)}_{\rm
Hx}[n_0^{\xi,\xi}]}{\partial \xi}\right|_{\xi=0}
\\
&\quad-\left.\dfrac{\partial E^{(\xi,\xi)}_{\rm
Hx}[n_0^{\xi,\alpha}]}{\partial \alpha}\right|_{\alpha=\xi=0}
-\left.\dfrac{\partial E^{(\xi,\xi)}_{\rm
Hx}[n_0^{\alpha,\xi}]}{\partial \alpha}\right|_{\alpha=\xi=0}
,
\end{split}
\ee
where we have introduced the double-weight $N$-centered ensemble KS density 
\be\label{eq:2ble_weight_dens_Nc}
n_0^{\alpha,\xi}(\bfr):=(1-2\alpha)n_{\Phi^{N,(\xi,\xi)}_0}(\bfr)+\alpha
n_{\Phi^{N-1,(\xi,\xi)}_0}(\bfr)+\alpha 
n_{\Phi^{N+1,(\xi,\xi)}_0}(\bfr), 
\ee
which reduces to $n_{\Psi^N_0}$ when $\alpha=\xi=0$. Thus, we can remove
all the contributions involving the derivative of
the ensemble density [see the second line of
Eq.~(\ref{eq:DDy_x_other_exp})] that are erroneously introduced by the
first term on the right-hand side of
Eq.~(\ref{eq:DDy_x_other_exp})]. We recall that, in the evaluation of
$\Delta^N_{\rm x}$, we must differentiate with respect to the weight for a fixed
density, as readily seen from Eq.~(\ref{eq:DDy_x_simp}). At first
sight, Eq.~(\ref{eq:DDy_x_other_exp}) is uselessly complicated, when
compared with Eq.~(\ref{eq:DDy_x_simp}), but it will actually enable us to obtain
simpler expressions. This trick has been introduced in the context of
GOK-DFT~\cite{loos2020weight,fromager2020individual}. First we need to 
realize that, according to the exact expression of the $N$-centered ensemble Hx functional
in Eq.~(\ref{eq:def_exact_ens_Hx_func_NeDFT}), 
\be\label{eq:Nc-EEXX_two_weight_energy}
\begin{split}
E^{(\xi,\xi)}_{\rm
Hx}[n_0^{\xi,\alpha}]
&=(1-2\xi)\expval{\hat{W}_{\rm
ee}}_{\Phi^{N,(\alpha,\alpha)}_0}+\xi\expval{\hat{W}_{\rm 
ee}}_{\Phi^{N-1,(\alpha,\alpha)}_0}
\\
&\quad+\xi\expval{\hat{W}_{\rm
ee}}_{\Phi^{N+1,(\alpha,\alpha)}_0}.
\end{split}
\ee
Moreover, we have 
\be
\dfrac{\partial E^{(\xi,\xi)}_{\rm
Hx}[n_0^{\alpha,\xi}]}{\partial \alpha}=\int d\bfr\dfrac{\delta E^{(\xi,\xi)}_{\rm Hx}[n_0^{\alpha,\xi}]}{\delta
n(\bfr)}\dfrac{\partial
n_0^{\alpha,\xi}(\bfr)}{\partial\alpha},
\ee  
where, in the $\alpha=\xi=0$ limit, the derivative of the density
\be\label{eq:deriv_two_weight_dens_Nc}
\begin{split}
\left.\dfrac{\partial
n_0^{\alpha,\xi}(\bfr)}{\partial\alpha}\right|_{\alpha=\xi=0}&=n_{\Phi^{N+1}_0}(\bfr)+n_{\Phi^{N-1}_0}(\bfr)-2n_{\Phi^{N}_0}(\bfr)
\\
&=\abs{\varphi^N_{N+1}(\bfr)}^2-\abs{\varphi^N_{N}(\bfr)}^2
\end{split}
\ee
can be evaluated from the regular $N$-electron KS frontier orbitals. 
By combining
Eqs.~(\ref{eq:Nc-EEXX_two_weight_energy})--(\ref{eq:deriv_two_weight_dens_Nc})
we finally obtain a simple (orbital-dependent) expression for the exchange-only
derivative discontinuity:
\be\label{eq:DDy_x-only}
\begin{split}
\Delta^N_{\rm x}&=\expval{\hat{W}_{\rm
ee}}_{\Phi^{N+1}_0}+\expval{\hat{W}_{\rm
ee}}_{\Phi^{N-1}_0}-2\expval{\hat{W}_{\rm
ee}}_{\Phi^{N}_0}
\\
&\quad-\int d\bfr\dfrac{\delta E_{\rm Hx}[n_{\Psi_0}]}{\delta
n(\bfr)}\left(\abs{\varphi^N_{N+1}(\bfr)}^2-\abs{\varphi^N_{N}(\bfr)}^2\right).
\end{split}
\ee
Adding this correction to the exact KS gap leads to the following 
approximate fundamental gap expression,
\be\label{eq:approx_gap_x-only}
\begin{split}
E_g^N\approx \varepsilon_{N+1}^N-\varepsilon_N^N+\Delta^N_{\rm x} 
&=
\expval{\hat{H}
}_{\Phi^{N+1}_0}+\expval{\hat{H}
}_{\Phi^{N-1}_0}-2\expval{\hat{H}
}_{\Phi^{N}_0}
\\
&\quad+\int d\bfr\dfrac{\delta E_{\rm c}[n_{\Psi_0}]}{\delta
n(\bfr)}\left(\abs{\varphi^N_{N+1}(\bfr)}^2-\abs{\varphi^N_{N}(\bfr)}^2\right),
\end{split}
\ee
where the physical interacting wave functions have been replaced by
the KS ones, and correlation is
introduced only through the correlation potential.\\

Note that a consistent implementation of Eq.~(\ref{eq:approx_gap_x-only}) on
the basis of Eq.~(\ref{eq:DDy_x_simp}) would in principle require using
{\it optimized effective potentials}
(OEPs)~\cite{gould2021ensemble,RMP08_Kronik_OEP}. Indeed, in
the present formulation of $N$-centered eDFT, the KS
orbitals are expected to be generated from a {\it local} (\ie, multiplicative) xc
potential [see Eq.~(\ref{eq:ensemble_KS_eq_xi})], unlike in
Hartree--Fock (HF)-based methods where the exchange
potential is nonlocal. In practice, we would have to consider a trial local
potential $v$ and determine the corresponding KS orbitals,
\be
\{\varphi_i\equiv
\varphi_i[v]\}&\longleftarrow&\left(-\dfrac{1}{2}\nabla^2_{\bfr}+v(\bfr)\right)\varphi_i(\bfr)=\varepsilon_i\varphi_i(\bfr),
\ee
thus ensuring that the KS wave functions in
Eq.~(\ref{eq:Nc-EEXX_two_weight_energy}) are eigenfunctions of a
non-interacting Hamiltonian, like in the exact theory. The ensemble
energy would
then be minimized with respect to $v$ rather than the orbitals (hence
the name OEP given to the method). Obviously, such a procedure induces a
substantial increase in computational complexity, even though simplifications can be made
in the optimization process~\cite{Nagy2001_GIC}. Alternative variational
evaluations of orbital-dependent ensemble exchange energies
exist~\cite{gould2021ensemble}.
Their practical advantages and drawbacks will be discussed in detail in Sec.~\ref{sec:EXX}.

\subsection{Connection between PPLB and $N$-centered pictures}\label{subsec:comparison_Ncentered}

Crossing an integer electron number, which is a key concept in PPLB, can
be described in the context of $N$-centered
eDFT by considering so-called left and right $N$-centered
ensembles~\cite{senjean2020n}. These ensembles are recovered when 
$\xi_+=0$ (electron removal only) and $\xi_-=0$ (electron addition
only), respectively. In the following, we will use the following
shorthand
notations, 
\be
\mbox{left $N$-centered ensemble:}\;\; (\xi_-,0)&\overset{notation}{\equiv}& \xi_-,
\\
\mbox{right $N$-centered ensemble:}\;\; (0,\xi_+)&\overset{notation}{\equiv}& \xi_+,
\ee
for convenience. For example, the right $N$-centered ensemble Hxc functional and 
KS orbital
energies will be denoted as $E^{\xi_+}_{\rm Hxc}[n]\equiv E^{(0,\xi_+)}_{\rm
Hxc}[n]$ and $\varepsilon^{\xi_+}_i\equiv \varepsilon^{(0,\xi_+)}_i$,
respectively. The exact left and right $N$-centered ensemble
densities read as [see Eq.~(\ref{eq:Nc_ens_dens_Bruno})]
\be\label{eq:left_Nc_dens}
n^{\xi_-}(\bfr)\equiv\left(1-\dfrac{(N-1)\xi_-}{N}\right)n_{\Psi^N_0}(\bfr)+\xi_-n_{\Psi^{N-1}_0}(\bfr),
\ee
and
\be\label{eq:right_Nc_dens}
n^{\xi_+}(\bfr)\equiv\left(1-\dfrac{(N+1)\xi_+}{N}\right)n_{\Psi^N_0}(\bfr)+\xi_+n_{\Psi^{N+1}_0}(\bfr),
\ee
respectively. Note that, with these notations, we have the following
equivalence relation,
\be\label{eq:equivalence_xi_minus_plus_zero}
\xi_-=0\Leftrightarrow \xi_+=0,
\ee  
as readily seen from Eqs.~(\ref{eq:left_Nc_dens}) and
(\ref{eq:right_Nc_dens}). When Eq.~(\ref{eq:equivalence_xi_minus_plus_zero}) is fulfilled, the system is
in its pure $N$-electron ground state which means, in the language of
PPLB, that it
contains exactly the integer number $N$ of electrons.\\

A clearer
connection between the two theories can be established by
comparing the two limits $\xi_+\rightarrow 0^+$ (which
describes the infinitesimal addition of an electron to the $N$-electron
system) and $\xi_+=0$ (or,
equivalently, $\xi_-=0$). For that purpose, we first need to realize
that, by analogy with PPLB (see
Appendix~\ref{app:v_xc_infty} for the proof in the simpler 1D case), the
exact IP/EA theorems of $N$-centered eDFT in Eqs.~(\ref{eq:IP_general}) and (\ref{eq:EA_general})
can be alternatively written as
follows~\cite{hodgson2021exact},   
\be\label{eq:EA_theo_Nc-eDFT} 
A_0^N=I_0^{N+1}\overset{\xi_+>0}{=}-\varepsilon^{\xi_+}_{N+1}+v^{\xi_+}_{\rm
xc}(\infty)
\ee
and
\be\label{eq:IP_theo_Nc-eDFT}
I_0^{N}\overset{\xi_-\geq0}{=}-\varepsilon^{\xi_-}_{N}+v^{\xi_-}_{\rm
xc}(\infty),
\ee
where we recall that $v^{\xi_\pm}_{\rm
xc}(\bfr)\equiv \left.\delta E^{\xi_\pm}_{\rm xc}[n]/\delta
n(\bfr)\right|_{n=n^{\xi_\pm}}$. 
Thus, from the explicit
expression of the LZ shift [see the second term on the right-hand side
of Eq.~(\ref{eq:LZshift_NeDFT})], we obtain the following exact
expressions for the asymptotic values of the right and left $N$-centered ensemble xc
potentials, respectively:   
\be\label{eq:link_pot_away_DDy_xiplus}
\begin{split}
v^{\xi_+}_{\rm
xc}(\infty) \overset{\xi_+>0}{=}&\left(\dfrac{\xi_+}{N}-1\right)\left.\dfrac{\partial E_{\rm
xc}^{\xi_+}[n]}{\partial \xi_+}\right|_{n=n^{\xi_+}}
\\
&-\dfrac{1}{N}\left(E_{\rm
Hxc}^{\xi_+}[n^{\xi_+}]-\int d\bfr\, v^{\xi_+}_{\rm
Hxc}(\bfr)n^{\xi_+}(\bfr)\right)
\end{split}
\ee 
and
\be\label{eq:link_pot_away_DDy_ximinus}
\begin{split}
v^{\xi_-}_{\rm
xc}(\infty) \overset{\xi_-\geq0}{=}&\left(\dfrac{\xi_-}{N}+1\right)\left.\dfrac{\partial E_{\rm
xc}^{\xi_-}[n]}{\partial \xi_-}\right|_{n=n^{\xi_-}}
\\
&-\dfrac{1}{N}\left(E_{\rm
Hxc}^{\xi_-}[n^{\xi_-}]-\int d\bfr\, v^{\xi_-}_{\rm
Hxc}(\bfr)n^{\xi_-}(\bfr)\right).
\end{split}
\ee 
We recall that the decomposition of
Eq.~(\ref{eq:H_andxc_decomp_Nc-eDFT}) is employed for a direct
comparison with PPLB. Let us now consider the
$\xi_+\rightarrow 0^+$ and $\xi_-=0$ limits in Eqs.~(\ref{eq:link_pot_away_DDy_xiplus}) and
(\ref{eq:link_pot_away_DDy_ximinus}), respectively. Since
\be
n^{\xi_+\rightarrow 0^+}(\bfr)&=&n^{\xi_-=0}(\bfr)=n_{\Psi^N_0}(\bfr),
\\
v^{\xi_+\rightarrow 0^+}_{\rm H}(\bfr)&=&v^{\xi_-=0}_{\rm H}(\bfr),
\\
\left.E_{\rm
Hxc}^{\xi_+}[n^{\xi_+}]\right|_{\xi_+\rightarrow 0^+}&=&\left.
E_{\rm
Hxc}^{\xi_-}[n^{\xi_-}]\right|_{\xi_-=0}=E_{\rm Hxc}[n_{\Psi^N_0}],
\ee
it comes, by subtraction,
\be\label{eq:link_deri_xi_DDy}
\displaystyle\int
\dfrac{d\bfr}{N}
\left(v^{\xi_+\rightarrow 0^+}_{\rm
xc}(\bfr)-v^{\xi_+=0}_{\rm
xc}(\bfr)\right)n_{\Psi^N_0}(\bfr)
&=v^{\xi_+\rightarrow 0^+}_{\rm
xc}(\infty)-v^{\xi_+=0}_{\rm
xc}(\infty)
+\Delta^N_{\rm xc},
\nonumber
\\
\ee
or, equivalently,
\be\label{eq:link_deri_xi_DDy_other}
\Delta^N_{\rm xc}
&=\displaystyle\int
\dfrac{d\bfr}{N} \Big[\left(v^{\xi_+\rightarrow 0^+}_{\rm
xc}(\bfr)-
v^{\xi_+\rightarrow 0^+}_{\rm
xc}(\infty)
\right)
-\left(v^{\xi_+=0}_{\rm
xc}(\bfr)
-v^{\xi_+=0}_{\rm
xc}(\infty)\right)\Big]n_{\Psi^N_0}(\bfr)
,
\nonumber
\\
\ee
where we
used Eq.~(\ref{eq:DDty_from_Nc-eDFT_weight0}) and the relation $v^{\xi_-=0}_{\rm
xc}(\bfr)=v^{\xi_+=0}_{\rm
xc}(\bfr)$, according to
Eq.~(\ref{eq:equivalence_xi_minus_plus_zero}). Note that, as readily
seen from Eq.~(\ref{eq:link_deri_xi_DDy_other}), $\Delta^N_{\rm
xc}$ is insensitive to constant shifts in the xc potential, as expected
from Eq.~(\ref{eq:PPLB_exp_gap_with_DDy}).\\

The connection that is made explicit in Eq.~(\ref{eq:link_deri_xi_DDy_other}) between the
$N$-centered ensemble weight derivative $\Delta^N_{\rm xc}$ of the xc density-functional
energy [see Eq.~(\ref{eq:DDty_from_Nc-eDFT_weight0})] and the xc potential
is an important result that was highlighted very recently in
Ref.~\cite{hodgson2021exact}. It proves that 
weight derivatives and derivative discontinuities are equivalent, thus
extending to charged excitations what was already known for neutral
excitations~\cite{levy1995excitation}. Indeed, if we
systematically {\it choose}
(but we do not have to in the $N$-centered formalism, unlike in PPLB) the xc
potential that asymptotically goes to zero, \ie, 
\be\label{eq:constraint_Nc-xc_pot_away}
v^{\xi_\pm}_{\rm
xc}(\infty){=}0,\;\; \xi_\pm\geq 0,
\ee
then we recover what looks like a Janak's theorem [see Eqs.~(\ref{eq:EA_theo_Nc-eDFT}) and
(\ref{eq:IP_theo_Nc-eDFT})] and, according to Eq.~(\ref{eq:link_deri_xi_DDy}),    
\be\label{eq:diff_int_pot_xiplus_zero-plus_zero}
\int d\bfr\left(v^{\xi_+\rightarrow 0^+}_{\rm
xc}(\bfr)-v^{\xi_+=0}_{\rm
xc}(\bfr)\right)n_{\Psi^N_0}(\bfr)=N\Delta^N_{\rm xc}\neq 0.
\ee 
It then becomes clear that, in the region of the system under study
(\ie, where the density $n_{\Psi^N_0}(\bfr)$ is nonzero), the xc
potentials
obtained in the $\xi_+\rightarrow 0^+$ and $\xi_+=0$ limits,
respectively, {\it cannot}
match. In order to fulfill the arbitrary constraint of
Eq.~(\ref{eq:constraint_Nc-xc_pot_away}), while still reproducing for
$\xi_+>0$ the
correct density in all regions of space [which includes a proper
description of the density's asymptotic
behavior (see Appendix~\ref{app:v_xc_infty})], the xc potential must be shifted
in the region the system, thus ensuring that the ground-state density
$n_{\Psi^N_0}(\bfr)$ is also correctly reproduced in that region. This has been
nicely illustrated in Ref.~\cite{hodgson2021exact} for an atom in 1D.
Thus, we recover a well-known result of PPLB: When an electron is
infinitesimally added (\ie, $\xi_+\rightarrow 0^+$) to a system with an
integer number of electrons ($\xi_+=0$ case), the xc potential exhibits
a jump (in the region of the system) which, according to
Eqs.~(\ref{eq:PPLB_exp_gap_with_DDy}) and (\ref{eq:diff_int_pot_xiplus_zero-plus_zero}), corresponds exactly
to the deviation in fundamental gap of the true system from the KS one.  

\subsection{Suppression of the derivative discontinuity}
\label{subsec:removing_dd_xc_pot}

The fundamental gap expression of
Eq.~(\ref{eq:fundgap_Ncentered_bmxi}), which has been derived within 
the $N$-centered
eDFT formalism, may intrigue PPLB practitioners. Indeed, it makes it possible to
describe charged excitations, in principle exactly, without invoking
explicitly
the concept of derivative discontinuity. Instead, all our
attention should be focused on the weight dependence of the $N$-centered
ensemble xc density functional.
Note that, despite this major difference between $N$-centered eDFT and PPLB, the xc
potential exhibits  
derivative discontinuities in both theories, as we have seen in Sec.~\ref{subsec:comparison_Ncentered}.
One may 
argue that modeling weight dependencies in ensemble xc density
functionals is actually easier than modeling functional derivative
discontinuities. Nevertheless, as discussed in further detail in Secs.~\ref{sec:EXX}
and \ref{sec:corr_energy}, designing weight-dependent
exchange and correlation DFAs from first principles raises several fundamental questions to which, up to now, no
definitive answers have been given.\\

From a conceptual point of view, the fact that we do not need 
anymore to put 
efforts into the explicit description of derivative discontinuities, once we have
moved from the
standard PPLB picture to the   
$N$-centered one, can be
interpreted as follows. Unlike in PPLB, the constraint in
Eq.~(\ref{eq:constraint_Nc-xc_pot_away}) is arbitrary because the KS
potential remains unique up to a constant when charged excitations occur
in $N$-centered eDFT, by construction. If, for simplicity, we keep this
constraint for $\xi_+=0$, \ie, we set $\tilde{v}^{\xi_+=0}_{\rm
xc}(\bfr)\equiv v^{\xi_+=0}_{\rm
xc}(\bfr)$ so that
$\tilde{v}^{\xi_+=0}_{\rm
xc}(\infty)=0$, which is likely to be fulfilled in a practical
$N$-electron DFT calculation, it can be relaxed as follows, when
$\xi_+\rightarrow 0^+$,   
\be\label{eq:moving_DDy_away}
v^{\xi_+\rightarrow 0^+}_{\rm
xc}(\bfr)&\rightarrow& \tilde{v}^{\xi_+\rightarrow 0^+}_{\rm
xc}(\bfr)=v^{\xi_+\rightarrow 0^+}_{\rm
xc}(\bfr)-\Delta^N_{\rm xc},
\ee
thus leading to $\tilde{v}^{\xi_+\rightarrow 0^+}_{\rm
xc}(\infty)=-\Delta^N_{\rm xc}$. We stress that $\tilde{v}^{\xi_+}_{\rm
xc}$ is as exact as ${v}^{\xi_+}_{\rm
xc}$. However, according to Eq.~(\ref{eq:link_deri_xi_DDy_other}), which also
holds for the new (shifted) potential $\tilde{v}^{\xi_+}_{\rm
xc}(\bfr)$, the relation in
Eq.~(\ref{eq:diff_int_pot_xiplus_zero-plus_zero}) now reads as 
\be\label{eq:no_DD_anymore}
\int d\bfr\left(\tilde{v}^{\xi_+\rightarrow 0^+}_{\rm
xc}(\bfr)-\tilde{v}^{\xi_+=0}_{\rm
xc}(\bfr)\right)n_{\Psi^N_0}(\bfr)=0.
\ee 
In other words, {\it via} the shifting procedure of
Eq.~(\ref{eq:moving_DDy_away}), we can simply move the derivative
discontinuity away from the system, \ie, in regions where the density
is essentially equal to zero. Consequently, with this change of paradigm, the absence of
derivative discontinuities in standard semi-local DFAs should not be considered
as an issue
anymore. The ability of the {\it local density approximation} (LDA) to reproduce relatively
accurate LZ-shifted KS orbital energies, as shown in a 1D atomic
model~\cite{hodgson2021exact}, is actually encouraging since the latter are central in the
evaluation of both the IP and the EA [see Eqs.~(\ref{eq:IP}) and
(\ref{eq:EA})]. On the other hand, the resulting charged excitation
energies are rather poor
because weight dependencies are completely absent from standard
LDA~\cite{hodgson2021exact}. We hope that, in the near future, (much)
more efforts will be put into the
design of weight-dependent DFAs. Recent
developments based on uniform electron gas models~\cite{loos2020weight,marut2020weight} are a first
and important step in this
direction.    

\section{The exact Hartree-exchange dilemma in eDFT}
\label{sec:EXX}

We have shown in Sec.~ \ref{sec:DD} that the infamous derivative
discontinuity problem, which appears in DFT when describing electronic
excitations,
can be bypassed, in principle exactly, {\it via} a proper modeling of the ensemble weight
dependence in the xc density functional. We focus in this section on the
design of weight-dependent exchange DFAs. Despite several (scarce
though)
attempts~\cite{nagy1996local,paragi2000investigation,paragi2001investigation,marut2020weight},
it is still unclear how weight dependencies can be
introduced into standard (semi-) local exchange functionals in a general
and systematically improvable way. The use of orbital-dependent exchange
functionals seems much more promising in this respect~\cite{filatov2015spin,loos2020weight,gould2021ensemble}.\\ 
 
Combining (a fraction of) orbital-dependent
HF-like
exchange energies with semi-local DFAs has been a key ingredient in the success of regular ground-state DFT 
in chemistry. This procedure finds its rigorous
foundation in the generalized KS theory of Seidl {\it et
al.}~\cite{seidl1996generalized}. As we will see in the following, its
extension to ensembles is nontrivial because different formulations that
have pros and cons are possible. The 
resulting dilemma is nicely summarized by the title ``{\it Ensemble generalized Kohn--Sham theory: The good, the bad, and the ugly}'' of a recent paper
by Gould and Kronik~\cite{gould2021ensemble}. Their discussion of the current situation will
serve as a guideline for this section. Following Lieb~\cite{LFTransform-Lieb}, we will show how
an in-principle-exact (OEP-free) hybrid
eDFT approach can be derived simply by exploiting the concavity (in
potential) of the state-averaged HF energy. For simplicity,      
we will focus on GOK ensembles but the discussion applies to other eDFTs
like, for example, PPLB or $N$-centered eDFT (see Sec.~\ref{sec:DD}).

\subsection{Extending the Hartree--Fock method to ensembles}

The reason why extending generalized KS-DFT~\cite{seidl1996generalized} to ensembles leads to
a dilemma has actually nothing to do with DFT. It is more a wave
function theory problem that arises at the HF level of approximation.
Therefore, for clarity, we will first discuss the 
extension of HF theory to ensembles.     

\subsubsection{Ensemble density matrix functional approach}

We start with a brief review of the procedure that is usually followed
by DFT practitioners for extending HF
to (GOK in the present case) ensembles. 
In the regular scheme, the ensemble HF energy is
evaluated variationally by inserting the {\it ensemble} (spin-summed
one-electron reduced) {\it density
matrix} (eDM) into the ground-state DM-functional HF energy. For that
reason, we refer to the approach as eDMHF. The corresponding
potential-functional
ensemble energy can be expressed as follows,  
\be\label{eq:eDMHF_ener_pot_func}
E^\bfw_{\rm eDMHF}[v]\equiv
\min_{\left\{\Phi_I\right\}}\left\{\sum_{I}\ttw_I\mel{\Phi_I}{\hat{T}+\hat{V}}{\Phi_I}
+\mathcal{E}_{\rm Hx}\left[\sum_{I}\ttw_I\bfD^{\Phi_I}\right]
\right\},
\ee
where $\hat{V}=\int d\bfr\,v(\bfr)\hat{n}(\bfr)$ is a local potential
operator (in practice it would correspond to the nuclear potential). The
eDM is evaluated from the trial orthonormal set $\left\{\Phi_I\right\}$
of {\it single-configuration} wave functions (\ie, Slater determinants or
configuration state functions). In an arbitrary orthonormal orbital basis
$\left\{\varphi_p\right\}$, the eDM reads in second
quantization as  
\be
\sum_{I}\ttw_I\bfD^{\Phi_I}\equiv \left\{\sum_{I}\ttw_I D^{\Phi_I}_{pq}\right\}=\left\{\sum_{I}\ttw_I\sum_{\tau=\uparrow,\downarrow}\expval{\hat{a}_{p\tau}^\dagger\hat{a}_{q\tau}}_{\Phi_I}\right\}
. 
\ee
In this context, the ensemble Hx energy is evaluated as follows,  
\be\label{eq:eDMHF_use_GS-HF_int_func}
\mathcal{E}_{\rm Hx}\left[\sum_{I}\ttw_I\bfD^{\Phi_I}\right]\equiv
W^{\rm HF}\left[\sum_{I}\ttw_I{\bm \gamma}^{\Phi_I}\right],
\ee
where the {\it ground-state} HF interaction functional reads as 
\be\label{eq:GS-HF_int_func}
W^{\rm HF}\left[{\bm \gamma}\right]=\dfrac{1}{2}\int d\bfr\int
d\bfr'\dfrac{\gamma(\bfr,\bfr)\gamma(\bfr',\bfr')-\frac{1}{2}\gamma^2(\bfr,\bfr')}{\abs{\bfr-\bfr'}},
\ee
and
\be
\gamma^{\Phi_I}(\bfr,\bfr')=\sum_{pq}\varphi_p(\bfr)\varphi_q(\bfr')D^{\Phi_I}_{pq}.
\ee
As shown in Appendix~\ref{appendix:eDMHF_eqs}, the orbitals
$\left\{\overline{\varphi}^\bfw_p\right\}$, from which the minimizing wave
functions $\left\{\overline{\Phi}^\bfw_I\right\}$ in
Eq.~(\ref{eq:eDMHF_ener_pot_func}) are constructed, fulfill the
following stationarity condition: 
\be\label{eq:stat_cond_eDMHF}
\left(\theta_p^\bfw-\theta_q^\bfw\right)f^\bfw_{qp}=0.
\ee
The (possibly fractional) occupation number $\theta_p^\bfw$ of the
orbital $\overline{\varphi}^\bfw_p$
within the ensemble is determined from the ensemble weights and the
(fixed) {\it
integer} occupation numbers $n^I_{p}$ of $\overline{\varphi}^\bfw_p$ in
each $\overline{\Phi}^\bfw_I$ as follows,   
\be\label{eq:GOK_frac_OON}
\theta^\bfw_p=\sum_I\ttw_In^I_p.
\ee
The ensemble Fock matrix elements $f^\bfw_{rs}\equiv f_{rs}(\bfD^\bfw)$
in Eq.~(\ref{eq:stat_cond_eDMHF})
are functionals of the eDM $\bfD^\bfw\equiv
\left\{D_{nl}^\bfw=\delta_{nl}\theta^\bfw_l\right\}$: 
\be
\begin{split}
f_{rs}(\bfD)=h_{rs}+\sum_{nl}\left(\bra{rn}\ket{sl}-\dfrac{1}{2}\bra{rn}\ket{ls}\right)D_{nl},
\end{split}
\ee
where $h_{rs}\equiv
\mel{\overline{\varphi}_r^\bfw}{\hat{h}}{\overline{\varphi}_s^\bfw}$ [with
$\hat{h}\equiv -\frac{1}{2}\nabla^2_\bfr+v(\bfr)$] and
\be
\bra{rn}\ket{sl}\equiv \int d\bfr\int
d\bfr'\overline{\varphi}_r^\bfw(\bfr)\overline{\varphi}_n^\bfw(\bfr')\overline{\varphi}_s^\bfw(\bfr)\overline{\varphi}_l^\bfw(\bfr')/\abs{\bfr-\bfr'}
\ee
are regular one- and two-electron integrals, respectively.\\

By analogy with the {\it complete active space} SCF (CASSCF) method~\cite{helgaker2014molecular}, we can
distinguish the doubly occupied (so-called inactive)
$\overline{\varphi}^\bfw_i,\overline{\varphi}^\bfw_j$ orbitals from the partially
occupied (so-called active)
$\overline{\varphi}^\bfw_u,\overline{\varphi}^\bfw_v$ and
unoccupied (virtual) $\overline{\varphi}^\bfw_a,\overline{\varphi}^\bfw_b$ ones. Thus, the
stationarity condition of Eq.~(\ref{eq:stat_cond_eDMHF}) can be detailed
as follows:
\be
f^\bfw_{ai}&=&f^\bfw_{au}=f^\bfw_{iu}=0
\ee
and
\be\label{eq:stat_cond_active_space_eDMHF}
(\theta_u^\bfw-\theta_v^\bfw)f^\bfw_{uv}&=&0.
\ee
Since $\theta_i^\bfw=\theta_j^\bfw=2$ and
$\theta_a^\bfw=\theta_b^\bfw=0$, there is no specific condition for the
inactive-inactive and virtual-virtual blocks of the Fock matrix, like in
a regular ground-state HF calculation. Therefore, we can freely rotate
the orbitals within the inactive and virtual orbital subspaces.
As readily seen from
Eq.~(\ref{eq:stat_cond_active_space_eDMHF}), this statement holds also
for active orbital subspaces in which the orbitals have the same
fractional occupation ($\theta_u^\bfw=\theta_v^\bfw$). However, if
$\theta_u^\bfw\neq\theta_v^\bfw$, then $f^\bfw_{uv}=0$. In conclusion,
an optimal set of orbitals can be determined by diagonalizing the
ensemble Fock matrix, \ie, by solving the (self-consistent) eigenvalue
equation   
\be
\hat{f}^\bfw\overline{\varphi}^\bfw_p(\bfr)=\overline{\varepsilon}^\bfw_p\overline{\varphi}^\bfw_p(\bfr).
\ee
The fact that standard SCF routines can be trivially recycled in this context is
the main reason why ensemble HF and, more generally, hybrid eDFT
calculations are performed this way. However, as discussed further in the
following, the eDMHF energy is unphysical in many ways. For example,
by construction, it varies quadratically with the ensemble
weights [see Eqs.~(\ref{eq:eDMHF_use_GS-HF_int_func}) and
(\ref{eq:GS-HF_int_func})] while the true physical ensemble energy is
expected to vary linearly. 

\subsubsection{Ghost interaction errors}
\label{subsec:eHF_ghost}

The most severe issue with the eDMHF energy expression of Eq.~(\ref{eq:eDMHF_ener_pot_func}) is that it
incorporates
unphysical interactions between the states of the ensemble.
These are known as {\it ghost
interactions} (GIs)~\cite{gidopoulos2002spurious}. The error, which is
inherent to the eDMHF method, simply originates
from the fact that, at the
HF level of approximation, the ground-state interaction energy is a quadratic
functional of the density matrix. More explicitly, we have 
\be\label{eq:detail_GI_errors}
\begin{split}
W^{\rm HF}\left[\sum_{I}\ttw_I{\bm \gamma}^{\Phi_I}\right]&=
\dfrac{1}{2}\sum_{IJ}\ttw_I\ttw_J
\int d\bfr\int
d\bfr'\dfrac{1}{\abs{\bfr-\bfr'}}
\\
&\quad\times\left(
{\gamma^{\Phi_I}(\bfr,\bfr)\gamma^{\Phi_J}(\bfr',\bfr')-\frac{1}{2}\gamma^{\Phi_I}(\bfr,\bfr')\gamma^{\Phi_J}(\bfr,\bfr')}
\right),
\end{split}
\ee
where, as readily seen, GI terms arise from all ``$I\neq J$'' pairs. Even though GI
corrections can be applied on top of the converged eDMHF
energies~\cite{loos2020weight,pastorczak2014ensemble}, the procedure is
not variational, thus making the evaluation of energy derivatives (and
therefore,
according to Eq.~(\ref{eq:indiv_EI}), of excited-state properties) less straightforward. Let us stress that,
in the original formulation of GOK-DFT~\cite{gross1988density},
the ensemble 
Hartree energy, which is evaluated from the standard (ground-state)
Hartree functional [see Eq.~(\ref{eq:original_decomp_H_and_xc})], includes GI errors [see the first term on the right-hand
side of Eq.~(\ref{eq:detail_GI_errors})]. In the exact theory, the
latter are supposed to be removed by the {\it weight-dependent} ensemble exchange functional.
It is not necessarily the case in practice
when, for example, standard (weight-independent) local or semi-local DFAs
are employed~\cite{pastorczak2014ensemble,senjean2015linear}.\\

For wave function theory practitioners, using eDMHF with {\it ad hoc}
GI corrections would probably seem uselessly
complicated. Indeed, substituting the (GI-free) weighted sum of individual
(single-configuration) 
interaction energies, which are evaluated from the individual
density matrices, for the HF interaction eDM functional of
Eq.~(\ref{eq:eDMHF_ener_pot_func}) looks, at least at first sight, like a 
simple and straightforward solution to the problem:
\be\label{eq:from_eDMHF_to_SAHF}
\begin{split}
\mathcal{E}_{\rm Hx}\left[\sum_{I}\ttw_I\bfD^{\Phi_I}\right]
\rightarrow\sum_{I}\ttw_I\expval{\hat{W}_{\rm ee}}_{\Phi_I}&\rightarrow 
\sum_{I}\ttw_I\mathcal{E}^I_{\rm Hx}\left[{\bf D}^{\Phi_I}\right]
\\
&=\sum_{I}\ttw_I\left(E_{\rm
H}[n_{\Phi_I}]+\mathcal{E}^I_{\rm x}\left[{\bf D}^{\Phi_I}\right]\right). 
\end{split}
\ee
Note that, in Eq.~(\ref{eq:from_eDMHF_to_SAHF}), we assumed that
individual interaction energies can be written as functionals of the
individual (one-electron reduced) density matrices, for simplicity.
The variational evaluation of state-averaged interaction energies is
discussed in detail in Sec.~\ref{subsec:SA_HF} on that basis.
Such a simplification is always valid for single Slater determinants.
For more general multideterminant (single configuration though) wave
functions, the simplification in Eq.~(\ref{eq:from_eDMHF_to_SAHF}) might
be used as an (additional) approximation. Alternatively, one may
evaluate exactly multideterminant interaction energies from the individual two-electron
reduced density matrices, by analogy with the state-averaged
CASSCF (SA-CASSCF) method~\cite{helgaker2014molecular}. The latter
approach is not
described further in the present review. Both (approximate anyway) strategies lead ultimately to an
in-principle-exact eDFT, once a proper complementary correlation ensemble density functional has been introduced (see
Sec.~\ref{subsubsec:sc-SAHF-based_eDFT}).\\
      
As discussed in the next section, the reason why the state-averaging 
of interaction energies is not as
popular as one would expect is that, as we switch from eDMHF to the
state-averaged energy paradigm, the
orbital optimization cannot be performed anymore with standard SCF
routines. An additional implementation work is needed in this case~\cite{filatov2015spin}. We stress that this statement holds even when individual
{\it one-electron} reduced density matrix-functional interaction energies are
employed [see Eq.~(\ref{eq:from_eDMHF_to_SAHF})], as assumed in the rest
of the review. 
   
\subsubsection{State-averaged Hartree--Fock approach}
\label{subsec:SA_HF}

The paradigm on the right-hand side of Eq.~(\ref{eq:from_eDMHF_to_SAHF})
can be seen as an adaptation of the SA-CASSCF method~\cite{helgaker2014molecular} to {\it
single-configuration} wave functions. While, in SA-CASSCF,
(correlated)
multiconfigurational wave functions are employed, we simply restrict, in the
present case, the energy
minimization to sets of (uncorrelated) single-configuration wave
functions. The 
resulting (GI-free) total ensemble energy, which is obtained from the eDMHF energy
expression and the substitution in Eq.~(\ref{eq:from_eDMHF_to_SAHF}),
will be referred to as {\it state-averaged} HF
(SAHF) energy in the following. It reads as follows, 
\be\label{eq:min_SAHF_ener}
\begin{split}
&E^{\bfw}_{\rm
SAHF}[v]\equiv\min_{\left\{\Phi_I\right\}}\Bigg\{\sum_{I}\ttw_I
\mel{\Phi_I}{\hat{T}+\hat{W}_{\rm ee}+\hat{V}}{\Phi_I}
\Bigg\}
\\
&=\min_{\left\{\Phi_I\right\}}\Big\{\sum_{I}\ttw_I\Big(\mel{\Phi_I}{\hat{T}
}{\Phi_I}+
E_{\rm
H}[n_{\Phi_I}]+\mathcal{E}^I_{\rm x}\left[{\bf D}^{\Phi_I}\right]
+
\int d\bfr\,v(\bfr)n_{\Phi_I}(\bfr)\Big)\Big\},
\end{split}
\ee
where, as already mentioned after Eq.~(\ref{eq:from_eDMHF_to_SAHF}), we
assume for simplicity that interaction energies can be
evaluated from the individual (one-electron reduced) density matrices.  
Let us denote $\left\{\tilde{\Phi}^\bfw_I\right\}$ (with a tilde symbol) the minimizing
single-configuration wave functions so
that they can be clearly distinguished from the eDMHF ones. As further discussed in
Appendix~\ref{appendix:SAHF_eqs}, these minimizing SAHF wave functions
are all constructed from the {\it same} set of
orthonormal molecular orbitals which can be optimized variationally through orbital
rotations. The minimizing orbitals
$\left\{\tilde{\varphi}^\bfw_p\right\}$ fulfill the following
stationarity conditions [see Appendix~\ref{appendix:SAHF_eqs}], 
\be\label{eq:derivat_SAHF_ener_zero_simplified}
 \left(\theta^\bfw_p-\theta^\bfw_q\right)\mel{\tilde{\varphi}^\bfw_p}{\hat{h}}{\tilde{\varphi}^\bfw_q}+\sum_I\ttw_I\left(n^I_p-n^I_q\right)\mel{\tilde{\varphi}^\bfw_p}{\hat{v}^{\bfw}_{{\rm
Hx},I}}{\tilde{\varphi}^\bfw_q}=0,
\ee
where the individual (non-local) density-matrix-functional Hx operators read as  
\be
\hat{v}^{\bfw}_{{\rm
Hx},I}=
\hat{v}_{\rm
H}[n_{\tilde{\Phi}^\bfw_I}]+\hat{v}^I_{{\rm x}}\left[{\bf
D}^{\tilde{\Phi}^\bfw_I}\right],
\ee
$\hat{v}_{\rm H}[n]\equiv v_{\rm H}[n](\bfr)\times=\delta E_{\rm
H}[n]/\delta n(\bfr)\times$ being the standard local (multiplicative) density-functional Hartree
potential operator and 
\be
\mel{\varphi_r}{\hat{v}^I_{{\rm x}}[\bfD]}{\varphi_s}\equiv\dfrac{
\partial \mathcal{E}^I_{\rm x}\left[{\bf D}\right]}{\partial D_{rs}}.
\ee
The major difference between
SAHF and eDMHF lies in the fact that, as we now   
employ individual
density matrices {\it separately} in the evaluation of the ensemble Hx energy
[see Eq.~(\ref{eq:from_eDMHF_to_SAHF})], differentiating with respect
to any variational orbital rotation parameter $\kappa_{pq}$ generates
{\it individual} Hx potentials,
\be\label{eq:deriv_kappa_pq_eDMHF_from_SAHF}
\begin{split}
\dfrac{\partial}{\partial \kappa_{pq}}\left(\mathcal{E}_{\rm
Hx}\left[\sum_{I}\ttw_I\bfD^{\Phi_I}\right]\right)&=
 \sum_{rs}\left(\sum_{I}\ttw_I\dfrac{\partial
D_{rs}^{{\Phi}_I}}{\partial\kappa_{pq}}\right)\left.\dfrac{\partial \mathcal{E}_{\rm Hx}[\bfD]}{\partial
D_{rs}}\right|_{\bfD=\sum_{I}\ttw_I\bfD^{{\Phi}_I}}
\\
\longrightarrow
\dfrac{\partial}{\partial \kappa_{pq}}\left(\sum_{I}\ttw_I\mathcal{E}^I_{\rm
Hx}\left[{\bf D}^{\Phi_I}\right]\right)&=
\displaystyle
\sum_{rs}
\sum_{I}\ttw_I\dfrac{\partial D_{rs}^{{\Phi}_I}}{\partial\kappa_{pq}}\left.\dfrac{\partial \mathcal{E}^I_{\rm Hx}[\bfD]}{\partial
D_{rs}}\right|_{\bfD=\bfD^{{\Phi}_I}}.
\end{split}
\ee
In eDMHF, the same (ensemble) Hx operator is systematically recovered by
differentiation [see the left-hand side of
Eq.~(\ref{eq:deriv_kappa_pq_eDMHF_from_SAHF})] and, from the stationarity
condition and the definition
in 
Eq.~(\ref{eq:GOK_frac_OON}) of the fractional orbital occupation numbers, 
a {\it unique} to-be-diagonalized Fock
operator can be extracted. This does not happen in
SAHF, as readily seen from the second term on the left-hand side of
Eq.~(\ref{eq:derivat_SAHF_ener_zero_simplified}) or, equivalently, on the
right-hand side of Eq.~(\ref{eq:deriv_kappa_pq_eDMHF_from_SAHF}). Still
we can introduce an {\it orbital-dependent} ensemble
Fock operator~\cite{filatov2015spin,gould2021ensemble,filatov2021description},    
\be\label{eq:orbital-dep_Fock-like_op}
\hat{\mathcal{F}}^\bfw_p:=
\hat{h}+\dfrac{1}{\theta^\bfw_p}
\sum_I\ttw_In_p^I\, 
\hat{v}^{\bfw}_{{\rm
Hx},I},
\ee
%
so that the stationarity condition of
Eq.~(\ref{eq:derivat_SAHF_ener_zero_simplified}) can be rewritten as
follows,
\be\label{eq:stationarity_cond_orb-dep_Fock_op}
\theta^\bfw_p\mel{\tilde{\varphi}^\bfw_p}{\hat{\mathcal{F}}^\bfw_p}{\tilde{\varphi}^\bfw_q}-\theta^\bfw_q\mel{\tilde{\varphi}^\bfw_p}{\hat{\mathcal{F}}^\bfw_q}{\tilde{\varphi}^\bfw_q}=0.
\ee
Let us stress that, unlike in GOK-DFT (see
Sec.~\ref{subsubsec:GOK-DFT}) or eDMHF, the minimizing orbitals are {\it a priori} not eigenfunctions of
their associated Fock operator. Indeed, if we consider two
fractionally occupied
orbitals $\tilde{\varphi}^\bfw_u$ and $\tilde{\varphi}^\bfw_v$, and
assume that
$\mel{\tilde{\varphi}^\bfw_u}{\hat{\mathcal{F}}^\bfw_u}{\tilde{\varphi}^\bfw_v}=0$,
Eq.~(\ref{eq:stationarity_cond_orb-dep_Fock_op}) would immediately imply that
$\mel{\tilde{\varphi}^\bfw_u}{\hat{\mathcal{F}}^\bfw_v}{\tilde{\varphi}^\bfw_v}=0$,
which is unlikely due to the orbital dependence of the Fock operator
[see Eq.~(\ref{eq:orbital-dep_Fock-like_op})].
As a result, the self-consistent SAHF equations will have the following
general
structure, 
\be\label{eq:Fock-like_eq_off-diag_epsi}
\theta^\bfw_p\hat{\mathcal{F}}^\bfw_p\tilde{\varphi}^\bfw_p(\bfr)=\sum_q\tilde{\varepsilon}^\bfw_{qp}\,\tilde{\varphi}^\bfw_q(\bfr),
\ee
or, more explicitly [see Eq.~(\ref{eq:orbital-dep_Fock-like_op})],
\be\label{eq:Fock-like_eq_off-diag_epsi_more_explicit}
\left(-\dfrac{1}{2}\nabla^2_{\bfr}+v(\bfr)+\dfrac{1}{\theta^\bfw_p}\sum_I\ttw_In_p^I\,
\hat{v}^\bfw_{{\rm
Hx},I}\right)\tilde{\varphi}^\bfw_p(\bfr)=\dfrac{1}{\theta^\bfw_p}\sum_q\tilde{\varepsilon}^\bfw_{qp}\,\tilde{\varphi}^\bfw_q(\bfr)
,
\ee
where off-diagonal one-electron energy couplings
$\left\{\tilde{\varepsilon}^\bfw_{qp}\right\}_{p\neq q}$ {\it cannot} be
removed. In practice, Eq.~(\ref{eq:Fock-like_eq_off-diag_epsi}) can be
solved with the coupling operator
technique~\cite{JCP73_Hirao_coupling_op_technique,filatov2021description} which consists in
diagonalizing repeatedly
$(\theta^\bfw_p\hat{\mathcal{F}}^\bfw_p-\theta^\bfw_q\hat{\mathcal{F}}^\bfw_q)/(\theta^\bfw_p-\theta^\bfw_q)$ until convergence
is reached. In the latter case, the matrix $\tilde{\bm
\varepsilon}^\bfw\equiv \left\{\tilde{\varepsilon}^\bfw_{qp}\right\}$
becomes hermitian, as a consequence of
Eqs.~(\ref{eq:stationarity_cond_orb-dep_Fock_op}) and (\ref{eq:Fock-like_eq_off-diag_epsi}), and the hermiticity of
the Fock operators:
\be\label{eq:Lagrange_multipliers_matrix_hermitian}
\tilde{\varepsilon}^\bfw_{qp}=
\mel{\tilde{\varphi}^\bfw_q}{\theta^\bfw_p\hat{\mathcal{F}}^\bfw_p}{\tilde{\varphi}^\bfw_p}
=
\mel{\tilde{\varphi}^\bfw_p}{\theta^\bfw_p\hat{\mathcal{F}}^\bfw_p}{\tilde{\varphi}^\bfw_q}
=\mel{\tilde{\varphi}^\bfw_p}{\theta^\bfw_q\hat{\mathcal{F}}^\bfw_q}{\tilde{\varphi}^\bfw_q}=
\tilde{\varepsilon}^\bfw_{pq}.
\ee 

\subsubsection{eDMHF versus SAHF}

Let us summarize what we have learned from the previous subsections. 
While the orbital optimization in eDMHF is relatively straightforward,
because standard SCF routines can be recycled in this context, it is
more involved in SAHF because no to-be-diagonalized Fock operator
emerges from the stationarity condition. On the other hand, the commonly
used eDMHF energy expression suffers from severe GI errors, while SAHF
is completely GI-free. The latter point is probably the strongest
argument for promoting SAHF over eDMHF. Note that both schemes would be
good starting points for turning the recently formulated {\it ensemble
reduced density matrix functional theory} ($\ttw$-RDMFT)~\cite{Schilling2021_eRDMFT}
into a practical method for the computation of low-lying
excited states. In the following, we will show
how eDMHF and SAHF can be merged rigorously with eDFT.      

\subsection{Concavity of approximate energies and Lieb maximization}\label{subsec:concavity_approx_ener}

In order to derive in-principle-exact hybrid eDFT schemes, where (a
fraction of) orbital-dependent exchange energies are combined with
ensemble density functionals, we need to ``exactify'' the eDMHF and SAHF
approximations reviewed previously. In a DFT perspective,
an approximation
becomes exact when it reproduces the exact density of the system under
study. This is how KS-DFT transforms an approximate non-interacting problem into an
exact one. In the present case, we want to extend the Hohenberg--Kohn
theorem to the more advanced eDMHF and SAHF approximations.
For that purpose, convex
analysis~\cite{LFTransform-Lieb,JCP14_Kvaal_MY_reg_in_DFT,PRL19_Penz_convergence_regularized_DFT}
turns out to be a
powerful mathematical tool because, as we will see, it allows for the
derivation of several exact eDFTs within the same
(unified) formalism.\\   

For the sake of generality, we will express the various approximate ensemble energies discussed previously as follows,  
\be\label{eq:general_VP_approx}
\mathscr{E}^\bfw_{\rm
approx.}[v]=\min_{\bfkap}\left\{\mathscr{F}_{\rm
approx.}^\bfw(\bfkap)+\int
d\bfr\,v(\bfr)n^\bfw(\bfkap,\bfr)\right\},
\ee
where $\bfkap$ denotes the collection of variational parameters (in the
present case, the latter will be orbital
rotation parameters). Each approximation (non-interacting (KS), eDMHF, or
SAHF) is characterized by a specific potential-independent function of
$\bfkap$: 
\be
\mathscr{F}^\bfw_{\rm
approx.}(\bfkap)&\overset{\rm
KS}{\equiv}&\sum_I\ttw_I\expval{\hat{T}}_{\Phi_I(\bfkap)},
\\
\mathscr{F}^\bfw_{\rm
approx.}(\bfkap)&\overset{\rm 
eDMHF}{\equiv}&
\sum_I\ttw_I\expval{\hat{T}}_{\Phi_I(\bfkap)}+\mathcal{E}_{\rm
Hx}\left[\sum_I\ttw_I\bfD^{\Phi_I(\bfkap)}\right],
\\
\label{eq:function_kappa_SAHF}
\mathscr{F}^\bfw_{\rm
approx.}(\bfkap)&\overset{\rm 
SAHF}{\equiv}& 
\sum_I\ttw_I\left[\expval{\hat{T}}_{\Phi_I(\bfkap)}+
E_{\rm
H}[n_{\Phi_I(\bfkap)}]+\mathcal{E}^I_{\rm x}\left[{\bf D}^{\Phi_I(\bfkap)}\right]
\right],
\ee
where single-configuration ground- and excited-state wave functions
$\left\{\Phi_I(\bfkap)\right\}$ are
employed. The potential-dependent contribution to the energy expression
of Eq.~(\ref{eq:general_VP_approx}) is determined from the ensemble
density $n^\bfw(\bfkap,\bfr)=\sum_I\ttw_In_{\Phi_I(\bfkap)}(\bfr)$.\\

From a mathematical point of view, the fact that the approximate
ensemble energies are evaluated
{\it variationally} has important implications. Even
though they are approximate, these energies still share a fundamental property with the exact 
ensemble energy, namely the
concavity with respect to the local potential $v$. Indeed, for two potentials $v_a$ and $v_b$, $\alpha$ in the range $0\leq
\alpha\leq 1$, and any set $\bfkap$ of variational parameters, we have
\be
\begin{split}
&\mathscr{F}_{\rm
approx.}^\bfw(\bfkap)+\int
d\bfr\,\left((1-\alpha)v_a(\bfr)+\alpha v_b(\bfr)\right)n^\bfw(\bfkap,\bfr)
\\
&=(1-\alpha)\left[\mathscr{F}_{\rm
approx.}^\bfw(\bfkap)+\int
d\bfr\,v_a(\bfr)n^\bfw(\bfkap,\bfr)\right]
\\
&\quad
+\alpha\left[\mathscr{F}_{\rm
approx.}^\bfw(\bfkap)+\int
d\bfr\,v_b(\bfr)n^\bfw(\bfkap,\bfr)\right]
\\
&\geq (1-\alpha)
\mathscr{E}^\bfw_{\rm
approx.}[v_a]+\alpha \mathscr{E}^\bfw_{\rm approx.}[v_b],
\end{split}
\ee
thus leading to 
\be
\mathscr{E}^\bfw_{\rm
approx.}[(1-\alpha)v_a+\alpha v_b]\geq (1-\alpha)
\mathscr{E}^\bfw_{\rm
approx.}[v_a]+\alpha \mathscr{E}^\bfw_{\rm approx.}[v_b].
\ee
Following Lieb~\cite{LFTransform-Lieb}, we
can now construct (thanks to this concavity property) an approximation
to the universal GOK density functional as
follows, 
\be\label{eq:approx_Lieb_func}
F^\bfw[n]\approx F^\bfw_{\rm
approx.}[n]=\max_v\left\{\mathscr{E}^{\bfw}_{\rm
approx.}[v]-\int d\bfr\,v(\bfr)n(\bfr)\right\},
\ee
where we assume, for simplicity, that a maximum is reached. An even more
rigorous definition (from a mathematical point of view~\cite{LFTransform-Lieb}) would actually be obtained by
using a ``$\sup$'' instead of a ``$\max$'' [and a ``inf'' instead of a
``min'', in Eq.~(\ref{eq:general_VP_approx})]. The maximizing potential
${v}_{\rm
approx.}^\bfw[n]$ in Eq.~(\ref{eq:approx_Lieb_func}) fulfills the
following stationarity condition:
\be\label{eq:stationary_cond_Lieb_max_approx}
\left.\dfrac{\delta 
\mathscr{E}^{\bfw}_{\rm
approx.}[v]}{\delta v(\bfr)}\right|_{v={v}_{\rm
approx.}^\bfw[n]}=n(\bfr).
\ee
We conclude from the Hellmann--Feynman theorem that,
when the approximate ensemble energy of Eq.~(\ref{eq:general_VP_approx})
is calculated with $v={v}_{\rm
approx.}^\bfw[n]$, the
minimizing single-configuration wave functions (which are determined
from the minimizing $\bfkap$) 
reproduce the desired ensemble density $n$. Thus, we automatically
extend the Hohenberg--Kohn theorem to eDMHF and SAHF ensembles. If we choose for $n$
the true physical ensemble density of a given system, both
approximations
become exact density wise, because they reproduce the correct density.
Exact ensemble energies can then be recovered from the approximate
ensembles once a complementary ensemble (x)c density functional
has been introduced. This final step will be discussed in Sec.~\ref{subsubsec:sc-SAHF-based_eDFT}\\
    
Let us finally focus on the OEP- and GI-free SAHF approximation. The
corresponding universal
density functional reads more explicitly as [see
Eqs.~(\ref{eq:general_VP_approx}), (\ref{eq:function_kappa_SAHF}), and (\ref{eq:approx_Lieb_func})]
\be\label{eq:SAHF_func_from_Lieb_to_Levy}
\begin{split}
{F}_{\rm SAHF}^\bfw[n]
&=\mathscr{E}^{\bfw}_{\rm
SAHF}[{v}_{\rm
SAHF}^\bfw[n]]-\int d\bfr\,{v}_{\rm SAHF}^\bfw[n](\bfr)n(\bfr)
\\
&\equiv \sum_{I}\ttw_I\mel{\tilde{\Phi}_I^\bfw[n]}{\hat{T}+\hat{W}_{\rm ee}}{\tilde{\Phi}_I^\bfw[n]}
,
\end{split}
\ee
where ${v}_{\rm SAHF}^\bfw[n]$ denotes the stationary (maximizing)
density-functional potential of
Eq.~(\ref{eq:stationary_cond_Lieb_max_approx}) in the particular case of the SAHF approximation. 
By analogy with the constrained-search formalism of
Levy~\cite{levy1979universal}, the SAHF functional can be rewritten as
follows,
\be\label{eq:SAHF_Levy_func}
{F}_{\rm SAHF}^\bfw[n]=\min_{\left\{\Phi_I\right\}\overset{\bfw}{\rightarrow}n}
\left\{\sum_{I}\ttw_I\mel{\Phi_I}{\hat{T}+\hat{W}_{\rm ee}}{\Phi_I}
\right\},
\ee
where the density constraint $\left\{\Phi_I\right\}\overset{\bfw}{\rightarrow}n$
imposed on the single-configuration wave functions $\left\{\Phi_I\right\}$ reads as
$\sum_I\ttw_In_{\Phi_I}(\bfr)=n(\bfr)$. Note that,
as illustrated in Sec.~\ref{subsec:insights_SAHF_Hubbard_dimer}, some densities may not be $v$-representable by
a single SAHF ensemble.
In this case, a more general Levy--Lieb-like~\cite{LFTransform-Lieb} functional, where
an ensemble of ensembles is considered, should
be employed:   
\be\label{eq:double_ens_SAHF_func}
{F}_{\rm SAHF}^\bfw[n]=
\min_{\left\{\left\{\Phi^{(i)}_I\right\},\alpha^{(i)}\right\}\overset{\bfw}{\rightarrow}n}
 \sum_i\alpha^{(i)}\left\{\sum_{I}\ttw_I\mel{\Phi^{(i)}_I}{\hat{T}+\hat{W}_{\rm ee}}{\Phi^{(i)}_I}
\right\},
\ee
where the density constraint reads as
\be\label{eq:double_ens_dens_constraint}
\sum_i\alpha^{(i)}\left(\sum_I\ttw_In_{\Phi^{(i)}_I}(\bfr)\right)&=&\sum_I\ttw_I\left(\sum_i\alpha^{(i)}n_{\Phi^{(i)}_I}(\bfr)\right)=n(\bfr),
\ee
with
\be
\sum_i\alpha^{(i)}&=&1.
\ee
We stress that, in the above more general definition, the additional ensemble
weights $\left\{\alpha^{(i)}\right\}$ are {\it not} given, unlike the GOK ensemble weights $\bfw$. They are
determined from the density constraint of Eq.~(\ref{eq:double_ens_dens_constraint}).  
An interesting feature of the constrained-search formalism is that it
applies to densities that
might be ensemble $N$-representable but not SAHF $v$-representable, \ie, densities
that cannot be generated from a given potential $v$ according to
Eq.~(\ref{eq:general_VP_approx}). For clarity, we will use in
Sec.~\ref{subsubsec:sc-SAHF-based_eDFT} the simpler density functional expression of Eq.~(\ref{eq:SAHF_Levy_func})
[rather than the one in Eq.~(\ref{eq:double_ens_SAHF_func})].
 
\subsection{Insights from the Hubbard dimer model}\label{subsec:insights_SAHF_Hubbard_dimer}

In order to compare SAHF with eDMHF, both methods are applied in this
section to the (two-electron) Hubbard
dimer model~\cite{carrascal2015hubbard,carrascal2016corrigendum,deur2017exact}. 
Despite its simplicity, it is nontrivial and has become in recent
years the model of choice for analyzing and understanding failures of
DFT or TD-DFT, but also for
exploring new concepts~\cite{senjean2017local,li2018density,carrascal2018linear,deur2018exploring,sagredo2018can,smith2016exact}.
In this
model, the {\it ab initio}
Hamiltonian is simplified as follows,
\be\label{eq:Hamil_Hubbard_dimer_model}
\hat{T} &\rightarrow& \hat{\mathcal{T}}=-t
\sum_{\tau=\uparrow\downarrow}(\hat{c}^\dagger_{0\tau}\hat{c}_{1\tau} +
\hat{c}^\dagger_{1\tau}\hat{c}_{0\tau}),\hspace{0.2cm} \hat{W}_{\rm ee}\rightarrow
\hat{U}=U\sum^1_{i=0}\hat{n}_{i\uparrow}\hat{n}_{i\downarrow},
\nonumber\\
\hat{V}&\rightarrow&\Delta v(\hat{n}_1 -
\hat{n}_0)/2,\hspace{0.3cm}\hat{n}_{i\tau}=\hat{c}^\dagger_{i\tau}\hat{c}_{i\tau},
\ee
where operators are written in second quantization and $\hat{n}_i=\sum_{\tau=\uparrow\downarrow}\hat{n}_{i\tau}$ is
the density operator on site $i$ ($i=0,1$). Note that the local
potential reduces to a single number $\Delta v$ which controls the
asymmetry of the dimer.
The density also reduces
to a single number $n=n_0$, which is the occupation of site 0, given that
$n_1 = 2-n$.\\

We consider in the following a two-electron singlet biensemble consisting of the ground
state and the singly-excited state. 

\subsubsection{SAHF and eDMHF energy expressions}

In (restricted) SAHF theory, both
ground- and excited-state wave functions are approximated by
configuration state functions. Following this view, trial
singlet ground-state ${\Phi_0}$
and first excited-state ${\Phi_1}$ wave functions of the two-electron Hubbard dimer read
as follows, in second quantization,
\be
\ket{\Phi_0}=
\ket{\sigma_0^2}\equiv\hat{c}_{{\sigma_0}\uparrow}^\dagger\hat{c}_{{\sigma_0}\downarrow}^\dagger\ket{\rm vac}
\ee
and
\be
\ket{\Phi_1}=\ket{\sigma_0\sigma_1}\equiv \dfrac{1}{\sqrt{2}}\left(\hat{c}_{{\sigma_1}\uparrow}^\dagger\hat{c}_{{\sigma_0}\downarrow}^\dagger
-\hat{c}_{{\sigma_1}\downarrow}^\dagger\hat{c}_{{\sigma_0}\uparrow}^\dagger\right)\ket{\rm
vac},
\ee
respectively,
where ${\sigma_0}$ and  ${\sigma_1}$ stand for the molecular orbitals (MOs) written in the basis
of the orthonormal local atomic orbitals (AOs) $a$ and $b$. In this
context, the latter simply correspond to site 0 ($\hat{c}^\dagger_{a\tau}\equiv\hat{c}^\dagger_{0\tau}$) and 1 ($\hat{c}^\dagger_{b\tau}\equiv\hat{c}^\dagger_{1\tau}$), respectively. Bonding ${\sigma_0}$ and
anti-bonding ${\sigma_1}$ MOs can be determined through orbital
rotation as follows,
\be\label{eq:MO_to_A0}
\begin{split}
\sigma_0 &= a \cos\left(\alpha\right) +
             b \sin\left(\alpha\right),
\\
\sigma_1 &= -a \sin\left(\alpha\right) +
             b \cos\left(\alpha\right),
\end{split}
\ee
where the angle $\alpha$ is the sole variational parameter in the model.
In SAHF, the energy that must be minimized, for a fixed ensemble weight
$\ttw$ in the range $0\leq \ttw\leq 1/2$, is constructed as follows, 
\be\label{eq:SAHF_ener_dimer}
E^{\ttw}_{\rm SAHF}\equiv
(1-\ttw)\expval{\hat{H}}_{\Phi_0}+\ttw\expval{\hat{H}}_{\Phi_1},
\ee
where
\be
\expval{\hat{H}}_{\Phi_0} = 2h_{\sigma_0\sigma_0} + J_{\sigma_0 \sigma_0} 
\ee
and
\be
\expval{\hat{H}}_{\Phi_1} = h_{\sigma_0\sigma_0} + h_{\sigma_1\sigma_1}+ J_{\sigma_0
\sigma_1} + K_{\sigma_0 \sigma_1}.
\ee
We use standard notations for Coulomb and exchange two-electron
integrals,
\be
J_{ij} = \left (ii,jj \right)=\bra{ij}\ket{ij},
\\
K_{ij} = \left (ij,ji \right)=\bra{ij}\ket{ji},
\ee
both expressed in the MO basis $\{\sigma_0,  \sigma_1 \}$
of Eq.~(\ref{eq:MO_to_A0}).
At this point, let us stress that the SAHF energy substantially differs
from that of a (truncated) configuration interaction calculation.
Indeed, in the present case, the weight $\ttw$ is
{\it fixed}. Moreover, the configurations $\Phi_0$ and
$\Phi_1$ are never coupled explicitly, whether the dimer is symmetric or
not. They are just mixed through the ensemble formalism. 
Note that, in the symmetric $\Delta v=0$ case, symmetry can in principle
be artificially broken, like in (spin) unrestricted calculations. This feature will be
discussed in further detail in the next section. 
For convenience, we denote 
\be
\theta = \frac{\pi}{4} - \alpha,
\ee
so that the symmetric solution corresponds to $\theta=0$. Consequently, the SAHF energy 
expression of Eq.~(\ref{eq:SAHF_ener_dimer}) reduces to
\be
\begin{split}
E^\ttw_{\rm SAHF}(\Delta v,\theta)&=-(1-\ttw)\left[2t\cos(2\theta)+\Delta v\sin (2\theta)\right]
\\
&\quad+\dfrac{U}{4}\left[3-\ttw+(3\ttw-1)\cos(4\theta)\right].
\end{split}
\ee
As readily seen from the above expression, the SAHF energy is
$\pi$-periodic, which
means that it is sufficient to vary $\theta$ in the range
$-\frac{\pi}{2}\leq \theta\leq\frac{\pi}{2}$.\\

Similarly, from the general expression in Eq.~(\ref{eq:eDMHF_ener_pot_func}), we obtain the following analytical expression for the eDMHF energy: 
\be
\begin{split}
E^{\ttw}_{\rm
eDMHF}(\Delta v,\theta)&=-(1-\ttw)\left[2t\cos(2\theta)+\Delta v\sin (2\theta)\right]
\\
&\quad
+\dfrac{U}{4}\left[\ttw^2-2\ttw+3-(1-\ttw)^2\cos(4\theta)\right].
\end{split}
\ee

Note that, according to Eq.~(\ref{eq:MO_to_A0}), the ensemble density
(on site 0) varies with the trial angle $\theta$ as follows,
\be\label{eq:trial_biens_dens_HD}
\begin{split}
n^\ttw(\theta)&\equiv(1-\ttw)\sum_{\tau=\uparrow\downarrow}\expval{\hat{c}_{0\tau}^\dagger\hat{c}_{0\tau}}_{\Phi_0}+\ttw\sum_{\tau=\uparrow\downarrow}\expval{\hat{c}_{0\tau}^\dagger\hat{c}_{0\tau}}_{\Phi_1}
\\
&=1+(1-\ttw)\sin(2\theta).
\end{split}
\ee

\subsubsection{Symmetric case}\label{eq:HD_symmetric_case_SAHF}

Let us concentrate on the symmetric dimer, for which $\Delta v =0$. In this case,
the dimer is a prototype for the H$_2$ molecule. If we denote
\be\label{eq:def_rho_fun}
\rho(\theta)=\cos(2\theta)=2\cos^2\theta-1,
\ee
then the SAHF energy reads as
\be
\label{eq:ensemble_energy}
E^\ttw_{\rm SAHF}(\Delta v=0,\theta)\equiv E^\ttw_{\rm SAHF}(\theta)=\mathcal{E^\ttw_{\rm
SAHF}}(\rho(\theta)),
\ee
where
\be
\mathcal{E^\ttw_{\rm
SAHF}}(\rho)=
\dfrac{U}{2}(3\ttw-1)\rho^2- 2t(1-\ttw)\rho + U(1-\ttw).
\ee
By taking its first derivative with respect to $\theta$, we obtain the
following stationarity condition,
\be\label{eq:stat_cond_SAHF_HD}
\begin{split}
\dfrac{dE^\ttw_{\rm SAHF}(\theta)}{d\theta} &= -2\sin(2\theta)
\left[(U(3\ttw-1)\rho(\theta)-2t(1-\ttw) \right] = 0.
\end{split}
\ee
Therefore, $\theta = 0$ is systematically an extremum where the
traditional in-phase and out-of-phase
linear combinations for  $\sigma_0$ and
$\sigma_1$  are recovered.
The nature (maximum or minimum) of this stationary point, which is
discussed in the following, emerges from 
a straightforward evaluation of the energy curvature:
\be
\label{eq:second_derivative}
  \left. \frac{d^2E^\ttw_{\rm
SAHF}(\theta)}{d\theta^2}\right|_{\theta=0} = -4 [U(3\ttw-1) -
2t(1-\ttw)].
\ee
Remembering that $\abs{\rho(\theta)} \leq 1$,
the stationarity condition of Eq.~(\ref{eq:stat_cond_SAHF_HD}) is also fulfilled for two additional (opposite) $\theta$ values
given by
\be
\label{eq:stationary_condition}
\rho(\theta)=\rho_0\equiv\dfrac{2t(1-\ttw)}{U(3\ttw-1)},
\ee
as long as
\be
\label{eq:3_value_stationary_condition}
\abs{\rho_0}  \leqslant 1.
\ee
Note that, with this notation, the successive energy derivatives can be expressed as
follows, 
\be\label{eq:gradient_SAHF_theta_zero}
\dfrac{dE^\ttw_{\rm SAHF}(\theta)}{d\theta}  
=-4t(1-\ttw)\sin(2\theta)\left[\dfrac{\rho(\theta)}{\rho_0}-1\right]
\ee
and
\be\label{eq:SAHF_curvature_at_zero}
\left. \frac{d^2E^\ttw_{\rm SAHF}(\theta)}{d\theta^2}\right|_{\theta=0}
=-8t(1-\ttw)\left[\dfrac{1}{\rho_0}-1\right].
\ee
The symmetric solution $\theta=0$ will not be the absolute minimum
anymore when the above curvature becomes
strictly negative, which implies $1/\rho_0>1$, as readily seen
from Eq.~(\ref{eq:SAHF_curvature_at_zero}). Obviously, this constraint
can only be fulfilled if $\ttw>1/3$, since $1/\rho_0$ must be strictly
positive. This is a necessary but not sufficient condition.
More precisely, for weights in the range $1/3<\ttw\leq1/2$, electron
correlation should be strong enough such that $\rho_0<1$ or,
equivalently,
\be\label{eq:HD_correlation_constraint_double-well}
\dfrac{U}{t}>\dfrac{2(1-\ttw)}{3\ttw-1}.
\ee
Interestingly, if we introduce effective weight-dependent hopping $\tilde{t} = t(1-\ttw)$
and on-site interaction $\tilde{U} = U(3\ttw-1)$ parameters, the
condition in Eq.~(\ref{eq:HD_correlation_constraint_double-well}) can be 
rewritten as 
\be
\dfrac{2\tilde{t}}{\tilde{U}}<1,
\ee
which resembles the usual definition of moderate (up to strong) electron
correlation in lattices. Actually, in the commonly used equiensemble case
($\ttw=1/2$), the effective ratio matches the physical one $2t/U$.\\   

Finally, when $\ttw\leq 1/3$, the SAHF energy becomes convex at
$\theta=0$ and, since it can have two additional (say $\theta_+>0$ and
$\theta_-=-\theta_+$) stationary points at most in the range
$-\pi/2\leq \theta\leq \pi/2$ [see
Eqs.~(\ref{eq:def_rho_fun}) and (\ref{eq:gradient_SAHF_theta_zero})], the symmetric solution has to
be the absolute minimum.     
The different possible scenarios are summarized in
Table~\ref{tab:ensembleHF_conditions}. 
\begin{table}[htbp]
  \caption{\label{tab:ensembleHF_conditions}
    $\theta$ values minimizing the SAHF energy of the symmetric Hubbard
    dimer. See text for further details.}
\begin{center}
  \begin{tabular}{cc|ccccccc}
    &&  $\ttw \leqslant 1/3$ &&   $\ttw > 1/3$  \\  \hline
    
       $\abs{\rho_0} > 1$    && $\theta = 0 $  &&  $\theta =  0$ \\
    $\abs{\rho_0} \leqslant 1$  && $\theta = 0 $ &&  $\theta \neq 0$ \\
    
    \end{tabular}
\end{center}
\end{table}
Note that a connection can be made with the (spin) unrestricted
energy of a symmetric dimer ({\it e.g.}, the H$_2$ molecule in the minimal basis)~\cite{Burton2021_frac_spin}.
For sufficiently large bond distances, the restricted solution becomes
a saddle point and two unrestricted lower-in-energy solutions 
emerge. Accordingly, the constraint in
Eq.~(\ref{eq:HD_correlation_constraint_double-well}) is compatible with a reduction
of the $t$ value featuring an increasing bond length.
Still, even in the strictly correlated $t\rightarrow 0$ limit, $\theta
= 0$ remains the global minimum
for weights in the range $0\leq \ttw\leq1/3$.\\

In order to illustate the above discussion, trial SAHF energies are plotted as functions of the rotation angle
$\theta$ in the top panel of Fig.~\ref{fig:ener_theta_SAHF-eDMHF} for
the strongly correlated $U/t=3.5$ dimer and various biensemble weight values.
As expected,
the symmetric $\theta=0$ solution gives systematically the lowest
ensemble energy as long as $\ttw\leq
1/3$. When $\ttw>1/3$, two scenarios can be observed. 
For example, when $\ttw=0.35$, which gives $2(1-\ttw)/(3\ttw-1)=
26\gg U/t$,
 the dimer is not strongly correlated enough to break the symmetry and $\theta=0$ is
still the global minimum. However, for the larger $\ttw=0.475$ weight
value 
[$2(1-\ttw)/(3\ttw-1)=2.47<U/t$ in this case] or in the commonly used
equiensemble case
[$\ttw=0.5$ and $2(1-\ttw)/(3\ttw-1)=2<U/t$], the energy has the
expected double-well
shape, thus leading to two degenerate (non-zero)
minima, both corresponding to asymmetric solutions. Interestingly, in
such situations, the popular eDMHF
approach always favors the symmetric solution, as shown in the bottom panel
of Fig.~\ref{fig:ener_theta_SAHF-eDMHF}.     

\subsubsection{Single SAHF ensemble $v$-representability issue}

\begin{figure}
\centering
\resizebox{\columnwidth}{!}{
\includegraphics[scale=1.5]{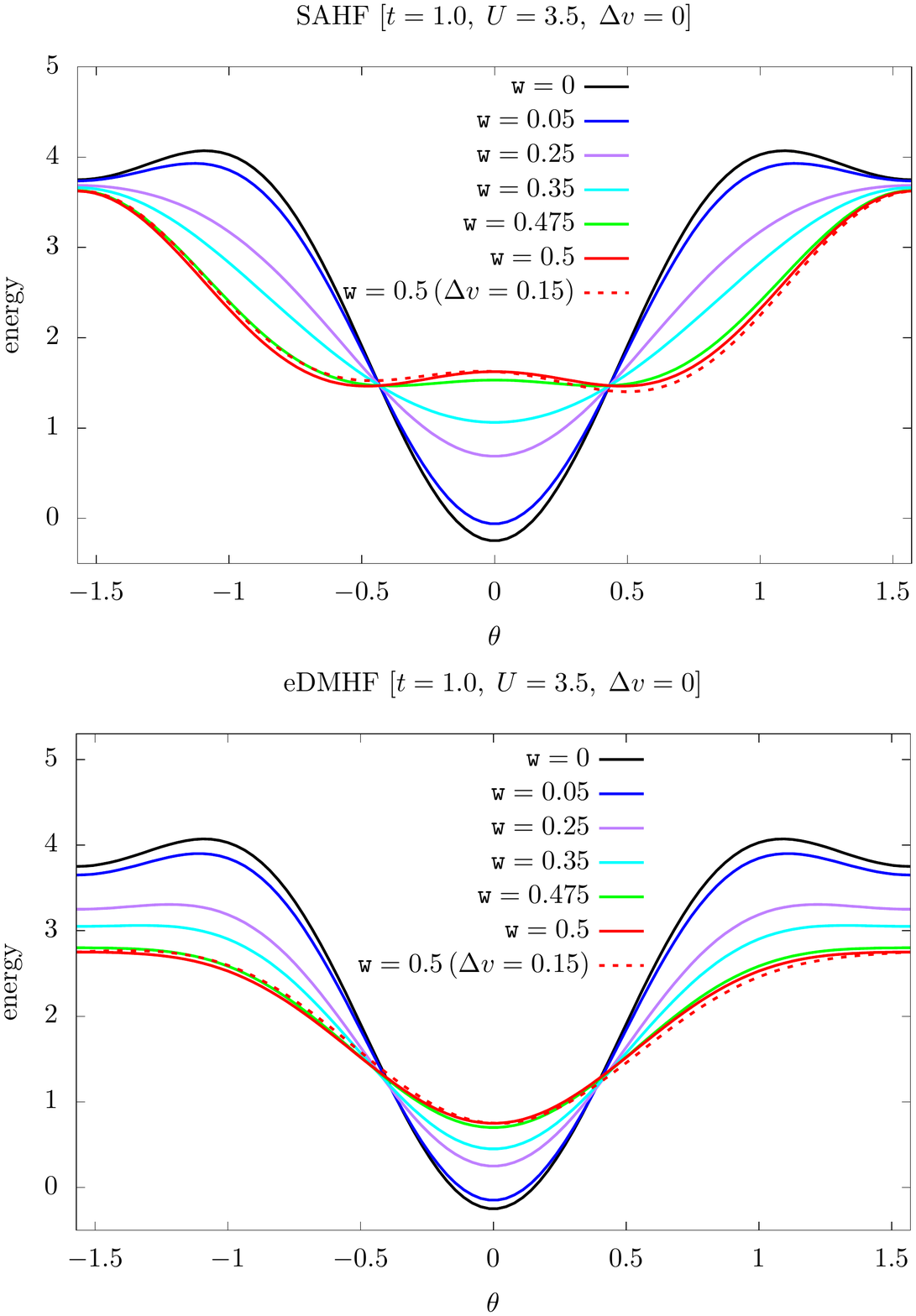}
}
\caption{Trial SAHF (top panel) and eDMHF (bottom panel) energies of the
symmetric Hubbard dimer plotted as functions of the orbital rotation
angle $\theta$ for $U/t=3.5$ and various ensemble weight values. In the
equiensemble ($\ttw=0.5$) case, results are also shown for a slightly
asymmetric ($\Delta v/t=+0.15$) dimer, for analysis purposes. In the latter case, a non-degenerate 
(positive) minimizing angle is recovered at the SAHF level (see the red
dashed curve in the
top panel), unlike
in the strictly symmetric $\Delta v=0$ case. See text for further
details. 
}
\label{fig:ener_theta_SAHF-eDMHF}
\end{figure}

\begin{figure}
\centering
\resizebox{\columnwidth}{!}{
\includegraphics[scale=1.5]{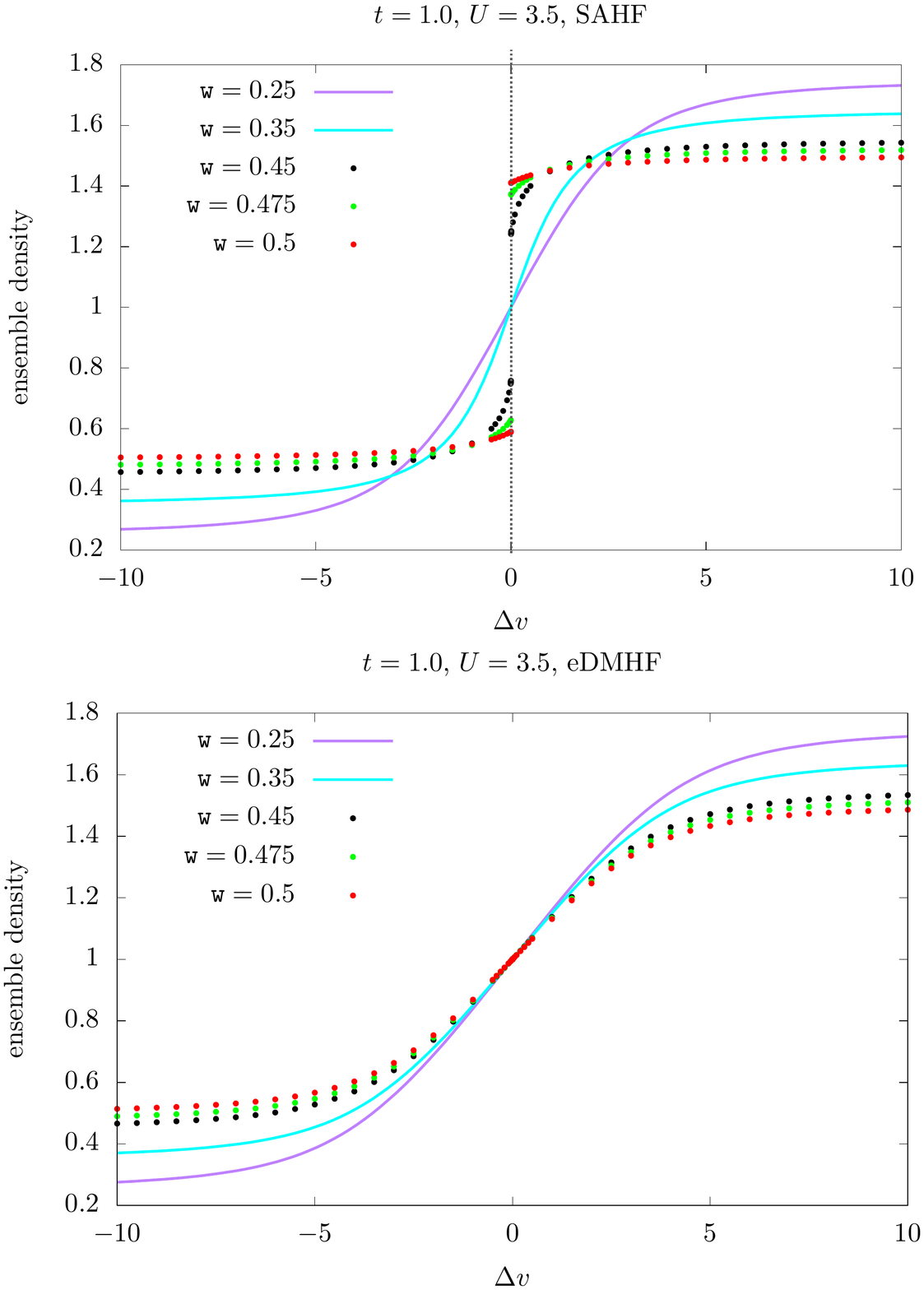}
}
\caption{Potential-ensemble-density maps generated for the Hubbard dimer at the SAHF (top panel) and eDMHF (bottom panel) levels of approximation for various ensemble weight values and $U/t=3.5$. See text for further details.
}
\label{fig:pot-dens_map_SAHF-eDMHF}
\end{figure}

\begin{figure}
\centering
\resizebox{\columnwidth}{!}{
\includegraphics[scale=1.5]{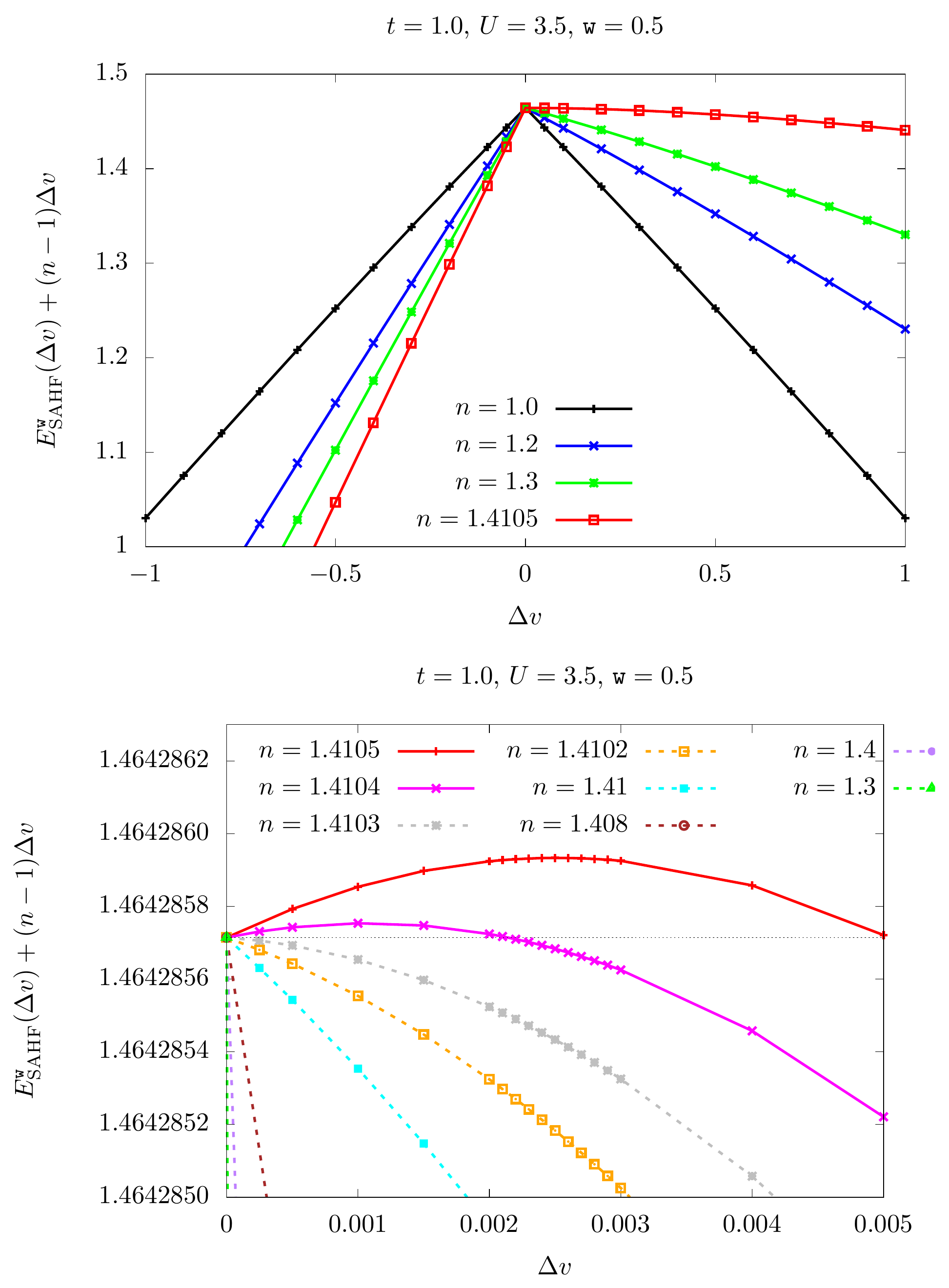}
}
\caption{(Top) To-be-maximized Lieb's potential functional (see
Refs.~\cite{deur2017exact,deur2018exploring} 
and the main text for further details) evaluated at the equal-weight
SAHF level of approximation and plotted as a function of $\Delta v$ for
various input densities $n$. (Bottom) Zoom around $\Delta v=0$ for
densities inside and outside the single SAHF-ensemble representability
domain. See text for further details.
}
\label{fig:Lieb_function_SAHF}
\end{figure}

We show in Fig.~\ref{fig:pot-dens_map_SAHF-eDMHF} the potential-ensemble-density maps
\be
\Delta v\rightarrow n^\ttw(\theta_{\rm min}(\Delta v))
\ee
generated from Eq.~(\ref{eq:trial_biens_dens_HD}) 
and
\be
\theta_{\rm min}(\Delta v)=\argmin_\theta\left\{E^\ttw_{\rm
approx.}(\Delta v,\theta)\right\},
\ee
at both eDMHF and SAHF levels
of approximation for the {\it fixed} $U/t=3.5$ interaction strength
value. Various scenarios are illustrated, in
particular those where broken symmetry SAHF solutions are obtained when
the dimer is symmetric (see
Sec.~\ref{eq:HD_symmetric_case_SAHF}). As we will see, what
happens in the symmetric case can play a crucial role in the
density-functional description of the {\it asymmetric} dimer. In cases
where $\ttw\leq 1/3$, or $\ttw>1/3$ and $2(1-\ttw)/(3\ttw-1)>U/t$, both approximations
give smooth density profiles. We note in passing 
the one-to-one correspondence between potentials and ensemble densities, as
expected from the concavity of the eDMHF and SAHF energies (see
Sec.~\ref{subsec:concavity_approx_ener}).
However, when $\ttw>1/3$ and $2(1-\ttw)/(3\ttw-1)<U/t$, the SAHF density
profile exhibits a discontinuity at $\Delta v=0$, unlike the eDMHF one. This
step in density can be interpreted as follows. If
$\ttw >1/3$ and the
constraint of Eq.~(\ref{eq:HD_correlation_constraint_double-well}) is fulfilled, as $\Delta v\rightarrow
0^\pm$, we will recover the SAHF biensemble solution
$\hat{\gamma}_\pm\equiv(1-\ttw)\ket{\Phi^\pm_0}\bra{\Phi^\pm_0}+\ttw\ket{\Phi^\pm_1}\bra{\Phi^\pm_1}$,
where $\Phi^\pm_I\equiv \Phi_I(\theta_\pm)$ and $\theta_\pm$ are the
minimizing angles associated to the broken-symmetry orbitals. Any
slight
deviation from $\Delta v=0$ will favor one of these solutions, depending
on its sign, as
illustrated in the top panel of
Fig.~\ref{fig:ener_theta_SAHF-eDMHF} (see the ``$\Delta
v=+0.15$'' curve which exhibits, in the equiensemble case, a single
absolute minimum in the vicinity
of $\theta_+$). Note that, in eDMHF, the
minimizing angle simply passes through $\theta=0$ when the potential
changes from $\Delta v=0^-$ to $\Delta v=0^+$ [see the ``$\Delta
v=+0.15$'' curve in the bottom panel of
Fig.~\ref{fig:ener_theta_SAHF-eDMHF}], and no discontinuity is observed
in the density profile.
The step
in density observed in SAHF covers the density range $n_-\leq  n\leq  n_+$, where
\be
n\pm\equiv1+(1-\ttw)\sin(2\theta_\pm)=1\pm(1-\ttw)\sqrt{1-\rho^2_0}.
\ee
In the equiensemble case ($\ttw=0.5$), we have $n\pm=1.0\pm0.410326$,
as readily seen from the top panel of 
Fig.~\ref{fig:pot-dens_map_SAHF-eDMHF}. We keep many digits for analysis
purposes (see the bottom panel of Fig.~\ref{fig:Lieb_function_SAHF}). 
It is important to
stress that none of the densities in the range $n_-<n<n_+$, 
which includes the {\it a
priori} simple symmetric $n=1$ case, can be represented by a {\it
single}
SAHF ensemble. This severe $v$-representability issue, which would deserve
further investigation at the {\it ab initio} level, for example, in the
stretched H$_2$ molecule, might be used as an argument for promoting the
eDMHF approach over the SAHF one in practical calculations. A
counter-argument is of course the presence of GI errors in eDMHF.
At the formal level, the $v$-representability issue can be solved easily
as follows. As illustrated in Fig.~\ref{fig:Lieb_function_SAHF}, Lieb's
maximization of Eq.~(\ref{eq:approx_Lieb_func}) systematically returns $\Delta v=0$ for input densities in
the range $n_-\leq  n\leq  n_+$. As soon as the input density leaves this
interval, a non-zero maximizing potential is obtained [see the bottom
panel of Fig.~\ref{fig:Lieb_function_SAHF}]. If we exploit the strict degeneracy of the
broken-symmetry solutions, we can write, for $\Delta v=0$ [see
Eq.~(\ref{eq:Hamil_Hubbard_dimer_model})],
\be
\begin{split}
E^\ttw_{\rm SAHF}(\theta_+)&=E^\ttw_{\rm SAHF}(\theta_-)
\\
&=(1-\alpha)E^\ttw_{\rm SAHF}(\theta_-)+\alpha E^\ttw_{\rm SAHF}(\theta_+)
\\
&=\Tr\left[\hat{\gamma}(\alpha)\left(\hat{\mathcal{T}}+\hat{U}\right)\right], 
\end{split}
\ee      
where $0\leq \alpha\leq 1$ and
\be
\hat{\gamma}(\alpha):=(1-\alpha)\hat{\gamma}_-+\alpha\hat{\gamma}_+
\ee 
is the convex combination of the two degenerate SAHF ensemble density
matrix operators. The
density (on site 0) of the
resulting ``ensemble of ensembles'' reads as
\be
n(\alpha)=\Tr\left[\hat{\gamma}(\alpha)\hat{n}_0\right]=(1-\alpha)n_-+\alpha
n_+,
\ee 
and, as readily seen, it can vary continuously from $n_-$ to $n_+$. The
generalization of this approach to the {\it ab initio} theory is
provided in Eqs.~(\ref{eq:double_ens_SAHF_func}) and (\ref{eq:double_ens_dens_constraint}).
  




\subsection{Exact self-consistent eDFT based on SAHF}\label{subsubsec:sc-SAHF-based_eDFT}

Let us continue with the general {\it ab initio} SAHF-based formulation of
eDFT that we left at the end of  
Sec.~\ref{subsec:concavity_approx_ener}. As already mentioned, the SAHF
universal density functional $F_{\rm SAHF}^\bfw[n]$ that is defined
in Eq.~(\ref{eq:SAHF_Levy_func}) is an approximation to the universal GOK
functional $F^\bfw[n]$. As readily seen from
Eq.~(\ref{eq:SAHF_Levy_func}), the former misses all correlation effects.
These effects can actually be introduced into the theory as a density-functional
complement:
\be\label{eq:c_func_SAHF}
\tilde{E}_{\rm c}^\bfw[n]:=F^\bfw[n]-F_{\rm
SAHF}^\bfw[n].
\ee 
As a result, according to Eqs.~(\ref{eq:Ew_GOK-DFT}) and
(\ref{eq:SAHF_Levy_func}), the exact ensemble energy can be calculated variationally
as follows,
\be
E^\bfw=
\min_n\Bigg\{
&&\min_{\left\{\Phi_I\right\}\overset{\bfw}{\rightarrow}n}
\left\{\sum_{I}\ttw_I\mel{\Phi_I}{\hat{T}+\hat{W}_{\rm
ee}}{\Phi_I}
\right\}+\tilde{E}_{\rm c}^\bfw[n]
\nonumber
\\
&&
+\int d\bfr\,v_{\rm
ext}(\bfr)n(\bfr)
\Bigg\},
\ee
or, equivalently,
\be
E^\bfw=
\min_n\left\{\min_{\left\{\Phi_I\right\}\overset{\bfw}{\rightarrow}n}
\left\{
\sum_{I}\ttw_I\mel{\Phi_I}{\hat{H}}{\Phi_I}
+\tilde{E}_{\rm c}^\bfw\left[\sum_I\ttw_In_{\Phi_I}\right]
\right\}
\right\},
\ee
thus leading to the final expression:
\be\label{eq:final_VP_SAH_eDFT}
E^\bfw&=&\min_{\left\{\Phi_I\right\}}\left\{\sum_{I}\ttw_I\mel{\Phi_I}{\hat{H}}{\Phi_I}
+
\tilde{E}_{\rm c}^\bfw\left[\sum_I\ttw_In_{\Phi_I}\right]
\right\}
.
\ee
By differentiating the density-functional correlation energy with
respect to any (orbital rotation) variational parameter $\kappa_{pq}$
[see Appendix~\ref{appendix:SAHF_eqs}], 
\be
\begin{split}
\dfrac{\partial 
}{\partial
\kappa_{pq}}
\left(\tilde{E}_{\rm
c}^\bfw\left[\sum_I\ttw_In_{\Phi_I({\bfkap})}\right]
\right)
&=
\int d\bfr\,\left.\dfrac{
\delta\tilde{E}^\bfw_{\rm c}[n]}{\delta
n(\bfr)}\right|_{n=\sum_I\ttw_In_{\Phi_I({\bfkap})}}
\\
&\quad\times 
\dfrac{\partial
}{\partial
\kappa_{pq}}
\left(
\sum_I\ttw_I
n_{\Phi_I({\bfkap})}(\bfr)
\right),
\end{split}
\ee
we realize that the minimizing orbitals in 
Eq.~(\ref{eq:final_VP_SAH_eDFT}), from which the single-configuration
wave functions that reproduce the exact ensemble density $n^\bfw$ are
constructed (we denote them $\tilde{\Phi}^\bfw_I$ for convenience), fulfill SAHF-like self-consistent equations [see
Eq.~(\ref{eq:Fock-like_eq_off-diag_epsi_more_explicit})] where the
density-functional
correlation potential $\left.\delta\tilde{E}^\bfw_{\rm c}[n]/{\delta
n(\bfr)}\right|_{n=\sum_I\ttw_In_{\tilde{\Phi}^\bfw_I}} 
$ is simply added to the local external one.\\  

In practice, the quantities of interest are usually the excitation energies and,
more generally, the ground- and excited-state energy levels (which are
needed, for example, for geometry optimizations). According to Eqs.~(\ref{eq:indiv_EI}) and
(\ref{eq:final_VP_SAH_eDFT}), and the
Hellmann--Feynman theorem, the latter can be evaluated, in principle
exactly, as follows,  
\color{black}
\be
\label{eq:ind_ener_from_SAHF_eDFT}
\begin{split}
E_I&=
\expval{\hat{H}}_{\tilde{\Phi}^\bfw_I}
+
\int d\bfr\,\tilde{\mathcal{V}}^\bfw_{\rm
c}(\bfr)n_{\tilde{\Phi}^\bfw_I}(\bfr)
+\sum_{J>0}(\delta_{IJ}-\ttw_J)\left.\dfrac{\partial
\tilde{E}^\bfw_{\rm
c}[n]}{\partial \ttw_J}\right|_{n=n^\bfw},
\end{split}
\ee
where  
the following relation has been used, 
\be
\sum_{J>0}(\delta_{IJ}-\ttw_J)\left(n_{\tilde{\Phi}^\bfw_J}(\bfr)-n_{\tilde{\Phi}^\bfw_0}(\bfr)\right)=n_{\tilde{\Phi}^\bfw_I}(\bfr)-n^\bfw(\bfr),
\ee   
and
$\tilde{\mathcal{V}}^\bfw_{\rm
c}\equiv \tilde{\mathcal{V}}^\bfw_{\rm
c}[n^\bfw]$ is the LZ-shifted~\cite{levy2014ground,loos2020weight}
ensemble correlation-only density-functional potential: 
\be
\tilde{\mathcal{V}}^\bfw_{\rm
c}[n](\bfr)=\dfrac{
\delta\tilde{E}^\bfw_{\rm c}[n]}{\delta
n(\bfr)}+
\dfrac{\tilde{E}_{\rm
c}^\bfw\left[n\right]
-\int d\bfr\,\frac{\delta\tilde{E}^\bfw_{\rm c}[n]}{\delta
n(\bfr)}n(\bfr)}{\int d\bfr\,n(\bfr)}.
\ee
Interestingly, an expression similar to that of Eq.~(\ref{eq:ind_ener_from_SAHF_eDFT}) has been derived in the context of
GOK-DFT~\cite{loos2020weight,fromager2020individual} (see also
Sec.~\ref{sec:corr_energy}) where,
unlike in the present case, a local ensemble exchange potential was
used.

\subsection{Connection with practical hybrid eDFT calculations}

By analogy with conventional (ground-state) hybrid functionals, we can
modify as follows the 
SAHF-based functional of Eq.~(\ref{eq:SAHF_func_from_Lieb_to_Levy}), in order to combine rigorously a fraction $\lambda$ of SAHF
exchange energy with a weighted sum of (approximate) individual density-functional
exchange
energies, thus leading to another (Hx-only) approximation to the GOK
functional:
\be\label{eq:SAHF_Lieb_func_lambda}
F^\bfw[n]\approx{F}^{\bfw,\lambda}_{\rm
hybrid}[n]=\max_v\left\{\mathscr{E}^{\bfw,\lambda}_{\rm
hybrid}[v]-\int d\bfr\,v(\bfr)n(\bfr)\right\},
\ee
where
\be\label{eq:min_SAHF_ener_lambda}
\mathscr{E}^{\bfw,\lambda}_{\rm
hybrid}[v]
=\min_{\left\{\Phi_I\right\}}\Bigg\{\sum_{I}\ttw_I\left[\expval{\hat{T}+\lambda\hat{W}_{\rm
ee}+\hat{V}}_{\Phi_I}+(1-\lambda)E_{\rm Hx}[n_{\Phi_I}]
\right]\Bigg\},
\ee
and $E_{\rm Hx}[n]$ is the regular {\it ground-state} Hx density functional of
KS-DFT.
The above approximate ensemble energy can be computed from SAHF routines
simply by scaling the
individual non-local exchange potentials and then adding to each of them
the complementary fraction
$(1-\lambda)$ of individual local density-functional exchange potential.    
As readily seen from Eq.~(\ref{eq:min_SAHF_ener_lambda}), in the present
scheme, the total
ensemble Hx energy remains GI-free, even in the limiting $\lambda=0$ case. On that
basis, an exact hybrid eDFT can be derived along the lines of
Sec.~\ref{subsubsec:sc-SAHF-based_eDFT} by considering the following
in-principle-exact decomposition of the universal GOK functional (where
correlation is described with an ensemble density functional): 
\be\label{eq:xc_func_SAHF_lambda}
F^\bfw[n]={F}^{\bfw,\lambda}_{\rm
hybrid}[n]+(1-\lambda)\Delta\tilde{{E}}_{\rm x}^{\bfw,\lambda}[n]+
\tilde{{E}}_{\rm c}^{\bfw,\lambda}[n].
\ee 
A density-functional correction
$\Delta\tilde{{E}}_{\rm x}^{\bfw,\lambda}[n]$ to the ensemble
exchange energy must in principle be introduced
since each individual exchange energy
$E_{\rm x}[n_{\Phi_I}]$ is evaluated, for both ground- and
excited-state densities, as a {\it ground-state}
exchange energy. This correction is usually neglected in practical
calculations~\cite{filatov2021description}. This
observation would actually hold also for approximate ensemble
correlation energies that are
constructed from the regular ground-state correlation functional $E_{\rm
c}[n]$ of KS-DFT (see below and Sec.~\ref{subsec:GS-ic_and_eLDA}).    
Note finally the $\lambda$-dependence of both $\Delta\tilde{{E}}_{\rm x}^{\bfw,\lambda}[n]$  
and $\tilde{{E}}_{\rm c}^{\bfw,\lambda}[n]$
functionals in Eq.~(\ref{eq:xc_func_SAHF_lambda}). It originates from the fact that the
ensemble xc energy is now evaluated from the (SAHF-like) $\lambda$-dependent single-configuration wave functions that reproduce the desired density
$n$.\\ 
 
Finally, as discussed in further detail in Sec.~\ref{sec:corr_energy}, it is
quite common to construct weight-dependent 
ensemble density-functional correlation energies by recycling the ground-state correlation
functional as follows~\cite{filatov2015spin},
\be\label{eq:intro_to_GS-ic}
\tilde{{E}}_{\rm
c}^{\bfw,\lambda}\left[\sum_I\ttw_In_{\Phi_I}\right]\approx
\sum_I\ttw_IE_{\rm c}[n_{\Phi_I}].
\ee
This standard approximation, which is referred to as {\it ground-state individual
correlations} (GS-ic) scheme in the following, can also be made formally exact within the present
formalism. Indeed, once we have introduced the following (approximate) potential-functional energy 
\be\label{eq:min_SA_ener_lambda}
\mathscr{E}^{\bfw,\lambda}_{{\rm
GS}-{\rm ic}}[v]
=\min_{\left\{\Phi_I\right\}}\Bigg\{
\sum_{I}\ttw_I\Big(
&&\expval{\hat{T}+\lambda\hat{W}_{\rm
ee}+\hat{V}}_{\Phi_I}
\nonumber
\\
&&
+(1-\lambda)E_{\rm Hx}[n_{\Phi_I}]
+E_{\rm c}[n_{\Phi_I}]
\Big)\Bigg\},
\ee
and the subsequent density functional
\be
F^{\bfw,\lambda}_{{\rm
GS}-{\rm ic}}[n]=\max_v\left\{\mathscr{E}^{\bfw,\lambda}_{\rm
{\rm
GS}-{\rm ic}}[v]-\int d\bfr\,v(\bfr)n(\bfr)\right\},
\ee
we only need to consider the alternative (but still exact) partitioning
of the GOK functional
\be
F^\bfw[n]=F^{\bfw,\lambda}_{{\rm
GS}-{\rm ic}}[n]+(1-\lambda)\Delta\check{{E}}_{\rm x}^{\bfw,\lambda}[n]+
\Delta\check{{E}}_{\rm c}^{\bfw,\lambda}[n],
\ee 
where complementary density-functional corrections to both exchange and
(GS-ic) correlation energies have been introduced. Note that, in practice, these
corrections are simply neglected~\cite{filatov2021description}. 
While SAHF is expected to provide, through its orbital dependence, a proper description
of the ensemble 
exchange energy, modeling the correlation density-functional correction
$\Delta\check{{E}}_{\rm c}^{\bfw,\lambda}[n]$ remains a necessary and
challenging task that has attracted too little attention until now. For
that purpose, it is essential to have a deeper understanding of how
individual correlation energies are connected to the ensemble one. This
is the main focus of the next section.

\section{Individual correlations within ensembles: An exact construction}
\label{sec:corr_energy}

While the previous section was dedicated to the description of orbital-
and weight-dependent ensemble Hx energies, this last section deals with correlation effects in
many-body ensembles. For convenience, we continue focusing on GOK
ensembles but the discussion applies to other types of
ensembles like, for example, $N$-centered~\cite{senjean2018unified} or thermal
ones~\cite{pastorczak2013calculation,PRL11_Pittalis_exact_conds_thermalDFT,PRB16_Pribram-Jones_AC_thermalDFT,smith2016exact}.
We will work within the original 
GOK-DFT formalism~\cite{gross1988density}, where a local multiplicative
ensemble-density-functional Hxc
potential is employed, but the discussion holds also when
orbital-dependent exchange energies are employed (see Sec.~\ref{sec:EXX}).

\subsection{State-of-the-art ensemble correlation DFAs and beyond}\label{subsec:GS-ic_and_eLDA}

To the best of our knowledge, very few works have addressed the
construction of weight-dependent ensemble correlation DFAs from first principles.
We can essentially distinguish two different general strategies. In the first and most
straightforward one, which was introduced in
Eq.~(\ref{eq:intro_to_GS-ic}) and that we referred to
as GS-ic, the (weight-independent) ground-state correlation functional is recycled as
follows,
\be\label{eq:GS-ic_in_GOK-DFT}
E^\bfw_{\rm c}[n^\bfw]\overset{\rm GS-ic}{\approx}\sum_I\ttw_IE_{\rm
c}[n_{\Phi^\bfw_I}],
\ee
where, in the exact theory, the KS wave functions
$\left\{\Phi^\bfw_I\right\}$ are expected to reproduce the true ensemble
density $n^\bfw$.\\

More recently~\cite{loos2020weight,marut2020weight}, Loos and coworkers explored another path. They designed a first generation of
weight-dependent ensemble
LDA (eLDA) correlation functionals where the regular ground-state LDA functional $E^{\rm LDA}_{\rm
c}[n]=\int d\bfr\,n(\bfr)\epsilon_{\rm
c}(n(\bfr))$, which is based on the {\it infinite} uniform electron gas (UEG)
model, is
combined with the density-functional correlation excitation energies of
a {\it finite} UEG (hence the acronym $f$LDA used below) as follows,  
\be\label{eq:eLDA_func}
E^{\bfw}_{\rm c}[n]\overset{\rm eLDA}{\approx}E^{\rm LDA}_{\rm
c}[n]+\sum_{I>0}\ttw_I\left(\mathcal{E}^{f{\rm LDA}}_{{\rm
c},I}[n]-\mathcal{E}^{f{\rm LDA}}_{{\rm c},I=0}[n]\right).
\ee
The individual correlation functional $\mathcal{E}^{f{\rm LDA}}_{{\rm
c},I}[n]=\int d\bfr\,n(\bfr)\epsilon^{f}_{{\rm
c},I}(n(\bfr))$ is constructed from the $I$th state of the finite UEG:
\be\label{eq:ind_corr_ener_eLDA}
N_f\,\epsilon^{f}_{{\rm
c},I}(n)\equiv\mel{\Psi_I(n)}{\hat{T}+\hat{W}_{\rm
ee}}{\Psi_I(n)}-\mel{\Phi_I(n)}{\hat{T}+\hat{W}_{\rm ee}}{\Phi_I(n)},
\ee
where $N_f$ is the number of electrons in the finite gas. If $N_f$ is
fixed, a parameterization of the correlation energy per particle $\epsilon^{f}_{{\rm
c},I}(n)$ as a function of the uniform density $n$ is obtained by varying the volume of the gas~\cite{loos2020weight}. Note
that, in a uniform system, there is no need to introduce weight
dependencies into the interacting and non-interacting (ground-
or excited-state) density-functional wave functions, unlike in the
general definition of Eq.~(\ref{eq:ens_corr_fun_components}). Indeed, all the eigenstates of the (interacting
or non-interacting) uniform gas have the same (uniform) density $n$,
which then becomes the ensemble density of the gas, whatever the
value of the ensemble
weights:
\be
\sum_{I}\ttw_I n_{\Psi_I(n)}=\sum_{I}\ttw_I
n_{\Phi_I(n)}=n\sum_{I}\ttw_I=n.
\ee
 Note also that,  
while the finite UEG allows for the incorporation of weight dependencies
into the correlation functional, the use of a regular LDA correlation functional
reduces 
finite-size errors. Refinements are possible, for example, by
including a dependence in the Fermi hole curvature~\cite{loos2017exchange}.\\

The strategies depicted in Eqs.~(\ref{eq:GS-ic_in_GOK-DFT}) and (\ref{eq:eLDA_func}) miss various correlation effects that we briefly
review below. More insight will be given in the next
subsections. Let us start with the GS-ic approximation. From the exact expression,
\be\label{eq:ind_GS_corr_ES_dens}
E_{\rm
c}[n_{\Phi^\bfw_I}]=\expval{\hat{T}+\hat{W}_{\rm
ee}}_{\Psi_0[n_{\Phi^\bfw_I}]}-\expval{\hat{T}+\hat{W}_{\rm
ee}}_{\Phi_0[n_{\Phi^\bfw_I}]},
\ee
where $\Psi_0[n]$ and $\Phi_0[n]$ are the interacting and
KS non-interacting {\it ground-state} density-functional wave functions
of regular KS-DFT,
respectively, we immediately identify two sources of errors. The first
one is related to the fact that, as already mentioned in
Sec.~\ref{subsubsec:indivGOK}, the individual KS density
$n_{\Phi^\bfw_I}$ does not necessarily match the interacting
individual one $n_{\Psi_I}$. This subtle point was recently highlighted by Gould
and Pittalis~\cite{gould2019density,gould2020density}. It induces what the authors referred to as
{\it density-driven} (DD) correlation effects. Even if the true
individual densities $\left\{n_{\Psi_I}\right\}$ (which can be extracted in principle exactly from
the KS ensemble, as shown in Eq.~(\ref{eq:indiv_dens_response}) and Ref.~\cite{fromager2020individual}) were inserted
into the expression of Eq.~(\ref{eq:ind_GS_corr_ES_dens}), we would
still not recover the correct individual excited-state correlation energies simply because
$\Psi_0[n_{\Psi_I}]$ will always be a ground-state wave function, even
when $n_{\Psi_I}$ is an excited-state density. The missing energy
contribution is connected to the concept of 
{\it state-driven} (SD) correlation~\cite{gould2019density}.
Interestingly, eLDA describes (approximately) SD correlations, as readily seen from
Eq.~(\ref{eq:ind_corr_ener_eLDA}). However, it completely misses DD
ones, simply because KS and interacting (ground- or excited-state) wave functions have the same
density in a uniform system.\\ 

Even though the physical meaning of DD and SD correlations is rather
clear, it is less obvious how their contributions to the total exact ensemble
correlation energy should be
defined
mathematically~\cite{gould2019density,fromager2020individual,gould2020density,gould2020approximately}. Addressing this fundamental question
is of primary importance for the design of more accurate and
systematically improvable ensemble correlation DFAs, which is probably the most challenging task in GOK-DFT.  
Up to now, we have discussed the concept of DD and SD correlations
in the light of the GS-ic approximation [see Eqs.~(\ref{eq:GS-ic_in_GOK-DFT}) and (\ref{eq:ind_GS_corr_ES_dens})]. We may actually
wonder if a proper definition can be (or should be) given without
referring explicitly to the GS correlation functional of KS-DFT. Indeed, the latter appears
naturally in GOK-DFT only in the limiting $\bfw=0$  case. 
Gould and Pittalis~\cite{gould2019density}, and then
Fromager~\cite{fromager2020individual},
recently addressed this SD/DD ensemble correlation energy decomposition issue from that
perspective. A detailed and
complemented review of the two approaches is presented in the following.\\ 

\subsection{Weight dependence of the KS wave functions in GOK-DFT}\label{subsec:difference_KS_true_dens}

Before proceeding with the extraction of individual correlation
energies from the GOK-DFT ensemble energy, which is
convenient for deriving in-principle-exact SD/DD
decompositions~\cite{fromager2020individual}, we would like to highlight the importance of 
weight dependencies in the KS wave functions. It might be
surprizing at first sight because the true ground and excited states of
the system under study are of course weight-independent. We 
explain below, with a simple argument, why it cannot be the case in the KS ensemble.\\

Since the KS and true ensemble densities match for any set of weights
$\bfw$, their derivatives with respect to the weights also match.
Therefore, if we consider the
ground-state $\bfw=0$ limit of GOK-DFT, it comes 
\be
\left.\dfrac{\partial}{\partial
\ttw_J}\left(\sum_I\ttw_In_{\Psi_I}(\bfr)\right)\right|_{\bfw=0}
\overset{J>0}{=}\left.\dfrac{\partial}{\partial
\ttw_J}\left(\sum_I\ttw_In_{\Phi_I^\bfw}(\bfr)\right)\right|_{\bfw=0},
\ee
or, equivalently,
\be
n_{\Psi_J}(\bfr)-n_{\Psi_0}(\bfr)=n_{\Phi_J}(\bfr)-n_{\Phi_0}(\bfr)+\left.\dfrac{\partial
n_{\Phi_0^\bfw}(\bfr)}{\partial \ttw_J}\right|_{\bfw=0},
\ee
where $\left\{\Phi_I\right\}$ denote here the ground- and excited-state
KS wave functions generated 
from a regular ground-state
KS-DFT calculation. In KS-DFT, the density constraint applies to the  
ground state only, \ie, $n_{\Phi_0}(\bfr)=n_{\Psi_0}(\bfr)$, not to the
excited states. Thus, we obtain the exact individual excited-state
density expression, which can be recovered from
Eq.~(\ref{eq:indiv_dens_response}) when
$\bfw=0$,
\be\label{eq:ind_ES_dens_GS-limit}
n_{\Psi_J}(\bfr)-n_{\Phi_J}(\bfr)\overset{J>0}{=}\left.\dfrac{\partial
n_{\Phi_0^\bfw}(\bfr)}{\partial \ttw_J}\right|_{\bfw=0}\neq 0,
\ee
where we readily see that, in GOK-DFT, the
KS wave functions (the ground-state one in the present case) 
are necessarily {\it weight-dependent}. 
This feature is central
in the design of DD correlation energies~\cite{fromager2020individual}, as discussed further in the
following. We refer to Eq.~ (\ref{eq:weight_dep_GS_KS_HD}) for an
illustrative example (based on the prototypical Hubbard dimer) of
weight-dependent KS ground-state density.

\subsection{Extraction of individual correlation energies}

In this section we revisit the derivation of the individual energy
levels in Eq.~(\ref{eq:Ek_GOK-DFT}) in order to construct individual correlation
energies within the ensemble under study. For that purpose, we start
from the exact relation between individual and ensemble energies in
Eq.~(\ref{eq:indiv_EI}), and the variational GOK-DFT ensemble energy
expression of Eq.~(\ref{eq:Ew_GOK-DFT_VP}), thus leading to, according to
the Hellmann--Feynman theorem,
\be
\begin{split}
E_J&=\sum_{I\geq 0}\ttw_I\mel{\Phi_I^\bfw}{\hat{T} + \hat{V}_{\rm
ext}}{\Phi_I^\bfw}+E^\bfw_{\rm Hxc}[n^\bfw]
\\
&\quad
+\sum_{I>0} (\delta_{IJ} - \ttw_I)\left[\mel{\Phi_I^\bfw}{\hat{T} + \hat{V}_{\rm
ext}}{\Phi_I^\bfw}-\mel{\Phi_0^\bfw}{\hat{T} + \hat{V}_{\rm
ext}}{\Phi_0^\bfw}\right]
\\
&\quad+\sum_{I>0} (\delta_{IJ} - \ttw_I)\left[\dfrac{\partial
E^\bfw_{\rm Hxc}[n^\bfw]}{\partial\ttw_I}-\left.\dfrac{\partial E^{\bmxi}_{\rm
Hxc}[n^{\bmxi,\bfw}]}{\partial
\ttw_I}\right|_{\bmxi=\bfw}\right],
\end{split}
\ee
or, equivalently,
\be\label{eq:ind_ener_GOK-DFT_a-bit-simplified}
\begin{split}
E_J&=\mel{\Phi_J^\bfw}{\hat{T} + \hat{V}_{\rm
ext}}{\Phi_J^\bfw}+E^\bfw_{\rm Hxc}[n^\bfw]
\\
&\quad+\sum_{I>0} (\delta_{IJ} - \ttw_I)\left[\dfrac{\partial
E^\bfw_{\rm Hxc}[n^\bfw]}{\partial\ttw_I}-\left.\dfrac{\partial E^{\bmxi}_{\rm
Hxc}[n^{\bmxi,\bfw}]}{\partial
\ttw_I}\right|_{\bmxi=\bfw}\right],
\end{split}
\ee
where, in analogy with Eq.~(\ref{eq:2ble_weight_dens_Nc}), the following double-weight ensemble KS density
has been introduced:
\be\label{eq:double_weight_KS_ens_dens}
n^{\bmxi,\bfw}(\bfr)=\sum_{I\geq 0}\xi_In_{\Phi_I^\bfw}(\bfr).
\ee
The last contribution (that is subtracted) on the right-hand side of Eq.~(\ref{eq:ind_ener_GOK-DFT_a-bit-simplified}) originates
from the Hellmann--Feynman theorem. In other words, derivatives of the
KS wave functions (and, therefore, of their densities) do not contribute to the
derivatives of the {\it total} ensemble energy, because the latter is
variational.\\
  
As shown in Refs.~\cite{loos2020weight,fromager2020individual}, the Hx contribution to the individual $J$th
energy level reduces to the expectation value of the
two-electron repulsion operator evaluated for the $J$th KS state, as one
would guess. Indeed, once we have realized that, for given weight values $\bmxi$, the ensemble KS potential
that reproduces $n^{\bmxi,\bfw}$ is simply the one that reproduces the true
ensemble density $n^\bfw$, we deduce from Eq.~(\ref{eq:def_exact_ens_Hx_func}) that 
\be\label{eq:trick_double_weight_Hx}
E^{\bmxi}_{\rm
Hx}[n^{\bmxi,\bfw}]=\sum_{K\geq 0}\xi_K\mel{\Phi_K^\bfw}{\hat{W}_{\rm
ee}}{\Phi_K^\bfw},
\ee
and, consequently,
\be
\left.\dfrac{\partial E^{\bmxi}_{\rm
Hx}[n^{\bmxi,\bfw}]}{\partial
\ttw_I}\right|_{\bmxi=\bfw}=\sum_{K\geq 0}\ttw_K\dfrac{\partial \mel{\Phi_K^\bfw}{\hat{W}_{\rm
ee}}{\Phi_K^\bfw}}{\partial \ttw_I}.
\ee
As a result, since $E^\bfw_{\rm Hx}[n^\bfw]=E^{\bfw}_{\rm
Hx}[n^{\bfw,\bfw}]$, it comes
\be
\dfrac{\partial
E^\bfw_{\rm Hx}[n^\bfw]}{\partial\ttw_I}-\left.\dfrac{\partial E^{\bmxi}_{\rm
Hx}[n^{\bmxi,\bfw}]}{\partial
\ttw_I}\right|_{\bmxi=\bfw}=\mel{\Phi_I^\bfw}{\hat{W}_{\rm
ee}}{\Phi_I^\bfw}-\mel{\Phi_0^\bfw}{\hat{W}_{\rm
ee}}{\Phi_0^\bfw},
\ee
thus leading to the expected result:
\be
\begin{split}
&E^\bfw_{\rm Hx}[n^\bfw]
+\sum_{I>0} (\delta_{IJ} - \ttw_I)\left[\dfrac{\partial
E^\bfw_{\rm Hx}[n^\bfw]}{\partial\ttw_I}-\left.\dfrac{\partial E^{\bmxi}_{\rm
Hx}[n^{\bmxi,\bfw}]}{\partial
\ttw_I}\right|_{\bmxi=\bfw}\right]
\\
&=\mel{\Phi_J^\bfw}{\hat{W}_{\rm
ee}}{\Phi_J^\bfw}.
\end{split}
\ee
We conclude from Eq.~(\ref{eq:ind_ener_GOK-DFT_a-bit-simplified})
that the energy levels can be evaluated exactly within GOK-DFT as
follows,
\be\label{eq:exact_ind_ener_with_corr}
E_J=\mel{\Phi_J^\bfw}{\hat{H}}{\Phi_J^\bfw}+E^\bfw_{{\rm c},J}[n^\bfw],
\ee  
where the individual correlation energy of the $J$th state 
is determined from the ensemble correlation density functional as follows,
\be\label{eq:ind_corr_ener_removedHx}
E^\bfw_{{\rm c},J}[n^\bfw]=E^\bfw_{\rm c}[n^\bfw]
+\sum_{I>0} (\delta_{IJ} - \ttw_I)\left[\dfrac{\partial
E^\bfw_{\rm c}[n^\bfw]}{\partial\ttw_I}-\left.\dfrac{\partial E^{\bmxi}_{\rm
c}[n^{\bmxi,\bfw}]}{\partial
\ttw_I}\right|_{\bmxi=\bfw}\right].
\ee  
In the following section, we will see how the concept of DD 
correlation emerges from Eq.~(\ref{eq:ind_corr_ener_removedHx}), once it
has been rewritten more explicitly in
terms of individual densities. 


\subsection{Individual correlations versus individual components}

According to the definition of the ensemble correlation functional in
GOK-DFT [see
Eq.~(\ref{eq:ens_corr_fun_components})], the exact ensemble
correlation energy can be decomposed as follows,   
\be\label{eq:decomp_ens_corr_components}
E^\bfw_{\rm c}[n^\bfw]=\sum_{J\geq 0}\ttw_J\mathcal{E}^\bfw_{{\rm c},J}[n^\bfw],
\ee
where the individual {\it components} read as
\be\label{eq:ind_components_ens_corr_ener}
\mathcal{E}^\bfw_{{\rm c},J}[n^\bfw]=\mel{\Psi_J}{\hat{T}+\hat{W}_{\rm
ee}}{\Psi_J}-\mel{\Phi_J^\bfw}{\hat{T}+\hat{W}_{\rm
ee}}{\Phi_J^\bfw}.
\ee 
Let us stress that these components do {\it not} match the
individual correlation energies of 
Eq.~(\ref{eq:ind_corr_ener_removedHx}). Indeed, unlike the latter [see
Eq.~(\ref{eq:exact_ind_ener_with_corr})], they do not give 
access to the exact individual energy levels,
\be\label{eq:ind_component_not_ind_corr}
\begin{split}
\mel{\Phi_J^\bfw}{\hat{H}}{\Phi_J^\bfw}+\mathcal{E}^\bfw_{{\rm c},J}[n^\bfw]&=\mel{\Psi_J}{\hat{T}+\hat{W}_{\rm
ee}}{\Psi_J}+\int d\bfr\,v_{\rm ext}(\bfr)n_{\Phi_J^\bfw}(\bfr)
\\
&\neq E_J,
\end{split}
\ee
simply because the KS density $n_{\Phi_J^\bfw}$ does not match, in
general, the true physical density $n_{\Psi_J}$. The concept of
DD correlation, which was introduced recently by Gould and
Pittalis~\cite{gould2019density}, originates from this observation. The
important property that the true individual correlation energies share with the
individual correlation components is that both of them can be used to construct the
total ensemble correlation energy, \ie,
\be\label{eq:decomp_ens_corr_from_exact_ind_corr}
E^\bfw_{\rm c}[n^\bfw]=\sum_{J\geq 0}\ttw_JE^\bfw_{{\rm c},J}[n^\bfw].
\ee
The above expression can be deduced from
Eq.~(\ref{eq:ind_corr_ener_removedHx}) and the fact that, 
for any $\left\{\Delta_I\right\}_{I>0}$,  
\be
\begin{split}
\sum_{J\geq 0}\ttw_J\left(\sum_{I>0} (\delta_{IJ} -
\ttw_I)\Delta_I\right)&=\sum_{I>0}\sum_{J\geq
0}\delta_{IJ}\ttw_J\Delta_I-\left(\sum_{J\geq 0}\ttw_J\right)\sum_{I>0}\ttw_I\Delta_I 
\\
&=\sum_{I>0}\ttw_I\Delta_I-\sum_{I>0}\ttw_I\Delta_I
\\
&=0.
\end{split}
\ee
From now on we will substitute the decomposition of
Eq.~(\ref{eq:decomp_ens_corr_from_exact_ind_corr}) for the more
conventional one of Eq.~(\ref{eq:decomp_ens_corr_components}). As shown
in the following, with this change of paradigm, DD-type correlation energy contributions will naturally emerge
from the derivation of a more explicit 
expression. 
Unlike in Ref.~\cite{gould2019density}, the approach of Ref.~\cite{fromager2020individual}, which is
reviewed in the next section, does not require additional (state-specific) KS systems, thus avoiding formal
issues such as the non-uniqueness of KS potentials for individual
excited states or 
$v$-representability issues~\cite{fromager2020individual}. 

\subsection{Density-driven ensemble correlation energy expression}

Let us now derive a more explicit expression for the ensemble
correlation energy, on the basis of Eq.~(\ref{eq:decomp_ens_corr_from_exact_ind_corr}). We
start with a simplification of the true individual correlation energy
expression of
Eq.~(\ref{eq:ind_corr_ener_removedHx}), where the standard decomposition
into components [see Eq.~(\ref{eq:decomp_ens_corr_components})] of the ensemble correlation
energy will be employed. On the one hand, we will have 
\be\label{eq:full_deriv_corr_ener}
\dfrac{\partial
E^\bfw_{\rm c}[n^\bfw]}{\partial\ttw_I}=\mathcal{E}^\bfw_{{\rm
c},I}[n^\bfw]-\mathcal{E}^\bfw_{{\rm c},0}[n^\bfw]+\sum_{K\geq
0}\ttw_K\dfrac{\partial \mathcal{E}^\bfw_{{\rm c},K}[n^\bfw]}{\partial \ttw_I}
,
\ee
where, according to Eq.~(\ref{eq:ind_components_ens_corr_ener}),
\be\label{eq:derivative_corr_component_simplified}
\begin{split}
\dfrac{\partial \mathcal{E}^\bfw_{{\rm c},K}[n^\bfw]}{\partial
\ttw_I}&=-\dfrac{\partial }{\partial
\ttw_I}\left[\mel{\Phi_K^\bfw}{\hat{T}+\hat{W}_{\rm
ee}}{\Phi_K^\bfw}\right]
\\
&=-2\mel{\Phi_K^\bfw}{\hat{T}+\hat{W}_{\rm
ee}}{\dfrac{\partial\Phi_K^\bfw}{\partial \ttw_I}}.
\end{split}
\ee 
As readily seen from
Eq.~(\ref{eq:derivative_corr_component_simplified}), the weight
derivatives of the individual
correlation components are evaluated solely from the KS wave
functions and their (static) linear response to variations in the
ensemble weights. The true interacting wave functions are not involved
since, unlike the KS wave functions, they do not vary with the ensemble weights
[see the comment that follows Eq.~(\ref{eq:ens_corr_fun_components}),
and Eq.~(\ref{eq:ind_components_ens_corr_ener})]. Combining
Eqs.~(\ref{eq:decomp_ens_corr_components}),
(\ref{eq:full_deriv_corr_ener}), and
(\ref{eq:derivative_corr_component_simplified}) leads to the following
expression for the first two contributions in Eq.~(\ref{eq:ind_corr_ener_removedHx}) to the true individual
correlation energy: 
\be\label{eq:simplified_part1_ind_corr_ener}
\begin{split}
&E^\bfw_{\rm c}[n^\bfw]
+\sum_{I>0} (\delta_{IJ} - \ttw_I)\dfrac{\partial
E^\bfw_{\rm c}[n^\bfw]}{\partial\ttw_I}
\\
&=\mathcal{E}^\bfw_{{\rm
c},J}[n^\bfw]-2\sum_{I>0}\sum_{K\geq 0} (\delta_{IJ} - \ttw_I)\ttw_K\mel{\Phi_K^\bfw}{\hat{T}+\hat{W}_{\rm
ee}}{\dfrac{\partial\Phi_K^\bfw}{\partial \ttw_I}}.
\end{split}
\ee
On the other hand, according to Eq.~(\ref{eq:double_weight_KS_ens_dens}), 
\be
\left.\dfrac{\partial E^{\bmxi}_{\rm
c}[n^{\bmxi,\bfw}]}{\partial
\ttw_I}\right|_{\bmxi=\bfw}=
\int d\bfr\dfrac{\delta E^\bfw_{\rm c}[n^\bfw]}{\delta n(\bfr)}\sum_{K\geq
0}\ttw_K\dfrac{\partial n_{\Phi^\bfw_K}(\bfr)}{\partial \ttw_I},
\ee
thus leading to [see Eq.~(\ref{eq:indiv_dens_response})]
\be\label{eq:simplified_part2_ind_corr_ener}
-\sum_{I>0} (\delta_{IJ} - \ttw_I)\left.\dfrac{\partial E^{\bmxi}_{\rm
c}[n^{\bmxi,\bfw}]}{\partial
\ttw_I}\right|_{\bmxi=\bfw}=\int d\bfr\dfrac{\delta E^\bfw_{\rm
c}[n^\bfw]}{\delta
n(\bfr)}\left(n_{\Phi^\bfw_J}(\bfr)-n_{\Psi_J}(\bfr)\right).
\ee
Finally, by combining Eqs.~(\ref{eq:ind_corr_ener_removedHx}), (\ref{eq:simplified_part1_ind_corr_ener}), and (\ref{eq:simplified_part2_ind_corr_ener}), we recover
the expression of Ref.~\cite{fromager2020individual} for the deviation of the true $J$th individual
correlation energy from the component $\mathcal{E}^\bfw_{{\rm
c},J}[n^\bfw]$,
\be
\begin{split}
E^\bfw_{{\rm c},J}[n^\bfw]-\mathcal{E}^\bfw_{{\rm
c},J}[n^\bfw]&=-2\sum_{I>0}\sum_{K\geq 0} (\delta_{IJ} - \ttw_I)\ttw_K\mel{\Phi_K^\bfw}{\hat{T}+\hat{W}_{\rm
ee}}{\dfrac{\partial\Phi_K^\bfw}{\partial \ttw_I}}
\\
&\quad
+\int d\bfr\dfrac{\delta E^\bfw_{\rm
c}[n^\bfw]}{\delta
n(\bfr)}\left(n_{\Phi^\bfw_J}(\bfr)-n_{\Psi_J}(\bfr)\right), 
\end{split}
\ee
thus leading [see Eq.~(\ref{eq:ind_components_ens_corr_ener})] to the
following exact expression for individual correlation
energies:
\be\label{eq:final_ind_corr_ener_exp}
\begin{split}
E^\bfw_{{\rm c},J}[n^\bfw]&=\mel{\Psi_J}{\hat{T}+\hat{W}_{\rm
ee}}{\Psi_J}-\mel{\Phi_J^\bfw}{\hat{T}+\hat{W}_{\rm
ee}}{\Phi_J^\bfw}
\\
&\quad
-2\sum_{I>0}\sum_{K\geq 0} (\delta_{IJ} - \ttw_I)\ttw_K\mel{\Phi_K^\bfw}{\hat{T}+\hat{W}_{\rm
ee}}{\dfrac{\partial\Phi_K^\bfw}{\partial \ttw_I}}
\\
&\quad
+\int d\bfr\dfrac{\delta E^\bfw_{\rm
c}[n^\bfw]}{\delta
n(\bfr)}\left(n_{\Phi^\bfw_J}(\bfr)-n_{\Psi_J}(\bfr)\right).
\end{split}
\ee
The above expression is a key result of Ref.~\cite{fromager2020individual}
which, as we will see, allows us to explore in-principle-exact SD/DD
correlation energy decompositions.\\

Let us now analyze the different contributions on the right-hand side of
Eq.~(\ref{eq:final_ind_corr_ener_exp}). While, on the first line, the
bare $J$th correlation energy component is
recovered, the additional terms on the second and third lines ensure
that the external potential energy is evaluated with the correct true density
(see Eqs.~(\ref{eq:exact_ind_ener_with_corr}) and (\ref{eq:ind_component_not_ind_corr}), and the supplementary
material of Ref.~\cite{fromager2020individual}). Interestingly, in the
summation (in $K$) over all the states that belong to the ensemble [see the second
line of Eq.~(\ref{eq:final_ind_corr_ener_exp})], one may separate the contribution of
the state under consideration (\ie, the $J$th state) from the others,
thus defining an individual SD correlation energy:        
\be\label{eq:def_SD_corr_ener}
\begin{split}
E^{\bfw,{\rm SD}}_{{\rm c},J}[n^\bfw]&=\mel{\Psi_J}{\hat{T}+\hat{W}_{\rm
ee}}{\Psi_J}-\mel{\Phi_J^\bfw}{\hat{T}+\hat{W}_{\rm
ee}}{\Phi_J^\bfw}
\\
&\quad-2\ttw_J\sum_{I>0}(\delta_{IJ} - \ttw_I)\mel{\Phi_J^\bfw}{\hat{T}+\hat{W}_{\rm
ee}}{\dfrac{\partial\Phi_J^\bfw}{\partial \ttw_I}}.
\end{split}
\ee
The above definition, which was denoted $\overline{\rm SD}$ in
Ref.~\cite{fromager2020individual} (the ``overline'' notation is dropped
in the present work, for simplicity), differs substantially from the definition of
Gould and Pittalis~\cite{gould2019density}. In the latter, an additional state-specific
KS wave function, which is expected to reproduce the true individual
density of the state under consideration, is introduced. In this case,
the name ``state-driven'' means that the correlation energy is evaluated
from interacting and non-interacting wave functions which share the {\it
same}
density. Here, no additional KS wave function is introduced, which is
obviously appealing from a computational point of view. One possible
criticism about the definition in Eq.~(\ref{eq:def_SD_corr_ener}) is its
arbitrariness. Indeed, we may opt for a more density-based definition,
in the spirit of what Gould and Pittalis proposed, by introducing, for
example, the following auxiliary wave functions:
\be\label{eq:Gould-Pittalis-like_KS_wfs}
\begin{split}
\overline{\Phi}^\bfw_J&=
\Phi_J^\bfw+\sum_{I>0}\sum_{K\geq 0} \sqrt{\abs{\delta_{IJ}
- \ttw_I}\ttw_K}\left({\rm sgn}(\delta_{IJ}
- \ttw_I)\,\Phi_K^\bfw+\dfrac{\partial\Phi_K^\bfw}{\partial
\ttw_I}\right).
\end{split}
\ee 
Note that, in the ground-state
$\bfw=0$ limit, $\overline{\Phi}^\bfw_0$ reduces to the
conventional KS wave function $\Phi_0^{\bfw=0}$ of KS-DFT.
What might be interesting in the (somehow artificial) construction of
the above individual auxiliary KS states is the possibility it gives
to recover, like in the Gould-Pittalis approach~\cite{gould2019density}, all the KS contributions (to the individual correlation
energy) that appear on the first two lines of
Eq.~(\ref{eq:final_ind_corr_ener_exp}) from a single expectation value, thus generating, on the other
hand, (several) additional terms that should ultimately be removed:  
\be
\begin{split}
&\mel{\overline{\Phi}_J^\bfw}{\hat{T}+\hat{W}_{\rm
ee}}{\overline{\Phi}_J^\bfw}
\\
&
=\mel{\Phi_J^\bfw}{\hat{T}+\hat{W}_{\rm
ee}}{\Phi_J^\bfw}
+2\sum_{I>0}\sum_{K\geq 0} (\delta_{IJ} - \ttw_I)\ttw_K\mel{\Phi_K^\bfw}{\hat{T}+\hat{W}_{\rm
ee}}{\dfrac{\partial\Phi_K^\bfw}{\partial \ttw_I}}
\\
&\quad
+\ldots
\end{split}
\ee
Moreover, according to Eq.~(\ref{eq:indiv_dens_response}), we recover (among other terms) the correct physical density:
\be
\begin{split}
\mel{\overline{\Phi}_J^\bfw}{\hat{n}(\bfr)}{\overline{\Phi}_J^\bfw}&=n_{\Phi^\bfw_J}(\bfr)+\sum_{I>0}\sum_{K\geq
0} (\delta_{IJ} - \ttw_I)\ttw_K\dfrac{\partial
n_{\Phi^\bfw_K}(\bfr)}{\partial \ttw_I}+\ldots
\\
&=n_{\Psi_J}(\bfr)+\ldots
\end{split}
\ee
On that basis, we could argue that the first two lines on the right-hand
side of Eq.~(\ref{eq:final_ind_corr_ener_exp}) should be interpreted as a 
SD correlation energy, while the third line would correspond to the
missing DD 
correlation energy. The issue with such a decomposition is that the  
individual DD correlation energies would then cancel out in the
weighted sum:
\be
\begin{split}
&\sum_{J\geq 0}\ttw_J\int d\bfr\dfrac{\delta E^\bfw_{\rm
c}[n^\bfw]}{\delta
n(\bfr)}\left(n_{\Phi^\bfw_J}(\bfr)-n_{\Psi_J}(\bfr)\right)
\\
&=\int d\bfr\dfrac{\delta E^\bfw_{\rm
c}[n^\bfw]}{\delta
n(\bfr)}\left(n^\bfw(\bfr)-n^\bfw(\bfr)\right)=0,
\end{split}
\ee
which means that the ensemble DD correlation energy would be zero.
As a result, with such an interpretation, the concept of DD correlation
would not be of any help in the development of correlation DFAs for
ensembles. This is of course not what we want~\cite{gould2019density}. In this
respect, the definition in Eq.~(\ref{eq:def_SD_corr_ener}) is much more
appealing. We will stick to this definition from now on. Consequently,
the complementary ensemble DD correlation energy
will read as [see Eqs.~(\ref{eq:decomp_ens_corr_components}), (\ref{eq:ind_components_ens_corr_ener}), and (\ref{eq:def_SD_corr_ener})]
\be\label{eq:DD_ener_full_minus_SD}
E_{\rm c}^{\bfw,{\rm DD}}[n^\bfw]&=&E_{\rm c}^\bfw[n^\bfw]-\sum_{J\geq 0}\ttw_JE^{\bfw,{\rm SD}}_{{\rm c},J}[n^\bfw]
\\
\label{eq:final_DD_ens_corr_ener}
&=&2\sum_{J\geq 0}\ttw^2_J\sum_{I>0}(\delta_{IJ} - \ttw_I)\mel{\Phi_J^\bfw}{\hat{T}+\hat{W}_{\rm
ee}}{\dfrac{\partial\Phi_J^\bfw}{\partial \ttw_I}}.
\ee
Thus, we recover another key result of Ref.~\cite{fromager2020individual}.
As readily seen from Eq.~(\ref{eq:final_DD_ens_corr_ener}), the exact
evaluation of the DD correlation energy only requires computing the static
linear response of the KS wave
functions that belong to the ensemble, which is computationally
affordable.\\

Finally, at a more formal level, we note that the ensemble
DD correlation energy expression of Eq.~(\ref{eq:final_DD_ens_corr_ener}) is related to the individual components 
$f_J^\bfw[n]=\mel{\Phi^\bfw_J[n]}{\hat{T}+\hat{W}_{\rm
ee}}{\Phi^\bfw_J[n]}$ of
the Hx-only approximation to the universal GOK functional [see
Eqs.~(\ref{eq:F_KS_GOK-DFT}), (\ref{eq:explicit_Tsw_func}), and (\ref{eq:def_exact_ens_Hx_func})]
\be\label{eq:ab_initio_Hx_GOK_fun}
f^\bfw[n]:=T^\bfw_{\rm s}[n]+E^\bfw_{\rm
Hx}[n]=\sum_K\ttw_K\mel{\Phi^\bfw_K[n]}{\hat{T}+\hat{W}_{\rm
ee}}{\Phi^\bfw_K[n]},
\ee         
as follows,
\be\label{eq:Ec_DD_Hxonly}
E_{\rm c}^{\bfw,{\rm DD}}[n^\bfw]=\sum_{J\geq 0}\ttw^2_J\sum_{I>0}(\delta_{IJ} - \ttw_I)
\dfrac{\partial f_J^\bfw[n^\bfw]}{\partial \ttw_I}.
\ee
We can even establish a direct connection with the total ensemble
functional $f^\bfw[n]$, by
analogy with Eq.~(\ref{eq:trick_double_weight_Hx}). Indeed, since
\be
f^{\bmxi}
\left[n^{\bmxi,\bfw}\right]=\sum_K\xi_K\mel{\Phi^\bfw_K}{\hat{T}+\hat{W}_{\rm
ee}}{\Phi^\bfw_K}
,
\ee
it comes
\be\label{eq:extraction_ind_X-only_ener}
f^\bfw_J\left[n^{{\bf
w}}\right]=f^\bfw\left[n^{{\bf
w}}\right]
+\sum_{I>0}\left(\delta_{IJ}-\ttw_I\right)
\left.\frac{\partial
f^{\bmxi}
\left[n^{\bmxi,\bfw}\right]
}{\partial \xi_I}
\right|_{\bmxi=\bfw}.
\ee

%

\subsection{Application to the Hubbard dimer}

The importance of DD correlations, which was revealed in
Ref.~\cite{gould2019density}, has been confirmed in
Ref.~\cite{fromager2020individual}, in the weakly asymmetric and stronly
correlated regime of the two-electron Hubbard dimer. We propose in the following to
complete the study of Ref.~\cite{fromager2020individual} by exploring all asymmetry and correlation regimes,
and comparing exact results with that of standard approximations. 

\subsubsection{Exact theory and approximations}

The Hubbard dimer has been introduced in
Sec.~\ref{subsec:insights_SAHF_Hubbard_dimer}. In this simple model
system,
exact (two-electron and singlet) biensemble
density-functional correlation energies $E^\ttw_{\rm c}(n)$ can be evaluated through Lieb
maximizations~\cite{deur2017exact,deur2018exploring} from the following
analytical expressions for the exact potential-functional interacting
energies~\cite{smith2016exact,carrascal2015hubbard,carrascal2016corrigendum}: 
\begin{equation}
E_{I}(\Delta v) =\frac{2U}{3} +\frac{2r}{3}\cos\left(\theta
+\frac{2\pi}{3}(I+1)\right),\hspace{0.2cm} I=0,1,
\label{eqn:8}
\end{equation}
where
\begin{equation}
r =\sqrt{3(4t^{2}+\Delta v^{2}) +U^{2}}
\label{eqn:9}
\end{equation}
and
\begin{equation}
\theta =\frac{1}{3}\arccos{\left[ \frac{9U(\Delta v^{2} \: - \: 2t^{2})
\: - \: U^{3}}{r^{3}}\right]}.
\label{eqn:10}
\end{equation}
Exact ground- and excited-state densities are then obtained from the 
Hellmann--Feynman theorem [see Eq.~(\ref{eq:Hamil_Hubbard_dimer_model})],
\begin{equation}
n_{\Psi_{I}} \: = \: 1 \: - \: \frac{\partial E_{I}(\Delta v)}{\partial
\Delta v},
\label{eqn:11}
\end{equation}
and the cubic polynomial equation that the energies fulfill (see the
Appendix of Ref.~\cite{deur2017exact}). The resulting biensemble density
reads as $n^\ttw=(1-\ttw)n_{\Psi_0}+\ttw\, n_{\Psi_1}$. 
The Hx-only GOK functional introduced in
Eq.~(\ref{eq:ab_initio_Hx_GOK_fun}) can be expressed analytically as
follows~\cite{deur2017exact},   
\be
f^\xi(n)
&=&T^\xi_{\rm s}(n)+E^\xi_{\rm Hx}(n)
\nonumber
\\
&=&-2t\sqrt{(1-\xi)^2-(1-n)^2}
+\dfrac{U}{2}\left[1+\xi-\dfrac{(3\xi-1)(1-n)^2}{(1-\xi)^2}\right],
\ee
so that, as shown in Appendix~\ref{app:DD_corr_ener_derivation},
the exact DD ensemble correlation energy reads explicitly as
\be\label{eq:final_DD_ens_corr_ener_dimer}
\begin{split}
E_{\rm c}^{\ttw,{\rm
DD}}(n^\ttw)&=-\ttw(n^\ttw-1)(n_{\Psi_1}-1)
\\
&\quad\times
\left[
\dfrac{2t}{\sqrt{(1-\ttw)^2-(1-n^\ttw)^2}}+\dfrac{U(1+\ttw)}{(1-\ttw)^2}
\right].
\end{split}
\ee
Since the KS excited-state density is always equal to 1 in this
model~\cite{deur2017exact}, the prefactor $(n_{\Psi_1}-1)$ matches the deviation in density of
the true physical excited state from the KS one:
\be
n_{\Psi_1}-1=n_{\Psi_1}-n_{\Phi^\ttw_1}.
\ee
Then it becomes clear that $E_{\rm c}^{\ttw,{\rm
DD}}(n^\ttw)$ is a DD correlation energy. We
also see from the expression of Eq.~(\ref{eq:final_DD_ens_corr_ener_dimer}) that, in the regular ground-state DFT limit
($\ttw=0$), this type of correlation disappears.\\ 

In the following we test two common DFAs: A
(weight-independent) {\it ground-state} density-functional description
of the {\it ensemble correlation} energy
(GS-ec)~\cite{senjean2015linear,deur2017exact}, 
\be
\label{eq:ens_corr_GS-ec_approx}
E_{\rm
c}^\ttw(n^\ttw)\overset{\rm GS-ec}{\approx}E_{\rm
c}(n^\ttw), 
\ee
where $E_{\rm
c}(n)=E^{\ttw=0}_{\rm
c}(n)$,
and the 
GS-ic approximation introduced in Sec.~\ref{subsec:GS-ic_and_eLDA}
which, in the present case, gives
\be
\label{eq:ens_corr_GS-ic_approx}
\begin{split}
E_{\rm
c}^\ttw(n^\ttw)&\overset{\rm GS-ic}{\approx} (1-\ttw)E_{\rm
c}(n_{\Phi^\ttw_0})+\ttw E_{\rm 
c}(n_{\Phi^\ttw_1})
\\
&=(1-\ttw)E_{\rm
c}(n_{\Phi^\ttw_0})+\ttw E_{\rm
c}(n=1).
\end{split}
\ee
Note that the KS ground-state density $n_{\Phi^\ttw_0}$ fulfills the constraint
$(1-\ttw)n_{\Phi_{0}^{\ttw}}+\ttw n_{\Phi_{1}^{\ttw}}=n^\ttw$, thus
leading to
\be\label{eq:weight_dep_GS_KS_HD}
n_{\Phi_{0}^{\ttw}}=\dfrac{n^\ttw-\ttw}{(1-\ttw)}=n_{\Psi_0}+\dfrac{\ttw
(n_{\Psi_1}-1)}{(1-\ttw)},
\ee
where we readily see that, in general, $n_{\Phi_{0}^{\ttw}}\neq
n_{\Psi_0}$. In the following, the local potential will be fixed. It
is then analogous to the external potential of {\it ab initio} calculations,
hence the notation $\Delta v=\Delta v_{\rm ext}$.

\subsubsection{Results and discussion}

Let us first discuss the strictly symmetric ($\Delta v_{\rm ext}=0$) dimer in which
simple analytical expressions can be derived for both exact and
approximate ensemble correlation energies. In this special case, ground- and
excited-state densities are equal to 1, in both KS and physical 
interacting systems. Consequently, the ensemble DD correlation energy
vanishes [see Eq.~(\ref{eq:final_DD_ens_corr_ener_dimer})]. Total and SD ensemble correlation energies are equal and vary linearly with the ensemble weight~\cite{deur2018exploring} as
\be
E_{\rm
c}^\ttw(n^\ttw=1)=2t(1-\ttw)\left(1-\sqrt{1+\dfrac{U^2}{16t^2}}\right)=(1-\ttw)E_{\rm
c}(n=1),
\ee
with the positive slope $-E_{\rm
c}(n=1)$. As readily seen, the excited state exhibits no correlation
effects in this density regime. Turning to the approximations, GS-ic erroneously assigns a (ground-state) correlation energy to the excited state
[see Eq.~(\ref{eq:ens_corr_GS-ic_approx})], thus leading to a total 
ensemble correlation energy that is wrong and equal to $E_{\rm
c}(n=1)$, like in GS-ec [see Eq.~(\ref{eq:ens_corr_GS-ec_approx})]. In conclusion, in the symmetric case, both GS-ic and GS-ec approximations completely miss the weight dependence of the ensemble correlation energy.
\\ 

\begin{figure}[!htb]
\includegraphics[scale=0.4]{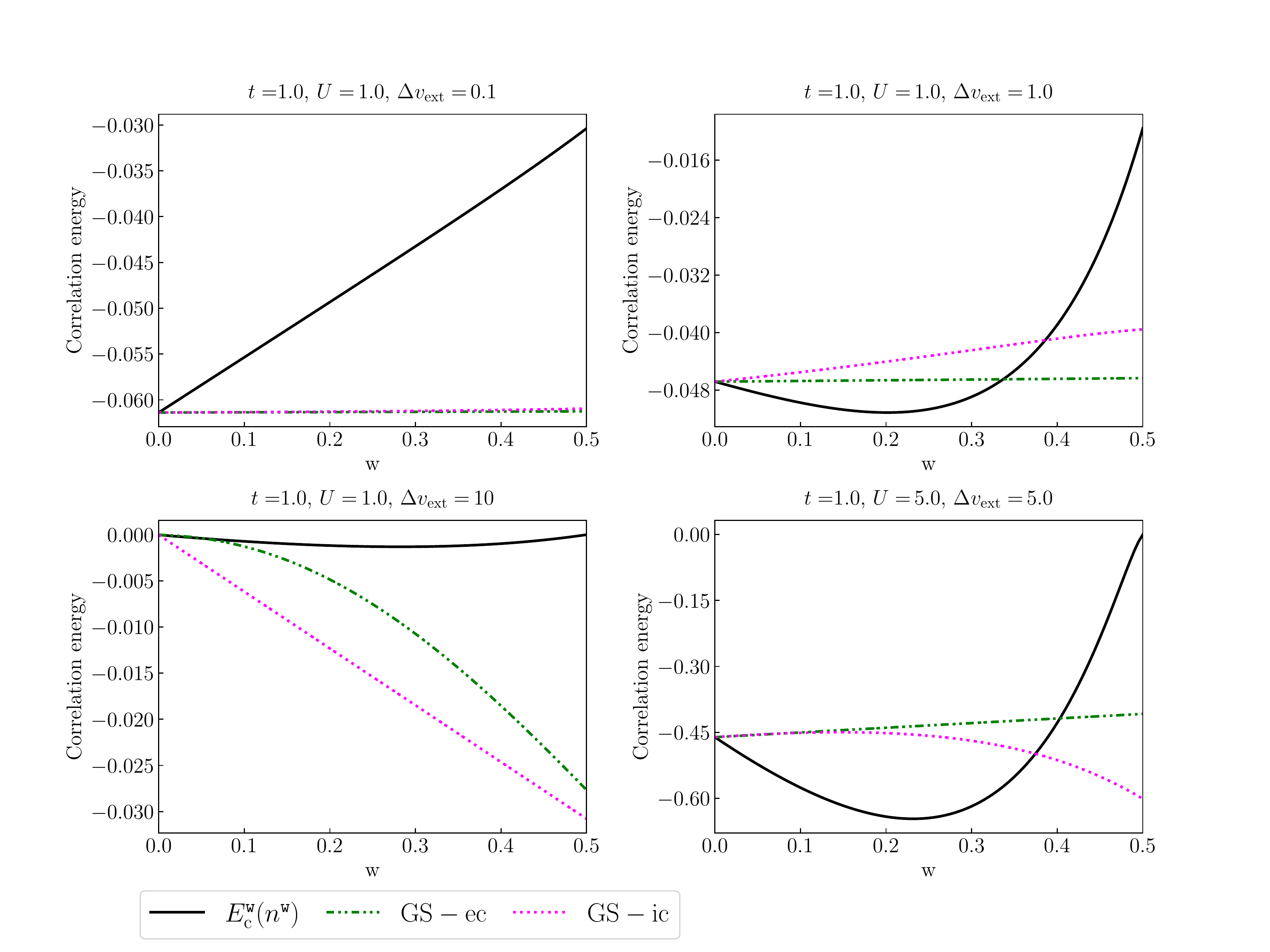}
\caption{Exact (solid black lines) and approximate ensemble correlation energies plotted as
functions of the biensemble weight for the Hubbard dimer in various
correlation and asymmetry regimes. See text for further details.}
\label{fig:ens_corr_ener_approx}
\end{figure}

We now discuss the performance of GS-ec and GS-ic in the asymmetric
dimer. Results are shown in Fig.~\ref{fig:ens_corr_ener_approx}. The
features described in the symmetric case are preserved in the weakly
asymmetric regime (see the top left panel of
Fig.~\ref{fig:ens_corr_ener_approx}). When the asymmetry is more
pronounced, both exact and approximate ensemble correlation energies
exhibit curvature.    
By construction, these energies all reduce to the same (ground-state) correlation energy when 
$\ttw=0$. 
They differ 
substantially by their slope in the ground-state limit
($\ttw=0$). Further insight into GS-ic, for example, is obtained from the following
analytical expression, 
\be\label{eq:slope_GS-ic}
\begin{split}
\left.\dfrac{\partial E_{\rm
c}^\ttw(n^\ttw)}{\partial \ttw }\right|_{\ttw=0}\overset{\rm GS-ic}{\approx}
&E_{\rm
c}(n=1)
-E_{\rm
c}(n=n_{\Psi_0})
\\
&+\left(n_{\Psi_1}-1\right)\left.\dfrac{\partial E_{\rm c}(n)}{\partial n}\right|_{n=n_{\Psi_0}},
\end{split}
\ee
where $E_{\rm
c}(n=1)
-E_{\rm
c}(n=n_{\Psi_0})\leq 0$, as readily seen from Fig.~4 of
Ref.~\cite{deur2018exploring}.
Interestingly, in the strongly asymmetric $\Delta v_{\rm ext}/U\gg 1$ regime,
the true ground- ($n_{\Psi_0}$) and excited-state ($n_{\Psi_1}$)
densities tend to 2 and 1, respectively (see Fig. 1 of
Ref.~\cite{deur2017exact}). This is the situation where the slope in
weight expressed in
Eq.~(\ref{eq:slope_GS-ic}) reaches its maximum (in absolute value), thus
inducing a large deviation from the exact slope, as shown in the bottom
left panel of Fig.~\ref{fig:ens_corr_ener_approx}. At the GS-ec level of approximation, the
situation is less critical, at least for small weight values. As
readily seen from the following expression [see Eq.~(\ref{eq:ens_corr_GS-ec_approx})],
\be\label{eq:slope_GS-ec}
\begin{split}
\left.\dfrac{\partial E_{\rm
c}^\ttw(n^\ttw)}{\partial \ttw }\right|_{\ttw=0}\overset{\rm GS-ec}{\approx}
\left(n_{\Psi_1}-n_{\Psi_0}\right)\left.\dfrac{\partial E_{\rm c}(n)}{\partial n}\right|_{n=n_{\Psi_0}}
,
\end{split}
\ee
when $\Delta v_{\rm ext}\sim U$, the slope (at $\ttw=0$) is relatively
small since
$n_{\Psi_1}\sim n_{\Psi_0}$ (see Fig. 1 of
Ref.~\cite{deur2017exact}). This is in agreement with the right panels
of Fig.~\ref{fig:ens_corr_ener_approx}. Note that, in this regime, GS-ic
can exhibit positive slopes (see the top right panel of
Fig.~\ref{fig:ens_corr_ener_approx}). In this case, the density
derivative contribution [second line of
Eq.~(\ref{eq:slope_GS-ic})], which is positive~\cite{deur2017exact,deur2018exploring}, is not negligible anymore
and it (more than) compensates the negative correlation energy difference [first line of Eq.~(\ref{eq:slope_GS-ic})]. When
the asymmetry of the dimer is more pronounced (\ie, $\Delta v_{\rm ext}\gg U$),
the GS-ec slope (in weight)   
remains negligible, as shown in the bottom left panel of
Fig.~\ref{fig:ens_corr_ener_approx}. Indeed, in this case, $n_{\Psi_0}$
tends to 2. Moreover, since the ground-state correlation functional expands as follows in the
weakly and strongly correlated regimes~\cite{deur2018exploring}, 
\be
E_{\rm c}(n)=E_{\rm
c}^{\ttw=0}(n)\overset{U/t\ll
1}{\approx}-\dfrac{U^2\left(1-(1-n)^2\right)^{5/2}}{16t},
\ee
and
\be
\label{eq:GS_Ec_fun_largeU}
E_{\rm
c}(n)\overset{U/t\gg 1}{\approx}U\left[\abs{n-1}-\dfrac{1}{2}(1+(n-1)^2)\right],
\ee
respectively, we immediately see that $\partial E_{\rm c}(n)/\partial
n\approx 0$ when $n$ approaches 2, whether $U/t$ is large or small. Note
finally that, as already mentioned, in regimes where the asymmetry is weaker than the
correlation, \ie, $\Delta v_{\rm ext}/t<t/U<1$ (see the
top left panel of Fig.~\ref{fig:ens_corr_ener_approx}), the slopes
obtained at $\ttw=0$ with GS-ec and
GS-ic are identical and relatively weak. This can now be understood from
the expressions in Eqs.~(\ref{eq:slope_GS-ic}) and
(\ref{eq:slope_GS-ec}), and the fact that
$\left.\partial E_{\rm c}(n)/\partial
n\right|_{n=1}=0$~\cite{deur2018exploring}, knowing that 
$n_{\Psi_0}\approx 1$~\cite{deur2017exact} in this case. 
\\  

We see in Fig.~\ref{fig:ens_corr_ener_approx} that the overall weight
dependence of the exact ensemble
correlation energy differs substantially from that of the GS-ec and GS-ic approximations. It is again instructive to look at the slope
at $\ttw=0$. It can be expressed exactly as follows,
\be
\begin{split}
\left.\dfrac{\partial E_{\rm
c}^\ttw(n^\ttw)}{\partial \ttw }\right|_{\ttw=0}=
\left(n_{\Psi_1}-n_{\Psi_0}\right)\left.\dfrac{\partial E_{\rm c}(n)}{\partial n}\right|_{n=n_{\Psi_0}}
+\left.\dfrac{\partial E^\ttw_{\rm c}(n_{\Psi_0})}{\partial
\ttw}\right|_{\ttw=0},
\end{split}
\ee
where we readily see from Eq.~(\ref{eq:slope_GS-ec}) that GS-ec neglects the derivative in weight 
of the ensemble correlation  
density functional. As highlighted in Eq.~(\ref{eq:Ek_GOK-DFT_w0}) [see
also Sec.~ \ref{subsec:comparison_Ncentered} for a more detailed discussion in the context of
charged excitations], the latter
contribution is connected to the derivative discontinuity that the xc
potential exhibits when an excited state is incorporated into the ensemble. Since, in the weakly
correlated regime~\cite{deur2018exploring}, 
\be
\begin{split}
E_{\rm
c}^\ttw(n)&\overset{U/t\ll 1}{\approx}-\dfrac{U^2\left((1-\ttw)^2-(1-n)^2\right)^{3/2}}{16t(1-\ttw)^2}
\\
&\quad\quad\times\left[1+\dfrac{(1-n)^2}{(1-\ttw)^2}\left(3-\dfrac{4(1-3\ttw)^2}{(1-\ttw)^2}\right)\right],  
\end{split}
\ee
it comes
\be\label{eq:exact_slope_weakly_corr}
\left.\dfrac{\partial E^\ttw_{\rm c}(n_{\Psi_0})}{\partial
\ttw}\right|_{\ttw=0}\overset{U/t\ll 1}{\approx}
\dfrac{U^2}{16t}\left(n_{\Psi_0}(2-n_{\Psi_0})\right)^{3/2}\left(1-12(n_{\Psi_0}-1)^2\right).
\ee
Therefore, as long as the ground state does not deviate too much from
the symmetric $n_{\Psi_0}=1$ density profile, which is the case when $\Delta
v_{\rm ext}\ll U$, the exact slope is not negligible, and it is
positive. This is in agreement with the top left panel of Fig.~\ref{fig:ens_corr_ener_approx}.  
Interestingly, in this density regime, this feature is preserved when
the strength of electron correlation increases (not shown). Indeed, in this case, the ensemble correlation functional
reads as~\cite{deur2018exploring} 
\be\label{eq:Ec_fun_Hubbard_suppmat}
E^\ttw_{\rm
c}(n)&\overset{U/t\gg 1,\,\abs{n-1}\leq\ttw }{\approx}&-\dfrac{U}{2}\left[(1-\ttw)-\dfrac{(3\ttw-1)(n-1)^2}{(1-\ttw)^2}\right],
\\
&\overset{n=n_{\Psi_0}}{\approx}&\dfrac{U}{2}(\ttw-1),
\ee
thus leading to $\partial E^\ttw_{\rm
c}(n_{\Psi_0})/\partial\ttw\approx U/2$.
When the dimer is strongly asymmetric, the ground-state density approaches 2 and, in this
case~\cite{deur2018exploring}, 
\be
\begin{split}
E^\ttw_{\rm
c}(n)\overset{U/t\gg 1,\,\ttw\leq\abs{n-1}\leq1-\ttw }{\approx}&
U\abs{n-1}
\\
&
-\dfrac{U}{2}\left[(1+\ttw)-\dfrac{(3\ttw-1)(n-1)^2}{(1-\ttw)^2}\right],
\end{split}
\ee
so that
\be\label{eq:exact_slope_strong_corr}
\left.\dfrac{\partial E^\ttw_{\rm
c}(n_{\Psi_0})}{\partial \ttw}
\right|_{\ttw=0}
&\overset{U/t\gg 1}{\approx}&-\dfrac{U}{2}\left(1-(n_{\Psi_0}-1)^2\right),
\ee
thus leading to $\left.\partial E^\ttw_{\rm
c}(n_{\Psi_0})/\partial \ttw\right|_{\ttw=0}\approx 0$. As readily seen
from Eq.~(\ref{eq:exact_slope_weakly_corr}), the same result is obtained
in the weakly correlated regime. This is in complete agreement with the
bottom left panel of Fig.~\ref{fig:ens_corr_ener_approx}. It also explains why the deviation of
GS-ec from the exact result 
drastically reduces when $\Delta
v_{\rm ext}$ increases for a fixed interaction strength $U$ and
relatively small weight values. Finally, in the particular case where $\Delta
v_{\rm ext}=U$, the computed ground-state densities equal 
$n_{\Psi_0}\approx1.30$ and $n_{\Psi_0}\approx1.46$ in
the moderately $U/t=1$ and strongly $U/t=5$ correlated regimes,
respectively. As expected from Eqs.~(\ref{eq:exact_slope_weakly_corr})
and (\ref{eq:exact_slope_strong_corr}), the exact slope will be
substantial and negative, which agrees with the right panels of
Fig.~\ref{fig:ens_corr_ener_approx}.   
\\
 
\begin{figure}[!htb]
\includegraphics[scale=0.4]{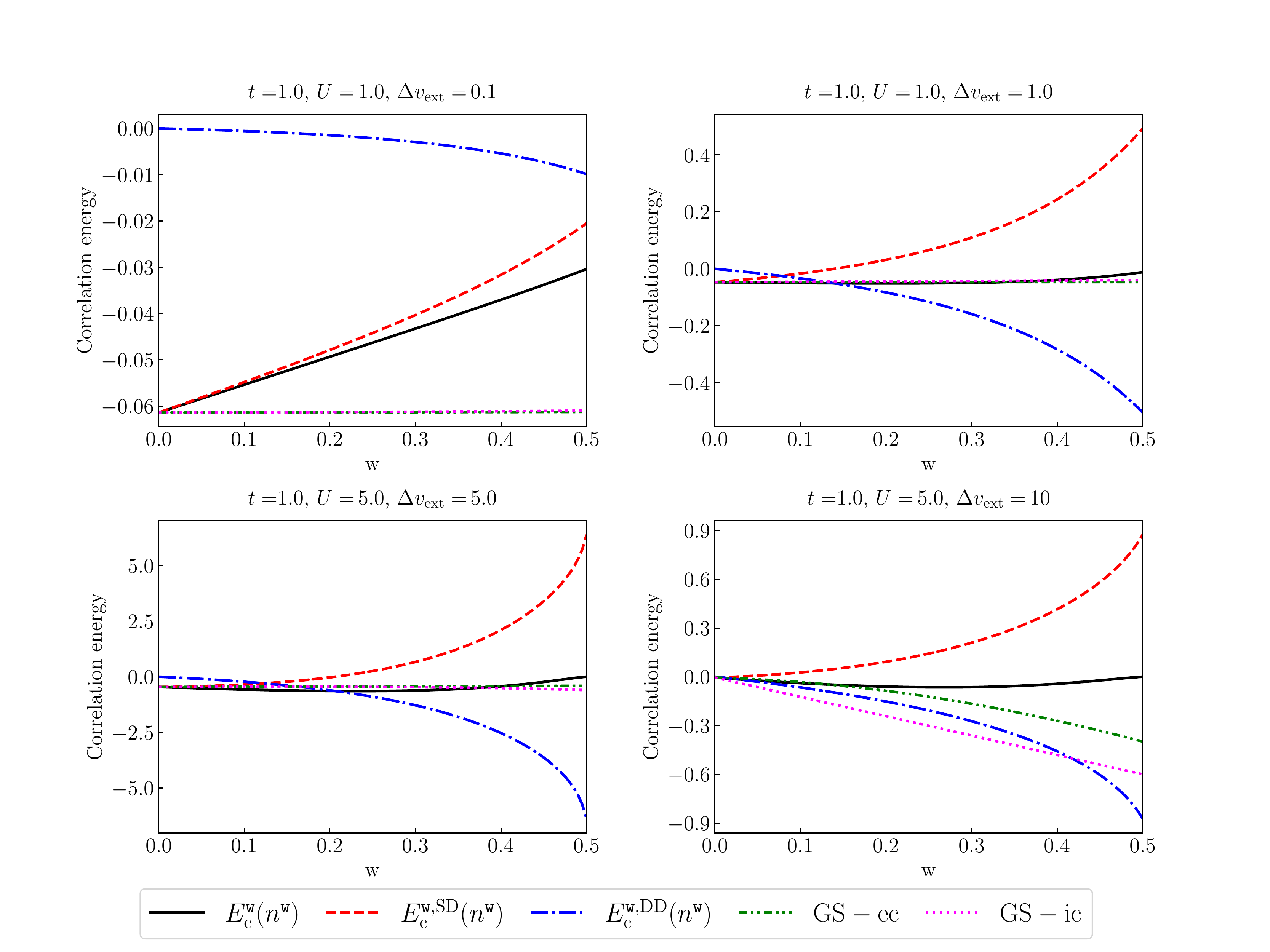}
\caption{Exact SD/DD decomposition of the ensemble correlation energy
plotted as a function of the biensemble weight $\ttw$ in various
asymmetry and correlation regimes. Comparison is made with the 
approximate GS-ec and GS-ic ensemble correlation energies, for analysis
purposes. See text for further
details.}
\label{fig:ens_corr_ener_approx_AND_SD-DD}
\end{figure}

We now focus on the exact SD/DD decomposition of the ensemble correlation
energy. Results are shown in
Fig.~\ref{fig:ens_corr_ener_approx_AND_SD-DD} for various correlation
and asymmetry regimes. In the ground-state limit, the slope of the DD ensemble
correlation energy reads as [see Eq.~(\ref{eq:final_DD_ens_corr_ener_dimer})]
\be
\begin{split}
\left.\dfrac{\partial E_{\rm
c}^{\ttw,{\rm DD}}(n^\ttw)}{\partial \ttw }\right|_{\ttw=0}
&=
-(n_{\Psi_0}-1)(n_{\Psi_1}-1)
\left[
\dfrac{2t}{\sqrt{1-(1-n_{\Psi_0})^2}}+U
\right].
\end{split}
\ee   
Interestingly, when $\Delta
v_{\rm ext}\approx U$ [$n_{\Psi_0}\approx n_{\Psi_1}$ in this case], the
slope is nonzero (and negative), whether the dimer is strongly
correlated or not,  
as seen from the top right and bottom left panels of Figs.~\ref{fig:ens_corr_ener_approx_AND_SD-DD}. 
In the strongly correlated regime, the DD
ensemble correlation energy essentially varies in $\ttw$ as [see Eq.~(\ref{eq:final_DD_ens_corr_ener_dimer})] 
\be\label{eq:DD_corr_SC-regime}
E_{\rm c}^{\ttw,{\rm
DD}}(n^\ttw)\overset{U/t\gg 1}{\approx}
-U(n_{\Psi_1}-1)(n_{\Psi_0}-1+\ttw(n_{\Psi_1}-n_{\Psi_0}))\dfrac{\ttw(1+\ttw)}{(1-\ttw)^2},
\ee
which means that, when approaching the equiensemble $\ttw=1/2$ case, 
it systematically decreases with the ensemble weight (because of the
term $(1-\ttw)^2$ in the denominator),
unlike the total ensemble correlation energy [see
Eq.~(\ref{eq:Ec_fun_Hubbard_suppmat})]. As long as the dimer remains
close to symmetric, which requires that $\Delta
v_{\rm ext}$ reduces as $U$ increases (see Fig.~1
of Ref.~\cite{deur2017exact}), the numerator in
Eq.~(\ref{eq:DD_corr_SC-regime}) will be small enough such that DD
correlations are at most equal to the total correlation energy. This
feature is actually observed in the moderately correlated $U/t=1$ regime
(see the top left panel
of Fig.~\ref{fig:ens_corr_ener_approx_AND_SD-DD}). However, in asymmetric
and strongly correlated regimes where $0<\Delta
v_{\rm ext}\ll U$ ($n_{\Psi_0}\approx 1$ and $n_{\Psi_1}>1$ in this
case) or $\Delta
v_{\rm ext}\approx U$ (\ie, $n_{\Psi_0}\approx n_{\Psi_1}\approx 1.5$)~\cite{deur2017exact},
the numerator is not negligible anymore and, consequently, the DD
ensemble correlation energy is significantly lower than the total one
(see the bottom left panel of
Fig.~\ref{fig:ens_corr_ener_approx_AND_SD-DD}; see also Ref.~\cite{fromager2020individual}).  
In such cases, the complementary SD ensemble correlation energy can be
large and
positive. This may look
unphysical at
first sight but, if we return to the definition of
Eq.~(\ref{eq:def_SD_corr_ener}), we see that the individual SD correlation
energies are not guaranteed to be negative. The reason is that, unlike the total ensemble correlation
energy, they are {\it not} evaluated variationally. Note finally that, when $\Delta
v_{\rm ext}>U\gg t$ ($n_{\Psi_0}\approx 2$ and $n_{\Psi_1}\approx 1$ in this
case),
the numerator in Eq.~(\ref{eq:DD_corr_SC-regime}) will be relatively small, because of the $(n_{\Psi_1}-1)$ prefactor, thus reducing the energy difference
between total and DD correlations (see the bottom right panel of Fig.~\ref{fig:ens_corr_ener_approx_AND_SD-DD}).\\   

In summary, with the present SD/DD decomposition [see Eqs.~(\ref{eq:DD_ener_full_minus_SD}) and
(\ref{eq:final_DD_ens_corr_ener})], both SD and DD correlation
energies become relatively large (when compared to the total ensemble correlation
energy), especially in the commonly used
equiensemble case, and they mostly compensate when the Hubbard
dimer has a pronounced asymmetry. This is clearly not a favorable situation for
the development of DFAs, which was the initial motivation for
introducing the SD/DD
decomposition~\cite{gould2019density,fromager2020individual}. 
The latter should definitely be implemented for
atoms and diatomics, for example, in order to get further insight. In
the case of stretched diatomics, the present study of 
the Hubbard dimer might be enlightening~\cite{fromager2015exact}. 
We should also stress that, in the asymmetric $\Delta
v_{\rm ext}=U$ case, standard GS-ic and GS-ec approximations give
ensemble correlation energies that are of the same order of magnitude as
the exact one, unlike the SD and DD correlation energies. 
As briefly mentioned in Sec.~\ref{subsec:GS-ic_and_eLDA},
exploring alternative SD/DD decompositions that rely explicitly on
GS-ic, which is maybe a better starting point, would be relevant
in this respect. Work is currently in progress in this direction. 




\section{Conclusions and perspectives}
\label{sec:conclu}

Despite the success of time-dependent post-DFT approaches for the
description of charged and neutral electronic excitations, various 
limitations (in terms of accuracy, computational cost, or physics) have
motivated in recent years the development of time-independent
alternatives. In the present review, we focused on GOK-DFT and
$N$-centered eDFT, which are two flavors of eDFT for neutral and
charged excitations, respectively. Their computational cost is
essentially that of a standard KS-DFT calculation, because they both
rely on self-consistent one-electron KS equations [see
Eqs.~(\ref{eq:ensemble_KS_eq}) and (\ref{eq:ensemble_KS_eq_xi})]. A
major difference though is that, in eDFT, the Hxc density functional is
ensemble weight-dependent.
This weight dependence is central in eDFT. It allows for the
in-principle-exact extraction, from the KS ensemble, of individual
(ground- and excited-state) energy
levels [see Eqs.~(\ref{eq:Ek_GOK-DFT}), (\ref{eq:ind_ener_from_N-eDFT_m}) and (\ref{eq:ind_ener_from_N-eDFT_p})]
and densities [Eq.~(\ref{eq:indiv_dens_response})]. 
We have also shown that the infamous derivative discontinuity problem
that must be addressed when computing fundamental (or optical) gaps [see
Eqs.~(\ref{eq:dEwdw_GOK-DFT}),(\ref{eq:fundgap_Ncentered_bmxi}) and
(\ref{eq:diff_int_pot_xiplus_zero-plus_zero})] can be bypassed, in
principle exactly, {\it via} a relocation of the derivative
discontinuity away from the system [see Eq.~(\ref{eq:no_DD_anymore})]
and a proper modeling of the ensemble weight
dependence in the xc density functional. Recent progress in the design
of weight-dependent xc DFAs have been reviewed. The pros and cons of
using an (orbital-dependent) ensemble density matrix functional exchange energy or
state-averaging individual exact exchange energies have been discussed in
detail. We reveal in passing that, in the latter case, severe (solvable
though) $v$-representability issues can occur when electron correlation
becomes strong. Turning to the design of DFAs for ensemble correlation
energies, state-of-the-art strategies have been discussed, in particular the
combination of finite and infinite uniform electron gas models as well
as the recycling of standard (ground-state) correlation DFAs through 
state-averaging. In the latter case, further improvements may emerge
from the concept of density-driven
correlation, which does not exist in ground-state KS-DFT. How to define
mathematically the corresponding correlation energy is an open question
to which we provided a tentative answer [see Eqs.~(\ref{eq:final_DD_ens_corr_ener}) and
(\ref{eq:Ec_DD_Hxonly})]. Test calculations on the Hubbard dimer reveal
how difficult it is to have a definition that is both rigorous and
useful for the development of approximations. Work
is currently in progress in other (briefly discussed) directions. Even
though it was not mentioned explicitly in the review, we
would like to stress that current formulations of eDFT do not give a
direct access to exact response properties such as oscillator strengths, Dyson
orbitals, or non-adiabatic couplings. Extending G\"{o}rling--Levy perturbation theory~\cite{gorling1994exact,IJQC95_Goerling-Levy_PT_density_constraints,JCP02_Ivanov_GL-PT_density_constraint} to
ensembles might be enlightening in this respect. We recently became aware of such
an extension~\cite{yang2021second} for the computation of excitation energies within the DEC
scheme~\cite{yang2017direct,sagredo2018can}, which is an important first step.
Nevertheless, a general quasi-degenerate density-functional
perturbation theory based on ensembles, where individual energy levels and properties can
be evaluated, is still highly desirable. Work is currently in progress
in this direction.\\

In conclusion, we have summarized in the present review recent efforts of a
growing community to put eDFT to the front of the scene. We 
highlighted several formal and practical aspects of the theory that should
be investigated further in the near future in order to turn eDFT into a
reliable and low-cost computational method for excited states.

\begin{acknowledgements}

E.F. would like to thank M. Levy, A. Savin, P.-F. Loos, T. Gould, M.~J.~P. Hodgson, and J.
Wetherell for fruitful discussions as well as LabEx CSC
(grant number: ANR-10-LABX-0026-CSC) for funding. 
E.F. is also grateful to Trygve Helgaker for his introductory lectures
on convex
analysis and DFT for fractional electron numbers. The authors also thank ANR
(CoLab project, grant number: ANR-19-CE07-0024-02) for funding. 
\end{acknowledgements}

%
\section*{Conflict of interest}

The authors declare that they have no conflict of interest.

\clearpage


\begin{appendices}
\appendix
\numberwithin{equation}{section}
\setcounter{equation}{0}


\section{Asymptotic behavior of the xc potential}\label{app:v_xc_infty}

Let us consider the simpler one-dimensional (1D) case in which the
KS-PPLB
equations read as
\be
-\dfrac{1}{2}\dfrac{d^2 \varphi^\alpha_i(x)}{dx^2}+\left(v_{\rm
ext}(x)+v^\alpha_{\rm
Hxc}(x)\right)\varphi^\alpha_i(x)=\varepsilon^\alpha_i\varphi^\alpha_i(x),
\ee 
thus leading to
\be
\dfrac{d^2 \varphi^\alpha_i(x)}{d
x^2}\underset{\abs{x}\rightarrow
+\infty}{=}-2\left(\varepsilon^\alpha_i-v^\alpha_{\rm
xc}(\infty)\right)\varphi^\alpha_i(x),
\ee
where we used the limits $v_{\rm
ext}(\infty)=v^\alpha_{\rm H}(\infty)=0$. Note that
$\abs{\varphi^\alpha_i(x)}$
is expected to decay as $\abs{x}\rightarrow
+\infty$, which implies $-2\left(\varepsilon^\alpha_i-v^\alpha_{\rm
xc}(\infty)\right)>0$. Therefore, 
\be
\varphi^\alpha_i(x)\underset{\abs{x}\rightarrow
+\infty}{\sim}e^{-\sqrt{-2(\varepsilon^\alpha_i-v^\alpha_{\rm
xc}(\infty))}\abs{x}},
\ee
and
\be\label{eq:KS_dens_1D_far_away}
n_{\hat{\gamma}_{\rm KS}^\alpha}(x)\underset{\abs{x}\rightarrow
+\infty}{\sim}
\alpha \abs{\varphi^\alpha_N(x)}^2
\sim\alpha\,e^{-2\sqrt{-2(\varepsilon^\alpha_N-v^\alpha_{\rm
xc}(\infty))}\abs{x}}.
\ee
In the true interacting system, the $N$-electron ground-state wave
function $\Psi^N_0$ fulfills 
\be
\begin{split}
&\left[\sum^N_{i=1}\left(-\dfrac{1}{2}\dfrac{\partial^2}{\partial x_i^2}+v_{\rm
ext}(x_i)\right)+\sum^N_{1\leq i<j}w_{\rm
ee}(\abs{x_i-x_j})\right]\Psi^N_0(x_1,\ldots,x_N)
\\
&=E^N_0\Psi^N_0(x_1,\ldots,x_N),
\end{split}
\ee   
where $w_{\rm ee}(\abs{x_i-x_j})$ is a well-behaved two-electron
repulsion energy in 1D. Let us consider the situation where $\abs{x_1}\rightarrow +\infty$ while
$x_2,\ldots,x_N$ remain in the region of the system, which corresponds
to an ionization process in the ground state. Since $w_{\rm
ee}(\abs{x_1-x_j})\rightarrow 0$, the (to-be-antisymmetrized) wave function and its density can be rewritten as
\be
\Psi^N_0(x_1,\ldots,x_N)\underset{\abs{x_1}\rightarrow +\infty}\sim
\varphi^{[N]}(x_1)\,\Psi^{N-1}_0(x_2,\ldots,x_N) 
\ee
and
\be\label{eq:dens_N_far_away_1D}
n_{\Psi^N_0}(x_1)\underset{\abs{x_1}\rightarrow +\infty}\sim \abs{\varphi^{[N]}(x_1)}^2,
\ee
respectively,
where
\be
\dfrac{d^2\varphi^{[N]}(x_1)}{d
x_1^2}\underset{\abs{x_1}\rightarrow +\infty}{\sim}-2\left(E^N_0-E^{N-1}_0\right)\varphi^{[N]}(x_1)=2I_0^N\varphi^{[N]}(x_1),
\ee
thus leading to the explicit expression
\be\label{eq:single_orb_far_away}
\varphi^{[N]}(x)\underset{\abs{x}\rightarrow +\infty}\sim
e^{-\sqrt{2I_0^N}\abs{x}}.
\ee
From the exact mapping of the ensemble PPLB density onto the KS system, we deduce from Eqs.~(\ref{eq:dens_N_far_away_1D}) and (\ref{eq:single_orb_far_away}) that 
\be\label{eq:KS_map_dens_PPLB_asymptotic}
\begin{split}
n_{\hat{\gamma}_{\rm KS}^\alpha}(x)&\underset{\abs{x}\rightarrow +\infty}\sim (1-\alpha)e^{-2\sqrt{2I_0^{N-1}}\abs{x}}+\alpha\,e^{-2\sqrt{2I_0^N}\abs{x}}\sim \alpha\,e^{-2\sqrt{2I_0^N}\abs{x}},
\end{split}
\ee
where we assumed that $E^{N-1}_g=I_0^{N-1}-I_0^{N}>0$. Thus, we conlude from Eq.~(\ref{eq:KS_dens_1D_far_away})
that
\be
I_0^N=-(\varepsilon^\alpha_N-v^\alpha_{\rm
xc}(\infty)).
\ee
Any constant shift in the xc potential $v^\alpha_{\rm
xc}(\bfr)$ does not affect the above expression. Since, according to
Janak's theorem, $I_0^N=-\varepsilon^\alpha_N$, the constant is imposed
in PPLB and 
\be
v^\alpha_{\rm
xc}(\infty)=0.
\ee

We now turn to the left and right formulations of $N$-centered eDFT. We
recall the shorthand notations $(\xi_-,0)\overset{notation}{\equiv}
\xi_-$ and $(0,\xi_+)\overset{notation}{\equiv} \xi_+$. When $\xi_+>0$, the right
$N$-centered ensemble density, which is mapped onto a non-interacting KS
ensemble, has the following asymptotic behavior [we just need to
substitute $N+1$ for $N$ in Eqs.~(\ref{eq:KS_dens_1D_far_away}),
(\ref{eq:dens_N_far_away_1D}), and (\ref{eq:single_orb_far_away})],
\be
\label{eq:asympt_behavior_Nc-dens_xi_plus_true}
n^{\xi_+}(x)&\underset{\abs{x}\rightarrow+\infty}{\sim}& \xi_+\,e^{-2\sqrt{2I_0^{N+1}}\abs{x}}
\\
&\sim&\xi_+\,e^{-2\sqrt{-2(\varepsilon^{\xi_+}_{N+1}-v^{\xi_+}_{\rm
xc}(\infty))}\abs{x}}.
\ee
Similarly, for $\xi_-\geq 0$, we have        
\be
n^{\xi_-}(x)&\underset{\abs{x}\rightarrow+\infty}{\sim}& \left(1-\dfrac{(N-1)\xi_-}{N}\right)\,e^{-2\sqrt{2I_0^{N}}\abs{x}}
\\
&\sim&\left(1-\dfrac{(N-1)\xi_-}{N}\right)\,e^{-2\sqrt{-2(\varepsilon^{\xi_-}_{N}-v^{\xi_-}_{\rm
xc}(\infty))}\abs{x}}.
\ee
Thus, we conclude that
\be
A_0^N=I_0^{N+1}\overset{\xi_+>0}{=}-\varepsilon^{\xi_+}_{N+1}+v^{\xi_+}_{\rm
xc}(\infty)
\ee
and
\be
I_0^{N}\overset{\xi_-\geq0}{=}-\varepsilon^{\xi_-}_{N}+v^{\xi_-}_{\rm
xc}(\infty).
\ee
\section{Derivation of the eDMHF equations}\label{appendix:eDMHF_eqs}

For convenience, we use the following exponential parameterization of
the single-configuration wave functions~\cite{helgaker2014molecular},   
\be\label{eq:exp_param}
\ket{\Phi_I}\equiv \ket{\Phi_I(\bfkap)}=e^{-\hat{\kappa}}\ket{\overline{\Phi}^\bfw_I},
\ee
where $\bfkap\equiv \left\{\kappa_{pq}\right\}_{p<q}$ are the
variational orbital
rotation parameters and $\hat{\kappa}$
is the corresponding real singlet rotation quantum operator. The latter reads as
follows in second quantization,
\be\label{eq:kappa_op}
\hat{\kappa}=\sum_{p<q}\kappa_{pq}(\hat{E}_{pq}-\hat{E}_{qp})=-\hat{\kappa}^\dagger,
\ee
where the index $p$ refers to the orbital $\overline{\varphi}^\bfw_p$ and
$\hat{E}_{pq}=\sum_{\tau=\uparrow,\downarrow}\hat{a}_{p\tau}^\dagger\hat{a}_{q\tau}$.
Therefore, the eDMHF energy becomes a function of $\bfkap$,
\be
E^\bfw_{\rm eDMHF}(\bfkap)=E_{\rm HF}\left(\bfD^\bfw(\bfkap)\right),
\ee
where $\bfD^\bfw(\bfkap)=\sum_I\ttw_I\bfD^{\Phi_I(\bfkap)}$ is a trial
ensemble density matrix, and $E_{\rm
HF}(\bfD)$ is the conventional ground-state HF density matrix functional
energy:
\be
E_{\rm
HF}(\bfD)=\sum_{mk}h_{mk}D_{mk}+\dfrac{1}{2}\sum_{klmn}\left(\bra{mn}\ket{kl}-\dfrac{1}{2}\bra{mn}\ket{lk}\right)D_{mk}D_{nl}.
\ee
By construction, the minimum is reached when $\bfkap=0$, and we denote
$\bfD^\bfw=\bfD^\bfw(\bfkap=0)$. Note that
\be\label{eqapp:integer_ON}
D^{\overline{\Phi}^\bfw_I}_{pq}=\mel{\overline{\Phi}^\bfw_I}{\hat{E}_{pq}}{\overline{\Phi}^\bfw_I}=\delta_{pq}n^I_{p},
\ee
where the occupation number $n^I_{p}$ is an {\it integer},
and
\be\label{eqapp:frac_ON}
D^\bfw_{pq}=\sum_I\ttw_ID^{\overline{\Phi}^\bfw_I}_{pq}=\delta_{pq}\sum_I\ttw_In^I_{p}=\delta_{pq}\theta^\bfw_p,
\ee
where $\theta^\bfw_p$ can be {\it fractional}. The
stationarity condition that is fulfilled by the minimizing eDMHF
orbitals can now be written explicitly as follows,
\be\label{eqapp:stat_cond_eDMHF}
\left.
\dfrac{\partial E^\bfw_{\rm eDMHF}(\bfkap)}{\partial \kappa_{pq}}
\right|_{\bfkap=0}
=\sum_{rs}
\left.\dfrac{\partial D^\bfw_{rs}(\bfkap)}{\partial \kappa_{pq}}\right|_{\bfkap=0}
\left.\dfrac{\partial E_{\rm HF}\left(\bfD\right)}{\partial D_{rs}}\right|_{\bfD=\bfD^\bfw}
=0,
\ee
where
\be
\begin{split}
\dfrac{\partial E_{\rm HF}\left(\bfD\right)}{\partial D_{rs}}&=h_{rs}+
\dfrac{1}{2}\sum_{nl}\left(\bra{rn}\ket{sl}-\dfrac{1}{2}\bra{rn}\ket{ls}\right)D_{nl}
\\
&\quad
+\dfrac{1}{2}\sum_{mk}\left(\bra{mr}\ket{ks}-\dfrac{1}{2}\bra{mr}\ket{sk}\right)D_{mk}
\\
&=h_{rs}+\sum_{nl}\left(\bra{rn}\ket{sl}-\dfrac{1}{2}\bra{rn}\ket{ls}\right)D_{nl}
\\
&\equiv f_{rs}(\bfD)
\end{split}
\ee
is the conventional density matrix functional Fock operator matrix
element, and
\be\label{eqapp:deriv_1RDM_kappa}
\begin{split}
\left.\dfrac{\partial D^\bfw_{rs}(\bfkap)}{\partial \kappa_{pq}}
\right|_{\bfkap=0}&=\sum_I\ttw_I
\expval{\left[\hat{E}_{pq}-\hat{E}_{qp},\hat{E}_{rs}\right]}_{\Phi^\bfw_I}
\\
&=\sum_I\ttw_I\left(\delta_{qr}\delta_{ps}n^I_{p}-\delta_{ps}\delta_{qr}n^I_{q}-\delta_{pr}\delta_{qs}n^I_{q}+\delta_{qs}\delta_{pr}n^I_{p}\right)
\\
&=\left(\delta_{qr}\delta_{ps}+\delta_{pr}\delta_{qs}\right)\sum_I\ttw_I\left(n^I_{p}-n^I_{q}\right)
\\
&=\left(\delta_{qr}\delta_{ps}+\delta_{pr}\delta_{qs}\right)(\theta^\bfw_p-\theta^\bfw_q),
\end{split}
\ee
where we used the relation
$[\hat{E}_{pq},\hat{E}_{rs}]=\delta_{qr}\hat{E}_{ps}-\delta_{ps}\hat{E}_{rq}$
(see Ref.~\cite{helgaker2014molecular})
with Eqs.~(\ref{eqapp:integer_ON}) and (\ref{eqapp:frac_ON}). If we
denote $f^\bfw_{rs}=f_{rs}(\bfD^\bfw)$, Eq.~(\ref{eqapp:stat_cond_eDMHF}) can be written in a
compact form as follows,
\be
\begin{split}
(\theta^\bfw_p-\theta^\bfw_q)\sum_{rs}\left(\delta_{qr}\delta_{ps}+\delta_{pr}\delta_{qs}\right)f^\bfw_{rs}=0,
\end{split}
\ee
thus leading to the final result: 
\be
\left(\theta_p^\bfw-\theta_q^\bfw\right)f^\bfw_{qp}=0.
\ee


\section{Derivation of the SAHF equations}\label{appendix:SAHF_eqs}

We use the same parameterization as in Appendix~\ref{appendix:eDMHF_eqs}, \ie,
\be\label{eq:exp_param_SAHF}
\ket{\Phi_I}\equiv \ket{\Phi_I(\bfkap)}=e^{-\hat{\kappa}}\ket{\tilde{\Phi}^\bfw_I},
\ee
where the indices $\{p\}$ in creation/annihilation operators (as well as
in one- and two-electron integrals) now refer
to the minimizing SAHF orbitals
$\left\{\tilde{\varphi}^\bfw_p\right\}$. The to-be-minimized SAHF energy
can be expressed as follows,
\be
E^\bfw_{\rm SAHF}(\bfkap)=\sum_I\ttw_I\left(\sum_{rs}h_{rs}D^{\Phi_I(\bfkap)}_{rs}
+E_{\rm
H}[n_{\Phi_I(\bfkap)}]+\mathcal{E}^I_{\rm x}\left[{\bf D}^{\Phi_I(\bfkap)}\right]
\right),
\ee
so that the stationarity condition reads as
\be\label{eqapp:stat_cond_SAHF}
\begin{split}
&\left.\dfrac{\partial E^\bfw_{\rm SAHF}(\bfkap)}{\partial
\kappa_{pq}}\right|_{\bfkap=0}
=0
\\
&=
\sum_I\ttw_I\left[\sum_{rs}\left(h_{rs}+\left[v^{\bfw}_{{\rm
x},I}\right]_{rs}\right)\dfrac{\partial D_{rs}^{{\Phi}_I(\bfkap)}}{\partial
\kappa_{pq}}
+\int d\bfr\, v_{\rm
H}[n_{\tilde{\Phi}^\bfw_I}](\bfr)\dfrac{\partial
n_{{\Phi}_I(\bfkap)}(\bfr)}{\partial
\kappa_{pq}}
\right]_{\bfkap=0},
\end{split}
\ee
where $\left[v^{\bfw}_{{\rm x},I}\right]_{rs}\equiv\left.
\partial \mathcal{E}^I_{\rm x}\left[{\bf D}\right]/\partial
D_{rs}\right|_{\bfD=\bfD^{\tilde{\Phi}^\bfw_I}}$ and $v_{\rm H}[n](\bfr)=\delta E_{\rm
H}[n]/\delta n(\bfr)$.
Note that the individual densities are recovered from the density
matrices as follows,
\be
n_{\Phi_I(\bfkap)}(\bfr)=\gamma^{\Phi_I(\bfkap)}(\bfr,\bfr)=\sum_{rs}\tilde{\varphi}^\bfw_r(\bfr)\tilde{\varphi}^\bfw_s(\bfr)D^{\Phi_I(\bfkap)}_{rs}.
\ee
Therefore, if we use the notation
\be
\mel{\tilde{\varphi}^\bfw_r}{\hat{h}+\hat{v}^{\bfw}_{{\rm
Hx},I}}{\tilde{\varphi}^\bfw_s}=h_{rs}
+\int d\bfr\,\tilde{\varphi}^\bfw_r(\bfr) v_{\rm
H}[n_{\tilde{\Phi}^\bfw_I}](\bfr)\tilde{\varphi}^\bfw_s(\bfr)
+\left[v^{\bfw}_{{\rm
x},I}\right]_{rs}
,
\ee
Eq.~(\ref{eqapp:stat_cond_SAHF}) can be rewritten in a compact form as follows, 
\be
\sum_I\ttw_I\sum_{rs}\mel{\tilde{\varphi}^\bfw_r}{\hat{h}+\hat{v}^{\bfw}_{{\rm
Hx},I}}{\tilde{\varphi}^\bfw_s}\left.\dfrac{\partial D_{rs}^{{\Phi}_I(\bfkap)}}{\partial
\kappa_{pq}}\right|_{\bfkap=0}=0.
\ee
We conclude from Eq.~(\ref{eqapp:deriv_1RDM_kappa}) that
\be
\begin{split}
0&=\sum_I\ttw_I\left(n^I_{p}-n^I_{q}\right)\sum_{rs}\left(\delta_{qr}\delta_{ps}+\delta_{pr}\delta_{qs}\right)
\mel{\tilde{\varphi}^\bfw_r}{\hat{h}+\hat{v}^{\bfw}_{{\rm
Hx},I}}{\tilde{\varphi}^\bfw_s}
\\
&=2\sum_I\ttw_I\left(n^I_{p}-n^I_{q}\right)\mel{\tilde{\varphi}^\bfw_p}{\hat{h}+\hat{v}^{\bfw}_{{\rm
Hx},I}}{\tilde{\varphi}^\bfw_q},
\end{split}
\ee
thus leading to the final result:
\be
(\theta^\bfw_p-\theta^\bfw_q)\mel{\tilde{\varphi}^\bfw_p}{\hat{h}
}{\tilde{\varphi}^\bfw_q}+\sum_I\ttw_I\left(n^I_{p}-n^I_{q}\right)\mel{\tilde{\varphi}^\bfw_p}{\hat{v}^{\bfw}_{{\rm
Hx},I}}{\tilde{\varphi}^\bfw_q}=0.
\ee

\section{Exact DD ensemble correlation energy in the Hubbard dimer}
\label{app:DD_corr_ener_derivation}

For convenience, we will use the following exact expression for the ensemble DD
correlation energy:
\be\label{eq_app:ens_DD_corr_ener_general}
E_{\rm c}^{\ttw,{\rm DD}}(n^\ttw)=-(1-\ttw)^2\ttw\dfrac{\partial
f_0^\ttw(n^\ttw)}{\partial\ttw}+\ttw^2(1-\ttw)\dfrac{\partial 
f_1^\ttw(n^\ttw)}{\partial\ttw}.
\ee
The individual Hx-only GOK energies are extracted from the
ensemble one,
\be
f^\xi(n)=&-2t\sqrt{(1-\xi)^2-(1-n)^2}
+\dfrac{U}{2}\left[1+\xi-\dfrac{(3\xi-1)(1-n)^2}{(1-\xi)^2}\right],
\ee
as follows,
\be\label{eq:exact_f0}
f^\ttw_{0}\left(n^\ttw\right)&=&f^\ttw\left(n^\ttw\right)-\ttw
\left.\dfrac{\partial f^\xi\left(n^{\xi,\ttw}\right)}{\partial
\xi}\right|_{\xi=\ttw},
\ee
and
\be
\label{eq:exact_f1}
f^\ttw_{1}\left(n^\ttw\right)
&=&
f^\ttw\left(n^\ttw\right)+(1-\ttw)\left.\dfrac{\partial f^\xi\left(n^{\xi,\ttw}\right)}{\partial
\xi}\right|_{\xi=\ttw},
\ee
where
\be
n^{\xi,\ttw}=(1-\xi) n_{\Phi_{0}^{\ttw}}+ \xi n_{\Phi_{1}^{\ttw}}.
\ee
Since $n_{\Phi_{1}^{\ttw}}=1$ and
\be
(1-\ttw)n_{\Phi_{0}^{\ttw}}+\ttw n_{\Phi_{1}^{\ttw}}=n^\ttw,
\ee
or, equivalently,
\be
n_{\Phi_{0}^{\ttw}}=\dfrac{n^\ttw-\ttw}{(1-\ttw)},
\ee
it comes 
\be
n^{\xi,\ttw}=(1-\xi)\frac{(n^{\ttw}-\ttw)}{1-\ttw}+ \xi
\ee
and
\be
\left.\dfrac{\partial n^{\xi,\ttw}}{\partial
\xi}\right|_{\xi=\ttw}=1-\frac{(n^{\ttw}-\ttw)}{1-\ttw}=\dfrac{1-n^\ttw}{1-\ttw}.
\ee
From the weight derivative expression
\be
\begin{split}
\left.\dfrac{\partial f^\xi\left(n^{\xi,\ttw}\right)}{\partial
\xi}\right|_{\xi=\ttw}=\left.\dfrac{\partial f^\xi(n^\ttw)}{\partial
\xi}\right|_{\xi=\ttw}+\left.\dfrac{\partial n^{\xi,\ttw}}{\partial
\xi}\right|_{\xi=\ttw}\times\left.\dfrac{\partial f^\ttw(n)}{\partial
n}\right|_{n=n^\ttw},
\end{split}
\ee
where
\be
\dfrac{\partial f^\xi(n)}{\partial \xi}=\dfrac{2t(1-\xi)}{\sqrt{(1-\xi)^2-(1-n)^2}}+\dfrac{U}{2}\left[1-\dfrac{(n-1)^2(1+3\xi)}{(1-\xi)^3}\right]
\ee
and
\be
\dfrac{\partial f^\ttw(n)}{\partial
n}=\dfrac{2t(n-1)}{\sqrt{(1-\ttw)^2-(1-n)^2}}+U\dfrac{(3\ttw-1)(1-n)}{(1-\ttw)^2},
\ee
thus leading to
\be
\begin{split}
\left.\dfrac{\partial f^\xi\left(n^{\xi,\ttw}\right)}{\partial
\xi}\right|_{\xi=\ttw}
&=\dfrac{2t(1-\ttw)}{\sqrt{(1-\ttw)^2-(1-n^\ttw)^2}}
\\
&\quad-\dfrac{2t(1-n^\ttw)^2}{(1-\ttw)\sqrt{(1-\ttw)^2-(1-n^\ttw)^2}}
\\
&\quad+
\dfrac{U}{2}\left[1-\dfrac{(n^\ttw-1)^2(1+3\ttw)}{(1-\ttw)^3}\right]
\\
&\quad
+U\dfrac{(3\ttw-1)(1-n^\ttw)^2}{(1-\ttw)^3},
\end{split}
\ee
or, equivalently,
\be
\left.\dfrac{\partial f^\xi\left(n^{\xi,\ttw}\right)}{\partial
\xi}\right|_{\xi=\ttw}=\dfrac{2t\sqrt{(1-\ttw)^2-(1-n^\ttw)^2}}{(1-\ttw)}
+\dfrac{U}{2}\left[1-\dfrac{3(1-n^\ttw)^2}{(1-\ttw)^2}\right],
\ee
it comes 
\be
f^\ttw_{0}\left(n^\ttw\right)=-\dfrac{2t\sqrt{(1-\ttw)^2-(1-n^\ttw)^2}}{(1-\ttw)}+\dfrac{U}{2}\left[1+\dfrac{(1-n^\ttw)^2}{(1-\ttw)^2}\right]
\ee
and
\be
f^\ttw_{1}\left(n^\ttw\right)=U\left[1-\dfrac{(1-n^\ttw)^2}{(1-\ttw)^2}\right].
\ee
As a result,
\be
\dfrac{\partial
f_0^\ttw(n^\ttw)}{\partial\ttw}=\dfrac{2t(n^\ttw-1)(n_{\Psi_1}-1)}{(1-\ttw)^2\sqrt{(1-\ttw)^2-(1-n^\ttw)^2}}+U\dfrac{(n^\ttw-1)(n_{\Psi_1}-1)}{(1-\ttw)^3}
\ee
and
\be
\dfrac{\partial
f_1^\ttw(n^\ttw)}{\partial\ttw}=-\dfrac{2U(n^\ttw-1)(n_{\Psi_1}-1)}{(1-\ttw)^3},
\ee
which leads, according to Eq.~(\ref{eq_app:ens_DD_corr_ener_general}),
to the final compact expression  
\be\label{eq:final_DD_ens_corr_ener_dimer_app}
\begin{split}
E_{\rm c}^{\ttw,{\rm
DD}}(n^\ttw)&=-\ttw(n^\ttw-1)(n_{\Psi_1}-1)
\\
&\quad\times
\left[
\dfrac{2t}{\sqrt{(1-\ttw)^2-(1-n^\ttw)^2}}+\dfrac{U(1+\ttw)}{(1-\ttw)^2}
\right].
\end{split}
\ee

\end{appendices}


\bibliographystyle{spphys}       


\newcommand{\Aa}[0]{Aa}

\end{document}